\documentclass[aps,pra,twocolumn,showpacs,showkeys]{revtex4-1}

\usepackage{amssymb}
\usepackage{graphicx}
\usepackage{placeins}
\usepackage{float}
\usepackage{subfig}
\usepackage{color}

\begin{document}

\title{How the electromagnetic field evolves in the presence of a mobile membrane}

\author{L. O. Casta\~{n}os$^{1*}$}
\author{R. Weder$^{2}$}
\email[1.]{ \ LOCCJ@yahoo.com}
\email[2.]{ \ weder@unam.mx}
\affiliation{1. Instituto Tecnol\'{o}gico y de Estudios Superiores de Monterrey, Campus Santa Fe, C.P. 01389, M\'{e}xico DF, M\'{e}xico.\\
2. Departamento de F\'{i}sica Matem\'{a}tica, Instituto de Investigaciones en Matem\'{a}ticas Aplicadas y en Sistemas, Universidad Nacional Aut\'{o}noma de M\'{e}xico, Apartado Postal 20-126, M\'{e}xico DF 01000, M\'{e}xico.}

\date{\today}

\begin{abstract}
We consider a one-dimensional cavity composed of two perfect, fixed mirrors and a mobile membrane in between.  Assuming that the membrane starts to move from rest and that the membrane moves appreciably in a time-scale much larger than the time-scale in which the cavity electromagnetic field evolves appreciably, we derive simple analytic formulas that describe to good approximation the evolution of the field and that provide an intuitive physical picture. These formulas take into account the position, velocity, and acceleration of the membrane and are valid for arbitrarily large displacements of the membrane along the cavity axis. Also, we deduce the conditions under which the field can be described to good approximation by a single instantaneous mode. 
\end{abstract}



\maketitle


\section{Introduction}

Optomechanics studies systems composed of a mechanical oscillator, such as a mobile mirror or membrane, and an electromagnetic field interacting by means of radiation pressure or thermal forces \cite{Marquardt}. Some of the main attractions of this area come from the facts that it provides systems that can interact with other quantum systems such as cold atoms \cite{Tech}, that can serve as microwave to optical converters in quantum information networks \cite{Andrews}, and that are very promising to study the quantum physics of macroscopic objects \cite {Marquardt}. In order to do the latter, one needs to create and manipulate quantum states of motion in the mechanical oscillator. In particular, the mechanical oscillator has already been prepared in its quantum ground state of motion in various optomechanical systems either by cryogenic cooling \cite{OConnell, Meenehan,Riedinger} or laser cooling using cavity fields \cite{Chan,Teufel}, and it has also been prepared in a squeezed state \cite{SqueezedM1, SqueezedM2, SqueezedM3}. Moreover, some experimental set ups have demonstrated single-phonon control in the mechanical oscillator using a superconducting qubit \cite{OConnell}, non-classical correlations with longer lifetimes between single photons and phonons from the mechanical oscillator \cite{Riedinger}, the generation of entanglement between a microwave field and a mechanical oscillator \cite{Entanglement}, and the generation of quantum squeezed states of light \cite{SqueezedO1, SqueezedO2}. Some of the technical difficulties of these experiments consist in the production of quantum states with sufficiently long lifetimes and that they operate at very low temperatures. The situation is now more encouraging because mechanical resonators for optomechanics that may allow studies in the quantum regime and at room temperature have recently been developed \cite{Room}.

Optomechanical systems are usually described in terms of a harmonic oscillator model \cite{Marquardt} where the mechanical oscillator can only have small displacements around an equilibrium position. Nevertheless, there are works that go beyond this and describe other physical phenomena arising from coupling the electromagnetic field with the motional degrees of freedom of the mechanical oscillator, see \cite{Law1,Law2,Scattering,PhysicaScripta,PRA,LAOP1,LAOP2,CS} and references therein. In particular, \cite{Scattering} develops a scattering theory that allows the description of a point-like scatterer (e.g., an atom or mobile mirror with infinitesimal thickness) coupled to the electromagnetic field through radiation pressure, while \cite{PhysicaScripta} presented a model from first principles to describe a mobile mirror with non-zero thickness interacting with the electromagnetic field. In this article we use the model of \cite{PhysicaScripta} to describe a one dimensional cavity composed of two perfect, fixed mirrors and a mobile membrane with non-zero thickness in between. This set up is frequently referred to in the literature as the \textit{membrane-in-the-middle optomechanical set up} \cite{Thompson}. In fact, it is one of the paradigmatic optomechanical systems because it has been noted to combine good properties of the cavity (e.g., highly reflective mirrors) with good properties of the mechanical oscillator (e.g., a very thin, light membrane that can be moved easily by radiation pressure). The objective of this work is to determine how the cavity electromagnetic field evolves in the presence of the mobile membrane. This has the purpose of improving and extending the understanding of the electromagnetic field in the presence of moving boundaries. In particular, the work has the novelties of using a model deduced from first principles, that the membrane is not restricted to small deviations around an equilibrium position, that we establish how, why, and when the electromagnetic field follows adiabatically a single cavity mode, and that we deduce and give a physical picture of how and why other cavity modes are excited. Moreover, we obtain simple analytic formulas that describe the cavity field to good approximation and that provide physical insight to its evolution. We note that some results were presented without giving any proofs in the conference \cite{LAOP2}.

The article is organized as follows. In Section II we present the physical system under consideration and we establish the equation governing the evolution of the electromagnetic field. It is a wave equation modified by terms that arise from the fact that the membrane's properties are altered when it is in motion. In Section III we review the evolution of the field when the membrane is fixed and, in particular, we introduce the modes of the cavity. In Sections IV and V we determine to good approximation the evolution of the field when the membrane can move and when only one mode is initially excited. In Section VI we extend the results to the case of several initially excited modes. Finally, a summary and the conclusions are given in Section VII. 


\section{The model}

\begin{figure}
\includegraphics[width=8cm]{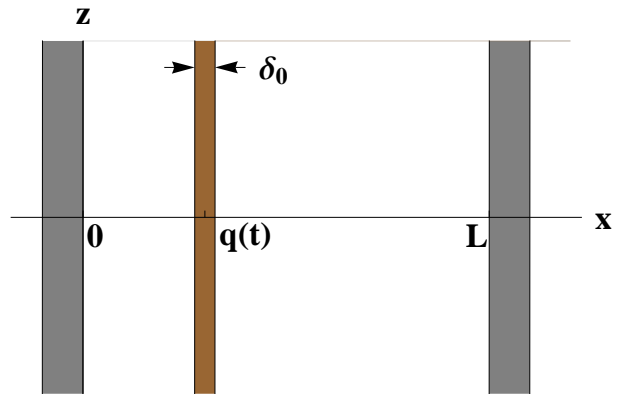}
\caption{\label{Figure1} Schematic representation of the system under consideration. The outer slabs are the perfect, fixed mirrors of the cavity, while the slab in the middle is the membrane. The interior of the cavity goes from $x=0$ to $x=L$ where the boundaries of the mirrors are located. The midpoint of the membrane along the $x$-axis at time $t$ is $q(t)$ and the membrane has thickness $\delta_{0}$. The $y$-axis points into the page.}
\end{figure}

We consider a one-dimensional cavity composed of two perfect, fixed mirrors and a mobile membrane in between, see Figure \ref{Figure1}. Here we are assuming that the mirrors and the membrane are slabs of infinite length and width that are parallel to each other and that there is vacuum between the membrane and the mirrors. This system is frequently referred to in the literature as the \textit{membrane-in-the-middle} optomechanical set up \cite{Thompson}. To describe it we consider a coordinate system such that the $x$-axis is perpendicular to the faces of the mirrors and the membrane, the interior of the cavity extends from $x=0$ to $x=L$ where the boundaries of the perfect mirrors are located, and the midpoint of the membrane along the $x$-axis at time $t$ is denoted by $q(t)$, see Figure \ref{Figure1}. In the rest of the article we refer to $q(t)$ as the \textit{position of the membrane} and we assume that the membrane is a linear, isotropic, non-magnetizable, non-conducting, and uncharged dielectric of thickness $\delta_{0}$ when it is at rest. Moreover, the electric susceptibility of the membrane is denoted by $\chi [x-q(t)]$ and we assume that it is piecewise continuous with a piecewise continuous derivative, non-negative, and that it vanishes outside of the membrane, that is, we assume that 
\begin{eqnarray}
\label{unoChi}
\chi[x-q(t)] \ = \ 0 \qquad \mbox{if} \ \vert x-q(t) \vert \geq \frac{\delta_{0}}{2} \ .
\end{eqnarray}
We emphasize that the membrane is free to move throughout the interior of the cavity, but only along the $x$-axis. In particular, the membrane is not restricted to small displacements around an equilibrium position. The motion can be due to an external agent or to the radiation pressure exerted by an electromagnetic field inside the cavity.

Inside the cavity there is an electromagnetic field and we use Gaussian units to describe it. We assume that it can be deduced from vector and scalar potentials of the form
\begin{eqnarray}
\label{potenciales}
\mathbf{A}(x,t) \ = \ A_{0}(x,t)\mathbf{\hat{z}} \ , \ \ V(x,t) \ = \ 0 \ ,
\end{eqnarray}
where $\mathbf{\hat{z}}$ is a unit vector in the direction of the positive $z$-axis. Hence, the electric and magnetic fields are given by 
\begin{eqnarray}
\label{Campos}
\mathbf{E}(x,t) &=&  -\frac{1}{c}\frac{\partial \mathbf{A}}{\partial t}(x,t) \ = \ -\frac{1}{c}\frac{\partial A_{0}}{\partial t}(x,t)\mathbf{\hat{z}} \ , \cr
&& \cr
\mathbf{B}(x,t) &=& \nabla \times \mathbf{A}(x,t) \ = \ -\frac{\partial A_{0}}{\partial x}(x,t)\mathbf{\hat{y}} \ .
\end{eqnarray} 
Here $c$ is the speed of light in vacuum and $\mathbf{\hat{x}}$ and $\mathbf{\hat{y}}$ are unit vectors in the direction of the positive $x$ and $y$ axes, respectively.

We now introduce the equation governing the dynamics of the potential $A_{0}(x,t)$. In order to do this we first assume that
\begin{enumerate}
\item $\lambda_{0}$ is the characteristic wavelength of the electromagnetic field,
\item $\nu_{0}$ is the characteristic frequency of the field and satisfies $\lambda_{0}\nu_{0} = c$,
\item $A_{00}$ is a characteristic value of $A_{0}(x,t)$,
\item $\nu_{\mbox{\tiny osc}}^{-1}$ is the time-scale in which $q(t)$ varies appreciably. 
\end{enumerate}
In the rest of the article we measure lengths in units of $\lambda_{0}$ and time in units of $\nu_{0}^{-1}$. Hence, the nondimensional position $\xi$ and time $\tau$ are given by
\begin{eqnarray}
\label{XTadimensional}
\xi = \frac{x}{\lambda_{0}} \ , \ \ \tau = \nu_{0}t \ .
\end{eqnarray}
These quantities allow one to introduce the following nondimensional quantities:
\begin{eqnarray}
\label{Adimensionales}
\tilde{A}_{0}(\xi , \tau ) &=& \frac{A_{0}(\lambda_{0}\xi , \nu_{0}^{-1}\tau )}{A_{00}} \ , \cr
&& \cr
\tilde{q}(\tau ) &=& \frac{q(\nu_{0}^{-1}\tau )}{\lambda_{0}} \ , \cr
&& \cr
\xi_{L} &=& \frac{L}{\lambda_{0}} \ , \cr
&& \cr
\tilde{\delta}_{0} &=& \frac{\delta_{0}}{\lambda_{0}} \ , \cr
&& \cr
\tilde{\chi}\left[ \xi -\tilde{q}(\tau) \right] &=& \chi \left\{ \lambda_{0}\left[ \xi -\tilde{q}(\tau) \right] \right\} \ , \cr
&& \cr
\tilde{\epsilon}\left[ \xi -\tilde{q}(\tau) \right] &=& 1 + 4\pi\tilde{\chi}\left[ \xi -\tilde{q}(\tau) \right] \ , \cr
&& \cr
\epsilon_{\mbox{\tiny pert}} &=& \frac{\nu_{\mbox{\tiny osc}}}{\nu_{0}} \ .
\end{eqnarray}
First observe that $\tilde{A}_{0}(\xi , \tau )$, $\tilde{q}(\tau )$, $\xi_{L}$, and $\tilde{\delta}_{0}$ are the non-dimensional potential, position of the membrane, length of the cavity, and thickness of the membrane, respectively. Moreover, recall that $\chi [ x - q(t)]$ is the electric susceptibility of the membrane and that we have assumed that it vanishes outside of the membrane, see (\ref{unoChi}). Hence, $\tilde{\chi}\left[ \xi -\tilde{q}(\tau) \right]$ and $\tilde{\epsilon}\left[ \xi -\tilde{q}(\tau) \right]$ are the electric susceptibility and the dielectric function of the membrane with a non-dimensional argument $[ \xi -\tilde{q}(\tau) ]$, respectively. 

We now discuss the interpretation of the parameter $\epsilon_{\mbox{\tiny pert}}$ in (\ref{Adimensionales}). Observe that the time-scale in which the electromagnetic field evolves appreciably can be taken to be $1/\nu_{0}$. Therefore, $\epsilon_{\mbox{\tiny pert}}$ is the time-scale $1/\nu_{0}$ in which the field evolves appreciably divided by the time-scale $1/\nu_{\mbox{\tiny osc}}$ in which the membrane moves appreciably. Since the electromagnetic field usually evolves more rapidly than the membrane, one usually has $1/\nu_{0} \ll 1/\nu_{\mbox{\tiny osc}}$ or, equivalently,  $0 \leq \epsilon_{\mbox{\tiny pert}} = \nu_{\mbox{\tiny osc}}/\nu_{0} \ll 1$. For example, using the parameters of the membrane-in-the-middle optomechanical set up in \cite{Thompson} one has $\nu_{0} = 2.82 \times 10^{14}$ Hz, $\nu_{\mbox{\tiny osc}} = 1.34 \times 10^{5}$ Hz, and, consequently, $\epsilon_{\mbox{\tiny pert}} =4.8 \times 10^{-10} $. 

Before proceeding it is important to give an interpretation of the first and second derivatives of $\tilde{q}(\tau)$ with respect to $\tau$. From the fact that $\lambda_{0}\nu_{0} = c$ and from (\ref{Adimensionales}) one has
\begin{eqnarray}
\label{Adimensionales2}
\tilde{q}'(\tau) &=& \frac{\dot{q}(\nu_{0}^{-1}\tau)}{c} \ , \cr
\tilde{q}''(\tau) &=& \frac{\ddot{q}(\nu_{0}^{-1}\tau)}{c\nu_{0}} \ . 
\end{eqnarray}
Here the $\tilde{q}'$ and $\tilde{q}''$ denote the first and second derivatives of $\tilde{q}$ with respect to its argument $\tau$, while $\dot{q}$ and $\ddot{q}$ denote the first and second derivatives of $q$ with respect to its argument $t = \nu_{0}^{-1}\tau$. In all the article $f'(x)$ denotes the derivative of $f(x)$ with respect to $x$. For example, $\omega_{N}'[\tilde{q}(\tau)]$ denotes the derivative of $\omega_{N}(\tilde{q})$ with respect to $\tilde{q}$ and evaluated at $\tilde{q}(\tau)$. 

Notice that (\ref{Adimensionales2}) indicates that $\tilde{q}'(\tau)$ is the velocity of the membrane divided by the speed of light in vacuum, while $\tilde{q}''(\tau)$ is the acceleration of the membrane divided by the product of the speed of light in vacuum times the characteristic frequency of the electromagnetic field. With this in mind, in the following we assume that $\tilde{q}'(\tau)$ and $\tilde{q}''(\tau)$ are quantities whose absolute value is small (that is, much less than $1$). In other words, we assume that the membrane has a speed much smaller than that of light in vacuum and that the membrane has an acceleration whose absolute value is much smaller than the product of the speed of light in vacuum times the characteristic frequency of the electromagnetic field. Notice that this is compatible with the usual situation where \ $\epsilon_{\mbox{\tiny pert}} \ll 1$.

Using a relativistic treatment, \cite{PhysicaScripta} established a general equation that governs the dynamics of $\tilde{A}_{0}(\xi,\tau)$ for all possible values of the velocity and acceleration of the membrane. In particular, it also showed that, to first order in $\tilde{q}'(\tau)$ and $\tilde{q}''(\tau)$, that equation is given by     
\begin{eqnarray}
\label{8}
\frac{\partial^{2}\tilde{A}_{0}}{\partial \xi^{2}}(\xi , \tau ) 
&=& \tilde{\epsilon}\left[ \xi - \tilde{q}(\tau) \right] \frac{\partial^{2}\tilde{A}_{0}}{\partial \tau^{2}}(\xi , \tau ) \cr
&& +8\pi \tilde{q}'(\tau )\tilde{\chi}\left[ \xi - \tilde{q}(\tau) \right] \frac{\partial^{2}\tilde{A}_{0}}{\partial \xi \partial \tau}(\xi , \tau ) \cr
&& +4\pi \tilde{q}''(\tau )\tilde{\chi}\left[ \xi - \tilde{q}(\tau) \right] \frac{\partial \tilde{A}_{0}}{\partial \xi }(\xi , \tau ) \ .
\end{eqnarray}
We note that \cite{Law2} and some references therein also present (\ref{8}) for the special case of a constant electric susceptibility. Notice that $\tilde{A}_{0}(\xi , \tau)$ satisfies a wave equation with a non-dimensional position- and time-dependent coefficient $\tilde{\epsilon}\left[ \xi - \tilde{q}(\tau) \right]$ and that is modified with terms proportional to the non-dimensional velocity $\tilde{q}'(\tau)$ and acceleration $\tilde{q}''(\tau)$ of the membrane. In (\ref{8}), the (non-dimensional) position of the membrane $\tilde{q}(\tau)$ can be determined either by an external agent or by an equation involving the radiation pressure exerted by the field \cite{PhysicaScripta}. Our results are valid for both cases because further below we assume that $\epsilon_{\mbox{\tiny pert}} \ll 1$ and this allows us to solve (\ref{8}) using the method of multiple scales \cite{Holmes}.

In order to take into account the perfect, fixed mirrors of the cavity, one must supplement the equation for $\tilde{A}_{0}(\xi ,\tau)$ in (\ref{8}) with the following boundary conditions:
\begin{eqnarray}
\label{BCo}
\tilde{A}_{0}(0,\tau) \ = \ 0 \ , \ \ \tilde{A}_{0}(\xi_{L},\tau) \ = \ 0 \ .
\end{eqnarray}
Recall that $\xi_{L}$ is the length $L$ of the cavity in units of the characteristic wavelength $\lambda_{0}$ of the electromagnetic field, see (\ref{Adimensionales}). The boundary conditions in (\ref{BCo}) come from applying the usual boundary conditions for electric fields at interfaces between different media \cite{Jackson}. Explicitly, they come from the following two requirements: (i) the electric field $\mathbf{E}(x,t)$ given in (\ref{Campos}) must vanish inside the perfect mirrors, and (ii) the component of the electric field tangent to the surface of each mirror must be continuous across the boundary of each mirror. 

In the next sections we solve the boundary value problem (\ref{8})-(\ref{BCo}) to good approximation. 


\section{Evolution with a fixed membrane}

In this section we briefly review the evolution of the field in the case where the membrane is fixed and, in particular, we introduce the modes of the cavity. The results of this section are important because they will be compared with those of the next sections where the case of a mobile membrane is considered. 

In this section and only this section we assume that 
\begin{eqnarray}
\label{Pi1}
\tilde{q}(\tau) = \tilde{q}_{0} \ \in \left[ \frac{\tilde{\delta}_{0}}{2} , \ \xi_{L} - \frac{\tilde{\delta}_{0}}{2} \right] \ .
\end{eqnarray}
Notice that $\tilde{q}_{0}$ is fixed in the interval shown in (\ref{Pi1}) because $\tilde{q}_{0}$ is the midpoint of the membrane along the $x$-axis and the membrane has (non-dimensional) thickness $\tilde{\delta}_{0}$.

The equation governing the evolution of $\tilde{A}_{0}(\xi , \tau)$ is obtained from (\ref{8}) by using (\ref{Pi1}). One obtains the well-known wave equation for $\tilde{A}_{0}(\xi , \tau)$:
\begin{eqnarray}
\label{8f}
\frac{\partial^{2}\tilde{A}_{0}}{\partial \xi^{2}}(\xi , \tau ) 
&=& \tilde{\epsilon}\left( \xi - \tilde{q}_{0} \right) \frac{\partial^{2}\tilde{A}_{0}}{\partial \tau^{2}}(\xi , \tau ) \ .
\end{eqnarray}
Solving (\ref{8f}) by separation of variables \cite{Zauderer}, imposing the boundary conditions in (\ref{BCo}), and using the results of Sturm-Liouville theory \cite{Folland, Ross} (in particular, see Chapter  8 of \cite{Folland}) one finds the following results:
\begin{enumerate}

\item The cavity has a countable set of (non-dimensional) angular frequencies \ $\{ \omega_{n}(\tilde{q}_{0}) \}_{n=1}^{+\infty}$ \ with \ $\omega_{n}(\tilde{q}_{0}) > 0$ \ for all $n$ and
\begin{eqnarray}
\label{OrdenModos}
\lim_{n\rightarrow + \infty}\omega_{n}(\tilde{q}_{0}) \ = \ + \infty \ .
\end{eqnarray}
Moreover, the frequencies can be ordered so that $\omega_{m}(\tilde{q}_{0}) < \omega_{n}(\tilde{q}_{0})$ if $m < n$. Observe that the notation indicates that $\omega_{n}(\tilde{q}_{0})$ depends on the (fixed, non-dimensional) position of the membrane $\tilde{q}_{0}$. Also, it can be shown that $\omega_{n}(\tilde{q}_{0})$ are continuously differentiable functions of $\tilde{q}_{0}$, see Appendix A.

\item The cavity has a set of modes $\{ G_{n}(\xi , \tilde{q}_{0}) \}_{n=1}^{+\infty}$ consisting of functions that satisfy the following boundary value problem
\begin{eqnarray}
\label{BVPmodos}
\frac{\partial^{2}G_{n}}{\partial \xi^{2}}(\xi , \tilde{q}_{0}) &=& -\omega_{n}(\tilde{q}_{0})^{2} \tilde{\epsilon}( \xi -\tilde{q}_{0} ) G_{n}(\xi , \tilde{q}_{0}) \ , \cr
&& \cr
G_{n}(0, \tilde{q}_{0}) &=& 0 \ , \cr
&& \cr
G_{n}(\xi_{L}, \tilde{q}_{0}) &=& 0 \ .
\end{eqnarray}
Here the notation indicates that $G_{n}(\xi , \tilde{q}_{0})$ depends on the (fixed, non-dimensional) position of the membrane $\tilde{q}_{0}$ and $\omega_{n}(\tilde{q}_{0})$ is the non-dimensional angular frequency associated with mode $G_{n}(\xi , \tilde{q}_{0})$. Also, it can be shown that $G_{n}(\xi , \tilde{q}_{0})$ are continuously differentiable functions of $\tilde{q}_{0}$, see Appendix A. 

\item The modes $\{ G_{n}(\xi , \tilde{q}_{0}) \}_{n=1}^{+\infty}$ can be chosen to be real-valued functions that satisfy the following orthonormalization relation
\begin{eqnarray}
\label{OrtonormalizacionO}
\int_{0}^{\xi_{L}}d\xi \ \tilde{\epsilon}(\xi - \tilde{q}_{0})G_{n}(\xi , \tilde{q}_{0})G_{m}(\xi , \tilde{q}_{0}) \ = \ \delta_{nm} \cr
&&
\end{eqnarray}
Here and in the following $\delta_{nm}$ is the Kronecker delta (equal to $0$ if $n \not= m$ and equal to $1$ if $n=m$).

\item  Each non-dimensional angular frequency $\omega_{n}(\tilde{q}_{0})$ is non-degenerate. In other words, there is only one linearly independent mode $G_{n}(x,\tilde{q}_{0})$ associated with each $\omega_{n}(\tilde{q}_{0})$.

\end{enumerate}

We expand $\tilde{A}_{0}(\xi ,\tau)$ in terms of the modes $G_{n}(\xi , \tilde{q}_{0})$ of the cavity
\begin{eqnarray}
\label{Pi2}
\tilde{A}_{0}(\xi , \tau ) &=& \sum_{n=1}^{+ \infty} c_{n}(\tau)G_{n}(\xi , \tilde{q}_{0}) \ .
\end{eqnarray}
One can reduce equation (\ref{8f}) for $\tilde{A}_{0}(\xi , \tau )$ to a set of uncoupled equations for the coefficients $c_{n}(\tau)$ by taking the following sequence of steps: (i) substitute the expansion of $\tilde{A}_{0}(\xi , \tau )$ given in (\ref{Pi2}) into the differential equation for $\tilde{A}_{0}(\xi , \tau )$ given in (\ref{8f}), (ii) use the differential equation for the modes in (\ref{BVPmodos}) to eliminate second order derivatives with respect to $\xi$, (iii) multiply the resulting equation by $G_{m}(\xi , \tilde{q}_{0})$, (iv) integrate with respect to $\xi$ from $\xi = 0$ to $\xi = \xi_{L}$ and, finally, (v) use the orthonormalization relation for the modes in (\ref{OrtonormalizacionO}) to get rid of the summations involved. One obtains that
\begin{eqnarray}
\label{Pi2b}
c_{m}''(\tau) + \omega_{m}(\tilde{q}_{0})^{2} c_{m}(\tau) \ = \ 0 \ .
\end{eqnarray}
Observe that these are the well-known independent harmonic oscillator equations for each of the coefficients $c_{m}(\tau)$ of the modes $G_{m}(\xi , \tilde{q}_{0})$ in (\ref{Pi2}). The equations in (\ref{Pi2b}) are easily solved \cite{Ross} to give
\begin{eqnarray}
\label{Pi3}
c_{m}(\tau) &=& \alpha_{1m} e^{-i \omega_{m}(\tilde{q}_{0})\tau } + \alpha_{2m} e^{i \omega_{m}(\tilde{q}_{0})\tau } \ ,
\end{eqnarray}
where $\alpha_{1m}$ and $\alpha_{2m}$ are complex constants that are determined by the initial conditions.

Substituting (\ref{Pi3}) in the expansion of $\tilde{A}_{0}(\xi , \tau )$ given in (\ref{Pi2}) and demanding $\tilde{A}_{0}(\xi ,\tau)$ to be a real quantity, one obtains that 
\begin{eqnarray}
\label{Pi3b}
\alpha_{2m} \ = \ \alpha_{1m}^{*}  \ .
\end{eqnarray}
Here and in the following the complex conjugate of a quantity $g$ is denoted by $g^{*}$. Consequently, 
\begin{eqnarray}
\label{Pi4}
\tilde{A}_{0}(\xi , \tau ) &=& \sum_{n=1}^{+ \infty} \left[ \alpha_{1n} e^{-i \omega_{n}(\tilde{q}_{0})\tau } + \alpha_{1n}^{*} e^{i \omega_{n}(\tilde{q}_{0})\tau } \right] G_{n}(\xi , \tilde{q}_{0})  \cr
&&
\end{eqnarray}

To end this section we assume that only mode $N$ is initially excited, that is, we assume that
\begin{eqnarray}
\label{Pi5}
c_{m}(0) \ = \ g_{0N}\delta_{mN} \ , \ \ c_{m}'(0) \ = \ g_{1N}\delta_{mN} \ . 
\end{eqnarray}
Here $g_{0N}$ and $g_{1N}$ have to be real quantities because, according to (\ref{Pi3}) and (\ref{Pi3b}), the $c_{m}(\tau)$ are real quantities.

Substituting (\ref{Pi3}) into (\ref{Pi5}) one obtains that
\begin{eqnarray}
\label{Pi6}
\alpha_{1m} &=& \left[ \frac{g_{0N}}{2} + \frac{ig_{1N}}{2\omega_{N}(\tilde{q}_{0})} \right] \delta_{mN} \ .
\end{eqnarray}
Notice that the Kronecker delta indicates that only the mode that is initially excited remains excited. 

Substituting the expressions for $\alpha_{1m}$ given in (\ref{Pi6}) into the expansion of $\tilde{A}_{0}(\xi , \tau)$ given in (\ref{Pi4}), one obtains
\begin{eqnarray}
\label{Pi7}
\tilde{A}_{0}(\xi , \tau ) &=& b_{N0} e^{-i \left[ \omega_{N}(\tilde{q}_{0})\tau -\Theta_{N0} \right] } G_{N}(\xi , \tilde{q}_{0}) \ + \ c.c.  \ , \ \ \
\end{eqnarray}
were we have taken $b_{N0}$ and $\Theta_{N0}$ to be the amplitude and a phase of $\alpha_{1N}$, that is, \ $\alpha_{1N} = b_{N0}e^{i\Theta_{N0}}$ \ with
\begin{eqnarray}
\label{Pi8}
b_{N0} &=& \left\vert \frac{g_{0N}}{2} + i \frac{g_{1N}}{2\omega_{N}(\tilde{q}_{0})} \right\vert \ , \cr
&& \cr
\Theta_{N0} &=& \mbox{arg}\left[ \frac{g_{0N}}{2} + i \frac{g_{1N}}{2\omega_{N}(\tilde{q}_{0})} \right] \ .
\end{eqnarray}
Here \textit{arg(z)} is an argument or phase of complex number $z$ and \textit{c.c.} indicates the complex conjugate.


\section{Expansion in terms of the instantaneous modes}

In this section we return to the case of a mobile membrane. First recall that we introduced in Section III the set of (non-dimensional) angular frequencies \ $\{ \omega_{n}(\tilde{q}_{0}) \}_{n=1}^{+\infty}$ of the cavity and the corresponding set of modes $\{ G_{n}( \xi , \tilde{q}_{0}) \}_{n=1}^{+\infty}$ in the case where the membrane is fixed at $\tilde{q}_{0}$. Letting the position $\tilde{q}(\tau)$ of the membrane vary, we obtain the instantaneous (non-dimensional) angular frequencies \ $\{ \omega_{n}[\tilde{q}(\tau) ] \}_{n=1}^{+\infty}$ \ of the cavity and the corresponding set of instantaneous modes \ $\{ G_{n}[ \xi , \tilde{q}(\tau) ] \}_{n=1}^{+\infty}$. In all that follows we refer to the instantaneous modes \ $\{ G_{n}[ \xi , \tilde{q}(\tau) ] \}_{n=1}^{+\infty}$ \ and angular frequencies \ $\{ \omega_{n}[\tilde{q}(\tau)] \}_{n=1}^{+\infty}$ \ of the cavity simply as the modes and frequencies of the cavity. Also, recall from items 1 and 2 in Section IV that $\omega_{n}[\tilde{q}(\tau)]$ and $G_{n}[ \xi , \tilde{q}(\tau) ]$ are continuously differentiable functions of $\tilde{q}(\tau)$. 

One can expand $\tilde{A}_{0}(\xi , \tau)$ in terms of the modes \ $\{ G_{n}[ \xi , \tilde{q}(\tau) ] \}_{n=1}^{+\infty}$ \ as follows:
\begin{eqnarray}
\label{ExpansionEnModos}
\tilde{A}_{0}(\xi , \tau) &=& \sum_{n=1}^{+\infty} c_{n}(\tau ) G_{n}[ \xi , \tilde{q}(\tau) ] \ . \ \ 
\end{eqnarray}

Before proceeding it is convenient to define the following set of mode-dependent quantities:
\begin{eqnarray}
\label{ModeDependent}
\Omega_{mn}\left[ \tilde{q}(\tau ) \right] &=& \int_{0}^{\xi_{L}} d\xi \ \tilde{\epsilon}\left[ \xi - \tilde{q}(\tau) \right] G_{m}\left[ \xi , \tilde{q}(\tau) \right] \times \cr
&& \qquad\qquad \times \frac{\partial G_{n}}{\partial \tilde{q}(\tau )}\left[ \xi , \tilde{q}(\tau) \right] \ , \cr
&& \cr
\theta_{mn}\left[ \tilde{q}(\tau ) \right] &=& \int_{0}^{\xi_{L}} d\xi \ 4 \pi \tilde{\chi}\left[ \xi - \tilde{q}(\tau) \right] G_{m}\left[ \xi , \tilde{q}(\tau) \right] \times \cr
&& \qquad\qquad \times\frac{\partial G_{n}}{\partial \xi }\left[ \xi , \tilde{q}(\tau) \right] \ , \cr
&& \cr
\Gamma_{mn}\left[ \tilde{q}(\tau ) \right] &=& \Omega_{mn}\left[ \tilde{q}(\tau ) \right] + \theta_{mn}\left[ \tilde{q}(\tau ) \right] \ .
\end{eqnarray}
Notice that these are all non-dimensional, real quantities, because we chose the modes to be real-valued functions.

We now deduce the equations for the coefficients $c_{n}(\tau )$ of the modes $G_{n}[ \xi , \tilde{q}(\tau) ]$ in (\ref{ExpansionEnModos}). One has to follow the same steps used in Section III for the case of a fixed membrane. Substituting the expansion of $\tilde{A}_{0}(\xi , \tau)$ given in (\ref{ExpansionEnModos}) into the equation for $\tilde{A}_{0}(\xi,\tau)$ given in (\ref{8}), neglecting terms of order $n \geq 2$ in $\tilde{q}'(\tau)$ and  $\tilde{q}''(\tau)$, and using the orthonormalization relation for the modes given in (\ref{OrtonormalizacionO}), one obtains
\begin{eqnarray}
\label{10}
&& c_{m}''(\tau) + \omega_{m}[ \tilde{q}(\tau) ]^{2} c_{m}(\tau ) \cr
&& \cr
&=& - \sum_{n=1}^{+\infty} \Gamma_{mn}[ \tilde{q}(\tau) ] \Big[ \ 2\tilde{q}'(\tau )c_{n}'(\tau) + \tilde{q}''(\tau )c_{n}(\tau) \ \Big] \ . \cr
&&
\end{eqnarray}
We note that equation (\ref{8}) for $\tilde{A}_{0}(\xi , \tau)$ was obtained by neglecting terms of order $n \geq 2$ in $\tilde{q}'(\tau)$ and  $\tilde{q}''(\tau)$. This is the reason why terms of these orders were neglected to obtain equation (\ref{10}). Therefore, (\ref{10}) is correct to first order in $\tilde{q}'(\tau)$ and $\tilde{q}''(\tau)$. 

The equations in (\ref{10}) reveal that all the coefficients $c_{m}(\tau)$ of the modes are coupled due to the term on the right-hand side. Nevertheless, notice that this coupling is small if both $| \tilde{q}'(\tau ) |$ and $|\tilde{q}''(\tau )|$ are small, which is the case under which the equation (\ref{8}) for $\tilde{A}_{0}(\xi , \tau)$ was deduced. In order to exhibit this \textit{smallness} explicitly it is convenient to first define the quantity
\begin{eqnarray}
\label{4b}
\tilde{\tilde{q}}( \tau_{\mbox{\tiny osc}} ) &=& \frac{q\left( \nu_{\mbox{\tiny osc}}^{-1}\tau_{\mbox{\tiny osc}} \right)}{\lambda_{0}} \ .
\end{eqnarray}
Recall that $\nu_{\mbox{\tiny osc}}^{-1}$ is the time-scale in which the membrane moves appreciably. Therefore, \ $\tau_{\mbox{\tiny osc}} = \nu_{\mbox{\tiny osc}}t$ \ is a non-dimensional time and $\tilde{\tilde{q}}(\tau_{\mbox{\tiny osc}})$ is the non-dimensional position of the membrane where time is measured in units of the time-scale $\nu_{\mbox{\tiny osc}}^{-1}$ in which the membrane moves appreciably. 

Using the definition of the non-dimensional time $\tau$ in (\ref{XTadimensional}) and the definitions of $\epsilon_{\mbox{\tiny pert}}$ in (\ref{Adimensionales}) and $\tilde{\tilde{q}}( \tau_{\mbox{\tiny osc}} )$ in (\ref{4b}),it follows that the relation between $\tilde{q}(\tau )$ and $\tilde{\tilde{q}}( \tau_{\mbox{\tiny osc}} )$ and their derivatives is given by
\begin{eqnarray}
\label{4}
\tilde{q}(\tau ) &=& \tilde{\tilde{q}}(\epsilon_{\mbox{\tiny pert}}\tau ) \ , \cr
&& \cr
\tilde{q}'(\tau ) &=& \epsilon_{\mbox{\tiny pert}} \tilde{\tilde{q}}'(\epsilon_{\mbox{\tiny pert}}\tau ) \ , \cr
&& \cr
\tilde{q}''(\tau ) &=& \epsilon_{\mbox{\tiny pert}}^{2} \tilde{\tilde{q}}''(\epsilon_{\mbox{\tiny pert}}\tau ) \ .
\end{eqnarray}
With (\ref{4}) we are all set to exhibit the \textit{smallness} we alluded to in the paragraph before equation (\ref{4b}). Expressing (\ref{10}) in terms of the quantities on the right-hand side of (\ref{4}) one obtains 
\begin{eqnarray}
\label{12}
&& c_{m}''(\tau) + \omega_{m}[ \tilde{\tilde{q}}(\epsilon_{\mbox{\tiny pert}} \tau) ]^{2} c_{m}(\tau) \cr
&& \cr
&=& - 2\epsilon_{\mbox{\tiny pert}} \tilde{\tilde{q}}'(\epsilon_{\mbox{\tiny pert}} \tau ) \sum_{n=1}^{+\infty} \Gamma_{mn}[ \tilde{\tilde{q}}(\epsilon_{\mbox{\tiny pert}} \tau) ] c_{n}'(\tau)  \cr
&& \cr
&& - \epsilon_{\mbox{\tiny pert}}^{2}\tilde{\tilde{q}}''(\epsilon_{\mbox{\tiny pert}} \tau ) \sum_{n=1}^{+\infty} \Gamma_{mn}[ \tilde{\tilde{q}}(\epsilon_{\mbox{\tiny pert}} \tau) ] c_{n}(\tau) \ .
\end{eqnarray}
We emphasize that the equations in (\ref{12}) are equivalent to the equations in (\ref{10}). Nevertheless, the equations in (\ref{12}) have a more transparent interpretation. They reveal that the coefficients $c_{m}(\tau)$ of the modes approximately satisfy uncoupled harmonic oscillator equations with a slowly-varying time-dependent frequency $\omega_{m}[ \tilde{\tilde{q}}(\epsilon_{\mbox{\tiny pert}} \tau) ]$ if $0 < \epsilon_{\mbox{\tiny pert}} \ll 1$, a result that should be expected because the $c_{m}(\tau)$ indeed satisfy harmonic oscillator equations when the membrane is fixed (see Section III). Also, it is explicitly seen that all of the $c_{m}(\tau)$ are weakly coupled due to the term on the right-hand side of (\ref{12}) if $0 < \epsilon_{\mbox{\tiny pert}} \ll 1$.

In the rest of the article we assume the following three conditions:  
\begin{enumerate}
\item $0 < \epsilon_{\mbox{\tiny pert}} \ll 1$. 

Since \ $\epsilon_{\mbox{\tiny pert}} = \nu_{\mbox{\tiny osc}}/\nu_{0}$ \ with $\nu_{0}$ the characteristic frequency of the field and $1/\nu_{\mbox{\tiny osc}}$ the time-scale in which the membrane evolves appreciably (see Section II), this condition indicates that there are two clearly defined and separate time-scales: a fast time-scale $1/\nu_{0}$ in which the electromagnetic field evolves and a slow time-scale $1/\nu_{\mbox{\tiny osc}}$ in which the membrane evolves. Also, as a consequence of this assumption, the equations in (\ref{12}) have coefficients that vary very slowly in the (non-dimensional) time $\tau$ and the $c_{m}(\tau)$ satisfy perturbed harmonic oscillator equations with a slowly varying frequency. These properties indicate that the system of differential equations in (\ref{12}) can be solved by the method of multiple scales \cite{Holmes}.

\item The initial conditions for the coefficients $c_{m}(\tau)$ of the modes are
\begin{eqnarray} 
\label{13}
c_{m}(0) &=& g_{0N}\delta_{mN} \ , \cr
c_{m}'(0) &=& g_{1N}\delta_{mN} \ ,
\end{eqnarray}
where $g_{0N}$ and $g_{1N}$ are real numbers and $N$ is a fixed, positive integer.

Notice that the initial conditions in (\ref{13}) indicate that only mode $N$ is initially excited. In Section VI (\ref{13}) will not hold because we consider the case of several initially excited modes.

\item The membrane is initially at rest, that is, $\tilde{q}'(0) = \epsilon_{\mbox{\tiny pert}}\tilde{\tilde{q}}'(0) = 0$.

Notice that this assumption combines with the one in the previous item so as to consider the following physical situation: The membrane is initially fixed and the field is found in one of the modes; afterwards, the membrane starts to move from rest either by an external agent or by radiation pressure. Moreover, we used (\ref{4}) to relate $\tilde{q}'(0)$ to $\tilde{\tilde{q}}'(0)$.

In Section VI we consider the case of several initially excited modes. In that section the following physical situation is considered: The membrane is initially fixed and the field is found in several modes; afterwards, the membrane starts to move from rest either by an external agent or by radiation pressure.
\end{enumerate}

Using the assumptions above and the method of multiple scales one can solve to good approximation the initial value problem posed in (\ref{12}) and (\ref{13}). This is what we do in the next section.


\section{Evolution with a mobile membrane}

In this section we first describe briefly how the initial value problem in (\ref{12}) and (\ref{13}) can be solved by the method of multiple scales. The details are provided in Appendix B for the case of two time-scales and in Appendix C for the case of three time-scales. We treat the case of three time-scales in order to have an approximate solution that is accurate for very long times, while the case of two time-scales is presented because some of its results are used in the case of three time-scales. In the subsequent subsections we only present the results and discuss them.

The method of multiple scales is applicable whenever there are several separate time-scales involved in a system, that is, when there are parts of a system that evolve much faster than others. The method consists in appropriately separating the parts of the system that evolve with the different time-scales and deducing the differential equations that govern the evolution of each part. For example, in the weakly-damped harmonic oscillator there are two clearly defined time-scales: the fast oscillation of the harmonic oscillator and the slow decay of its amplitude. In the method of multiple scales one separates the original differential equation for the harmonic oscillator into two differential equations, one for the fast oscillation and one for the slow decay of the amplitude. Since \ $\epsilon_{\mbox{\tiny pert}} \ll 1$ \ (see item 1 in Section IV), we have a similar situation in the system under consideration. The analogue of the fast oscillation of the harmonic oscillator is the evolution of the electromagnetic field with a fixed membrane, while the analogue of the slow decay of the amplitude is the dependence on the motion of the membrane. From (\ref{12}) it is clear that the slow time-scale in our system is given by
\begin{eqnarray}
\label{EscalaLenta}
t_{2}(\tau) &=& \epsilon_{\mbox{\tiny pert}}\tau \ ,
\end{eqnarray}
since all of the coefficients of $c_{m}(\tau)$ in (\ref{12}) are functions of $\epsilon_{\mbox{\tiny pert}}\tau$. The deduction of the fast time-scale is a bit more elaborate, but using the assumption that only mode $N$ is initially excited (see item 2 in Section IV), it can be shown (see Appendix B) that it is given by 
\begin{eqnarray}
\label{EscalaRapida}
t_{1N}(\tau) &=& \int_{0}^{\tau} \omega_{N}\left[ \tilde{q}(\tau') \right] d\tau' \ .
\end{eqnarray}
Physically this result is very reasonable as we now explain. First imagine that the membrane is fixed at $\tilde{q}_{0}$ and that only mode $N$ is initially excited. In this case the exact solution was calculated in (\ref{Pi7}). Observe from that equation that the time-dependence of $\tilde{A}(\xi ,\tau)$ is of the form
\begin{eqnarray}
\label{EscalaRapidab}
\omega_{N}(\tilde{q}_{0})\tau \ = \ \int_{0}^{\tau} \omega_{N}(\tilde{q}_{0}) d\tau' \ . 
\end{eqnarray}
If the membrane can move, then $\omega_{N}(\tilde{q}_{0})$ becomes position-dependent and changes to $\omega_{N}\left[ \tilde{q}(\tau) \right]$. As a consequence, (\ref{EscalaRapidab}) changes to (\ref{EscalaRapida}).  

With the two time-scales $t_{1N}(\tau)$ and $t_{2}(\tau)$ one can solve to good approximation the initial value problem in (\ref{12}) and (\ref{13}), and this is done in Appendix B. The problem with just considering these two time-scales is that the approximate solution obtained might not be accurate for very long times. In fact, the theory of multiple scales \cite{Holmes} indicates that the approximate solutions obtained with the two time-scales are accurate at least for times
\begin{eqnarray}
\label{AccurateResultbis}
0 \ \leq \ \epsilon_{\mbox{\tiny pert}} \tau \ \leq \ \mathcal{O}\left( 1 \right) \ \Leftrightarrow \ 0 \ \leq \ t \ \leq \ \mathcal{O}\left(  \frac{1}{\nu_{\mbox{\tiny osc}}} \right) \ .
\end{eqnarray}
Here and in the following $\mathcal{O}$ is the \textit{Big Oh} \cite{Holmes} and (\ref{AccurateResultbis}) means that \ $0 \leq t \leq \mathcal{D}/\nu_{\mbox{\tiny osc}}$ \ for some $\mathcal{D}>0$.

To remedy this, we introduce an even slower time-scale $t_{3}(\tau)$ given by 
\begin{eqnarray}
\label{EscalaMasLenta}
t_{3}(\tau) &=& \epsilon_{\mbox{\tiny pert}}^{2}\tau \ .
\end{eqnarray}
These three time-scales allow one to obtain an approximate solution of the initial value problem in (\ref{12}) and (\ref{13}) that is accurate for long times. This is done in Appendix C where it is shown that the approximate solution of (\ref{12}) and (\ref{13}) obtained by the method of multiple scales is accurate at least for times \ $\tau = \nu_{0}t$ \ such that 
\begin{eqnarray}
\label{AccurateResult}
0 \ \leq \ \epsilon_{\mbox{\tiny pert}}^{2} \tau \ \leq \ \mathcal{O}\left( 1 \right) &\Leftrightarrow&  0 \ \leq \ t \ \leq \ \mathcal{O}\left(  \frac{1}{\epsilon_{\mbox{\tiny pert}} \nu_{\mbox{\tiny osc}}} \right) \ . \ \
\end{eqnarray}
Recall that \ $0 \leq \epsilon_{\mbox{\tiny pert}} \ll 1$ \ and that $1/\nu_{\mbox{\tiny osc}}$ is the time-scale in which the mirror moves appreciably, see item 4 in Section II and item 1 in Section IV. Hence, (\ref{AccurateResult}) indicates that the approximate solutions obtained using the method of multiples scales with the three time-scales $t_{1N}(\tau)$, $t_{2}(\tau)$, and $t_{3}(\tau)$ will be accurate for long times. Explicitly, at least for times $t$ greater than $0$ and less than or equal to \ $\mathcal{D}/(\epsilon_{\mbox{\tiny pert}} \nu_{\mbox{\tiny osc}})$ \ with \ $\mathcal{D}>0$.

With the three time-scales $t_{1N}(\tau)$, $t_{2}(\tau)$, and $t_{3}(\tau)$ defined above, one then proposes an asymptotic expansion for each of the $c_{m}(\tau)$ as follows:
\begin{eqnarray}
\label{ExpansionAsintotica}
c_{m}(\tau) &\sim& Y_{m0}\left[ t_{1N}(\tau), t_{2}(\tau), t_{3}(\tau)  \right] \cr
&& \cr
&& + \epsilon_{\mbox{\tiny pert}} Y_{m1}\left[ t_{1N}(\tau), t_{2}(\tau), t_{3}(\tau)  \right]  \cr
&& \cr
&& + \epsilon_{\mbox{\tiny pert}}^{2} Y_{m2}\left[ t_{1N}(\tau), t_{2}(\tau), t_{3}(\tau)  \right]  \ + \ ... \ .
\end{eqnarray}
The next step is to substitute (\ref{ExpansionAsintotica}) into the initial value problem given in (\ref{12}) and (\ref{13}). One then equates equal powers of $\epsilon_{\mbox{\tiny pert}}$ and solves the resulting initial value problems for the new variables $t_{1N}$, $t_{2}$, and $t_{3}$. If one only retains the first $n$ terms in the asymptotic expansion in (\ref{ExpansionAsintotica}), then one speaks of an $n$-term approximation. Therefore, \ $c_{m} \simeq Y_{m0}$ \ is a first-term approximation, \ $c_{m} \simeq Y_{m0} + \epsilon_{\mbox{\tiny pert}} Y_{m1}$ \ is a two-term approximation, so on and so forth. Moreover, the $(n+1)$-term approximation is going to be more accurate than the $n$-term approximation. We mention that the value of the multiple-scales method is that it allows one to identify the contributions to $c_{n}(\tau)$ in order of importance, that is, it permits one to obtain more accurate approximations by adding more terms from the asymptotic expansion and this allows one to identify which physical phenomena are introduced and have larger effects than others.

We now present the results for the case of three time-scales. As mentioned above, the deduction of these is presented in detail in Appendix C. Moreover, we note that the one- and two-term approximations using three time-scales are exactly the same as the one- and two-term approximations using two-time-scales. This suggests that (at least) the one-term approximation is accurate for times longer than (\ref{AccurateResult}).

\subsection{The first-term approximation}

The first-term approximation $c_{n}^{(1)}(\tau)$ to $c_{n}(\tau)$ is given by
\begin{eqnarray}
\label{Aprox1termino}
c_{N}^{(1)}(\tau) &=&
 \alpha_{N}\left[ \tilde{q}(\tau) \right] b_{N0} e^{-i[ t_{1N}(\tau) - \Theta_{N0}] } \ + \ c.c \ , \cr
&& \cr
&& \cr
c_{m}^{(1)}(\tau) &=& 0  \qquad (m\not= N).
\end{eqnarray}
with $t_{1N}(\tau)$ given in (\ref{EscalaRapida}) and
\begin{eqnarray}
\label{Aprox1terminoCoef}
\alpha_{N} \left[ \tilde{q}(\tau) \right] &=& \sqrt{ \frac{\omega_{N}\left[ \tilde{q}(0) \right]}{\omega_{N}\left[ \tilde{q}(\tau) \right]} } \ \mbox{exp}\left\{ - \int_{\tilde{q}(0)}^{\tilde{q}(\tau )} dy \ \Gamma_{NN}(y) \right\} \ , \cr
&& \cr
b_{N0} &=& \left\vert \frac{g_{0N}}{2} + \frac{i g_{1N}}{2\omega_{N}\left[ \tilde{q}(0) \right]}  \right\vert \ , \cr
&& \cr
\Theta_{N0} &=& \mbox{arg}\left[ \frac{g_{0N}}{2} + \frac{i g_{1N}}{2\omega_{N}\left[ \tilde{q}(0) \right]} \right] \ .
\end{eqnarray} \ 
Recall that (\ref{Aprox1termino}) is accurate at least for the times given in (\ref{AccurateResult}).

Notice that $b_{N0}$ and $\Theta_{N0}$ are exactly the same real quantities as in (\ref{Pi8}) if \ $\tilde{q}(0) = \tilde{q}_{0}$. Also, recall that $g_{0N}$ and $g_{1N}$ are real constants given by the initial conditions in (\ref{13}). Observe that they define the initial amplitude $b_{N0}$ and phase $\Theta_{N0}$ of mode $N$. As before, \textit{arg(z)} denotes an argument (or phase) of complex number $z$ and $c.c$ indicates the complex conjugate. Finally, notice that $\alpha_{N} \left[ \tilde{q}(\tau) \right]$ is a real quantity.

From the expansion of $\tilde{A}_{0}(\xi , \tau)$ in terms of the modes given in (\ref{ExpansionEnModos}) and the approximate value of $c_{n}(\tau)$ given in (\ref{Aprox1termino}), it follows that the first-term approximation $\tilde{A}_{0}^{(1)}(\xi , \tau)$ to $\tilde{A}_{0}(\xi , \tau)$ is given by
\begin{eqnarray}
\label{A01}
\tilde{A}_{0}^{(1)}(\xi , \tau) &=& 
\alpha_{N} \left[ \tilde{q}(\tau) \right] b_{N0} e^{-i[ t_{1N}(\tau) - \Theta_{N0}] } G_{N}[ \xi , \tilde{q}(\tau) ] \cr
&& + \ c.c. 
\end{eqnarray}

Comparing (\ref{A01}) with the result for $\tilde{A}_{0}(\xi, \tau)$ in the case of a fixed membrane given in (\ref{Pi7}) and (\ref{Pi8}), we find the following:
\begin{enumerate}
\item[(a)] The electromagnetic field follows mode $G_{N}[ \xi , \tilde{q}(\tau) ]$ with a frequency and an amplitude that depend on the position of the membrane. 
\\
\\
Notice that (\ref{A01}) embodies the physically expected result: an electromagnetic field initially in mode $N$ will follow mode $N$ if the membrane moves slowly.

\item[(b)] The amplitude changes from $b_{N0}$ to \ $\alpha_{N}\left[ \tilde{q}(\tau) \right] b_{N0}$. 
\\
\\
Since the membrane moves, the amplitude of mode $N$ depends on the position of the membrane. Also, observe that \ $\alpha_{N}\left[ \tilde{q}(\tau) \right] b_{N0}$ \ reduces to $b_{N0}$ if $\tilde{q}(\tau) = \tilde{q}(0)$ for all $\tau$, that is, if the membrane is fixed.

\item[(c)] The phase of the electromagnetic field changes from \ $[ \omega_{N}(\tilde{q}_{0})\tau - \Theta_{N0} ]$ \ to  \ $[ t_{1N}(\tau) -\Theta_{N0} ]$ \ with $t_{1N}(\tau)$ given in (\ref{EscalaRapida}). 
\\
\\
Notice that this is physically reasonable because the field follows the instantaneous mode $G_{N}[ \xi , \tilde{q}(\tau) ]$ and the frequency $\omega_{N}[\tilde{q}(\tau)]$ associated with it depends on the position of the membrane, see also the discussion in the paragraph following equation (\ref{EscalaRapida}).
\end{enumerate}

We now turn to the two-term approximation. It gives a more accurate description of the electromagnetic field and it provides us with a criterion to determine when the first term approximation is an accurate description of the field. 

\subsection{The two-term approximation }

The two-term approximation $c_{n}^{(2)}(\tau)$ to $c_{n}(\tau)$ is given by
\begin{eqnarray}
\label{Aprox2Terminos}
c_{N}^{(2)}(\tau) &=& \alpha_{N} \left[ \tilde{q}(\tau) \right] b_{N0} e^{-i\left[ t_{1N}(\tau) - \Theta_{N0} \right] } \times \cr
&& \cr
&& \times \left\{ 1 + i\tilde{q}'(\tau) \frac{\omega_{N}'\left[ \tilde{q}(\tau) \right]}{4 \omega_{N}\left[ \tilde{q}(\tau) \right]^{2}} \right\} \cr
&& \cr
&& + \ c.c \ , \cr
&& \cr
&& \cr
c_{m}^{(2)}(\tau) &=& \alpha_{N} \left[ \tilde{q}(\tau) \right] b_{N0} e^{-i\left[ t_{1N}(\tau) - \Theta_{N0} - \pi/2 \right] } \times \cr
&& \cr
&& \times \tilde{q}'(\tau) \frac{2 \Gamma_{mN}\left[ \tilde{q}(\tau) \right] \omega_{N} \left[ \tilde{q}(\tau) \right] }{\omega_{m}\left[ \tilde{q}(\tau ) \right]^{2} - \omega_{N}\left[ \tilde{q}(\tau ) \right]^{2}} \cr
&& \cr
&& + \ c.c \qquad (m\not= N), 
\end{eqnarray}
where $\alpha_{N} [\tilde{q}(\tau)]$, $b_{N0}$, and $\Theta_{N0}$ are given in (\ref{Aprox1terminoCoef}) and $t_{1N}(\tau)$ is given in (\ref{EscalaRapida}). Recall that (\ref{Aprox2Terminos}) is accurate at least for the times given in (\ref{AccurateResult}). Moreover, during that time interval it is more accurate than the first-term approximation.

From the expansion of $\tilde{A}_{0}(\xi , \tau)$ in terms of the modes given in (\ref{ExpansionEnModos}) and the approximate value of $c_{n}(\tau)$ given in (\ref{Aprox2Terminos}), it follows that the two-term approximation $\tilde{A}_{0}^{(2)}(\xi , \tau)$ to $\tilde{A}_{0}(\xi , \tau)$ is given by
\begin{eqnarray}
\label{A02}
\tilde{A}_{0}^{(2)}(\xi , \tau) &=&  \alpha_{N} \left[ \tilde{q}(\tau) \right] b_{N0} e^{-i\left[ t_{1N}(\tau) - \Theta_{N0} \right] } G_{N}[ \xi , \tilde{q}(\tau) ] \times \cr
&& \times \left\{ 1 + i\tilde{q}'(\tau) \frac{\omega_{N}'\left[ \tilde{q}(\tau) \right]}{4 \omega_{N}\left[ \tilde{q}(\tau) \right]^{2}} \right\} \cr
&& \cr
&& \cr
&& + \alpha_{N} \left[ \tilde{q}(\tau) \right] b_{N0} e^{-i\left[ t_{1N}(\tau) - \Theta_{N0} - \pi/2 \right] } \tilde{q}'(\tau) \times \cr
&& \times \sum_{n=1 \atop n\not=N}^{+\infty}  \frac{2 \Gamma_{nN}\left[ \tilde{q}(\tau) \right] \omega_{N} \left[ \tilde{q}(\tau) \right] }{\omega_{n}\left[ \tilde{q}(\tau ) \right]^{2} - \omega_{N}\left[ \tilde{q}(\tau ) \right]^{2}}  G_{n}[ \xi , \tilde{q}(\tau) ] \cr
&& \cr
&& \cr
&& + c.c.
\end{eqnarray}

Comparing the two-term approximation of $\tilde{A}_{0}(\xi ,\tau)$ given in (\ref{A02}) with its first-term approximation given in (\ref{A01}), we find the following facts:
\begin{enumerate}
\item[(a)] All other modes \ $m\not= N$ \ of the electromagnetic field are (weakly) excited. Only modes \ $m$ \ with $\omega_{m}[\tilde{q}(\tau)]$ in a small band around  $\omega_{N}[\tilde{q}(\tau)]$ can have a non-negligible excitation.
\\
\\
The most evident change from the first-term approximation to the two-term approximation is that all the other modes \ $m\not= N$ \ of the electromagnetic field are now excited. Nevertheless, this excitation must be small if our results are to hold. The reason for this is that the order of the asymptotic expansion in (\ref{ExpansionAsintotica}) must be preserved, that is, each term \ $\epsilon_{\mbox{\tiny pert}}^{n+1}Y_{m(n+1)}$ must be smaller than the term \ $\epsilon_{\mbox{\tiny pert}}^{n}Y_{mn}$ before it. Since the first-term approximation $c_{m}^{(1)}(\tau)$ with \ $m\not=N$ is zero (see (\ref{Aprox1termino})), it must occur that the two-term approximation  $c_{m}^{(2)}(\tau)$ with \ $m\not=N$ must be small. We now explain why it does indeed happen that all modes \ $m \not=N$ \ have negligible excitation except for those with frequency $\omega_{m}[\tilde{q}(\tau)]$ in a small band around  $\omega_{N}[\tilde{q}(\tau)]$.

Notice that the coefficient of $G_{n}[ \xi , \tilde{q}(\tau) ]$ $(n \not= N)$  in (\ref{A02}) has a factor of the form
\begin{eqnarray}
\label{Resonancia}
\frac{2 \Gamma_{nN}\left[ \tilde{q}(\tau) \right] \omega_{N} \left[ \tilde{q}(\tau) \right] }{\omega_{n}\left[ \tilde{q}(\tau ) \right]^{2} - \omega_{N}\left[ \tilde{q}(\tau ) \right]^{2}}
\end{eqnarray}
Since \ $\omega_{n}[\tilde{q}(\tau)] \rightarrow +\infty$ \ as \ $n\rightarrow +\infty$ \ (see item 1 of Section III), it follows that the excitation of each mode \ $m\not= N$ \ is negligible except in the case where $\omega_{m}[\tilde{q}(\tau)] \simeq \omega_{N}[\tilde{q}(\tau)]$. Therefore, only modes \ $n\not= N$ \ with $\omega_{n}[\tilde{q}(\tau)]$ in a small band around  $\omega_{N}[\tilde{q}(\tau)]$ can have a non-negligible excitation. These quasi-resonant modes are weakly excited if the membrane moves \textit{sufficiently slowly}. This \textit{sufficiently slow motion of the membrane} is quantified in item (e) below.

\item[(b)] All other modes \ $m\not= N$ \ evolve with the same frequency as the original mode $N$, but with a $\pm\pi/2$ phase shift.
\\
\\
Since all the factors multiplying $G_{n}[ \xi , \tilde{q}(\tau) ]$ $(n \not=N)$ in (\ref{A02}) are real, except for the complex exponential $e^{-i\left[ t_{1N}(\tau) - \Theta_{N0} - \pi/2 \right] }$, it follows that all other modes \ $n\not= N$ \ have a phase $[t_{1N}(\tau) -\Theta_{N0} \mp \pi/2]$, where $-\pi/2$ $(+\pi/2)$ is chosen if the product of the real factors is positive (negative). Since the phase of the complex exponential multiplying $G_{N}[\xi, \tilde{q}(\tau)]$ in (\ref{A02}) is $[t_{1N}(\tau) -\Theta_{N0} ]$ and \ $t_{1N}'(\tau) = \omega_{N}[\tilde{q}(\tau)]$, it follows that all other modes $n\not= N$ evolve with the frequency $\omega_{N}[\tilde{q}(\tau)]$ of mode $N$ but with a $\pm\pi/2$ phase shift.

\item[(c)] A Doppler-like phase shift appears in the original mode.
\\
\\
Since all of our results are correct to first order in $\tilde{q}'(\tau)$ and $\tilde{q}''(\tau)$, one can express the coefficient $c_{N}(\tau)$ of $G_{N}[ \xi , \tilde{q}(\tau) ] $ given in (\ref{Aprox2Terminos}) as follows:
\begin{eqnarray}
\label{Aprox2TerminosCoef}
c_{N}^{(2)}(\tau ) &=& \alpha_{N} \left[ \tilde{q}(\tau) \right] b_{N0} e^{-i[ t_{1N}(\tau) - \Theta_{N0}]} \times \cr
&& \times \mbox{exp}\left\{ i\tilde{q}'(\tau) \frac{\omega_{N}'\left[ \tilde{q}(\tau) \right]}{4 \omega_{N}\left[ \tilde{q}(\tau) \right]^{2}} \right\} \cr
&& \cr
&& + \ c.c.
\end{eqnarray}
Comparing the first-term approximation of $c_{N}(\tau )$ given in (\ref{Aprox1termino}) with its two-term approximation given in (\ref{Aprox2TerminosCoef}), it follows that the phase of $c_{N}(\tau )$ changes from \ $[ t_{1N}(\tau) - \Theta_{N0} ]$ \ to 
\begin{eqnarray}
\label{Doppler}
 t_{1N}(\tau) - \Theta_{N0} -\tilde{q}'(\tau) \frac{\omega_{N}'\left[ \tilde{q}(\tau) \right]}{4 \omega_{N}\left[ \tilde{q}(\tau) \right]^{2}} \ .
\end{eqnarray}
Hence, a Doppler-like phase shift appears in mode $N$, since it depends on the velocity $\tilde{q}'(\tau)$ of membrane.

\item[(d)] The two-term approximation includes information about the velocity of the membrane.

Comparing $c_{n}^{(1)}(\tau)$ with $c_{n}^{(2)}(\tau)$, see (\ref{Aprox1termino}) and (\ref{Aprox2Terminos}), one readily notices that the corrections to $c_{n}^{(1)}(\tau)$ involve terms proportional to $\tilde{q}'(\tau)$. Moreover, in the previous items we  observed that the aforementioned terms represent a Doppler-like phase shift in the original mode and the excitation of the other modes.

\item[(e)] The two-term approximation tells us that the first term approximation is accurate whenever the following two conditions are satisfied:
\begin{eqnarray}
\label{ValidezAprox1Termino}
\left\vert \tilde{q}'(\tau) \frac{\omega_{N}'\left[ \tilde{q}(\tau) \right]}{4 \omega_{N}\left[ \tilde{q}(\tau) \right]^{2}}  \right\vert &\ll & 1 \ , \cr
&& \cr
&& \cr
\left\vert \tilde{q}'(\tau) \frac{2 \Gamma_{mN}\left[ \tilde{q}(\tau) \right] \omega_{N} \left[ \tilde{q}(\tau) \right] }{\omega_{m}\left[ \tilde{q}(\tau ) \right]^{2} - \omega_{N}\left[ \tilde{q}(\tau ) \right]^{2}} \alpha_{N} \left[ \tilde{q}(\tau) \right] b_{N0} \right\vert &\ll & 1 \ , \cr 
&& \cr
(m\not= N). && 
\end{eqnarray}
\\
\\
The first condition in (\ref{ValidezAprox1Termino}) comes from demanding that the first-term approximation $c_{N}^{(1)}(\tau)$ of $c_{N}(\tau)$ must be much larger than the added term in the two-term approximation $c_{N}^{(2)}(\tau)$, that is, from demanding that
\begin{eqnarray}
\label{Comparacion1}
\left\vert c_{N}^{(2)}(\tau) - c_{N}^{(1)}(\tau) \right\vert &\ll& \left\vert c_{N}^{(1)}(\tau) \right\vert \ .
\end{eqnarray}
See (\ref{Aprox1termino}) and (\ref{Aprox2Terminos}) for the definitions of $c_{N}^{(1)}(\tau)$ and $c_{N}^{(2)}(\tau)$.

The second condition in (\ref{ValidezAprox1Termino}) comes from demanding that the two-term approximation $c_{m}^{(2)}(\tau)$ of $c_{m}(\tau)$ \ $(m \not=N)$ must be much smaller than $1$ (recall that $c_{m}^{(2)}(\tau)$ is a non-dimensional quantity), that is, from demanding that
\begin{eqnarray}
\label{Comparacion2}
\left\vert c_{m}^{(2)}(\tau) \right\vert &\ll& 1 \ \ (m\not= N) .
\end{eqnarray}
See (\ref{Aprox2Terminos}) for the definition of $c_{m}^{(2)}(\tau)$.

We note that we only used the first term in the definition of $c_{m}^{(2)}(\tau)$ given in (\ref{Aprox2Terminos}) to derive the conditions in (\ref{ValidezAprox1Termino}), that is, we omitted the complex conjugate ($c.c$) part. 

From the second condition in (\ref{ValidezAprox1Termino}) one immediately notices that it cannot be fulfilled if \ $\omega_{m}\left[ \tilde{q}(\tau ) \right] = \omega_{N}\left[ \tilde{q}(\tau ) \right]$ \ for some $\tau$. It turns out that this \textit{crossing} is not possible, see Appendix A for a rigorous mathematical proof of this fact. Also, Figure \ref{Figure2} illustrates this result.

The frequencies $\omega_{n}[\tilde{q}(\tau)]$ of the modes will vary as the membrane moves. Although they cannot cross, some may approach the frequency $\omega_{N}[\tilde{q}(\tau)]$ of the original mode, see Figure \ref{Figure2}. Hence, the factors \ $1/\{ \omega_{n}[ \tilde{q}(\tau ) ]^{2} - \omega_{N}[ \tilde{q}(\tau ) ]^{2} \}$ \ in the expression for $\tilde{A}_{0}^{(2)}(\xi ,\tau)$ can become quite large. Nevertheless, the first-term approximation  $\tilde{A}_{0}^{(1)}(\xi , \tau)$ given in (\ref{A01}) will remain an accurate description of $\tilde{A}_{0}(\xi ,\tau )$ if the \textit{membrane moves sufficiently slowly} so that the conditions in (\ref{ValidezAprox1Termino}) hold for all time $\tau$. 

\end{enumerate}

\begin{figure}
\includegraphics[width=8cm]{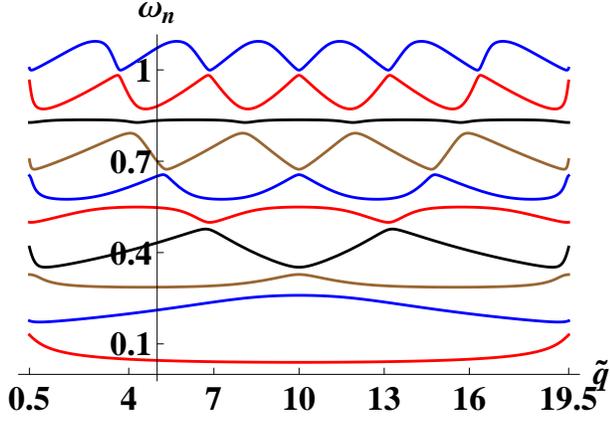}
\caption{\label{Figure2} The figure illustrates the $10$ lowest (non-dimensional) angular frequencies $\omega_{n}$ of the cavity as a function of the position $\tilde{q}$ of the midpoint of the membrane for the electric susceptibility in (\ref{Susceptibilidad}). The frequencies were obtained by substituting the parameters in (\ref{ParametrosF}) into (\ref{EcTrascendente}) and obtaining the zeros of the resulting equation numerically. Notice that $\tilde{q}$ ranges from $\tilde{q} = 0.5$ to $\tilde{q} = 19.5$ because the (non-dimensional) length of the cavity is $\xi_{L} = 20$ and the membrane has (non-dimensional) thickness $\tilde{\delta}_{0} = 1$.}
\end{figure}

\subsection{The three-term approximation}

The three-term approximation $c_{n}^{(3)}(\tau)$ to $c_{n}(\tau)$ is given by

\begin{eqnarray}
\label{Aprox3Terminos}
c_{N}^{(3)}(\tau) &=& \alpha_{N} \left[ \tilde{q}(\tau) \right] b_{N0} e^{-i\left[ t_{1N}(\tau) - \Theta_{N0} \right] } \Big\{ 1 + \sigma(\tau) \cr
&& \left. + i\tilde{q}'(\tau) \frac{\omega_{N}'\left[ \tilde{q}(\tau) \right]}{4 \omega_{N}\left[ \tilde{q}(\tau) \right]^{2}}  \ + \ \epsilon_{\mbox{\tiny pert}}^{2}\frac{b _{N2}(0, \epsilon_{\mbox{\tiny pert}}^{2}\tau )}{b_{N0}e^{i\Theta_{N0}}} \right\} \cr
&& + \ c.c \ , \cr
&& \cr
&& \cr
c_{m}^{(3)}(\tau) &=& \alpha_{N} \left[ \tilde{q}(\tau) \right] b_{N0} e^{-i\left[ t_{1N}(\tau) - \Theta_{N0} - \pi/2 \right] } \times \cr
&& \times \left\{ \tilde{q}'(\tau) \frac{2 \Gamma_{mN}\left[ \tilde{q}(\tau) \right] \omega_{N} \left[ \tilde{q}(\tau) \right] }{\omega_{m}\left[ \tilde{q}(\tau ) \right]^{2} - \omega_{N}\left[ \tilde{q}(\tau ) \right]^{2}} + i \mathcal{D}(\tau) \right\} \ \cr
&& \cr
&& \cr
&& + \epsilon_{\mbox{\tiny pert}}^{2} \alpha_{m}\left[ \tilde{q}(\tau) \right] e^{-it_{1m}(\tau)} b_{m2}(0, \epsilon_{\mbox{\tiny pert}}^{2}\tau ) \ , \cr
&& \cr
&& + \ c.c \qquad (m\not= N), 
\end{eqnarray}
where $\alpha_{N} [\tilde{q}(\tau)]$, $b_{N0}$, and $\Theta_{N0}$ are given in (\ref{Aprox1terminoCoef}), $t_{1N}(\tau)$ is given in (\ref{EscalaRapida}), and $\alpha_{m} \left[ \tilde{q}(\tau) \right]$, $t_{1m}(\tau)$, $\sigma (\tau)$, and $\mathcal{D}(\tau)$ are defined as follows:
\begin{eqnarray}
\alpha_{m}\left[ \tilde{q}(\tau) \right] &=& \sqrt{\frac{\omega_{N}\left[ \tilde{q}(0)\right]}{\omega_{N}\left[ \tilde{q}(\tau)\right]}} \mbox{exp}\left\{ -\int_{\tilde{q}(0)}^{\tilde{q}(\tau)}dy \Gamma_{mm}(y) \right\} \ , \nonumber
\end{eqnarray}
\begin{eqnarray}
\label{Aprox3TerminosCoef}
t_{1m}(\tau) &=& \int_{0}^{\tau} d\tau' \omega_{m}\left[ \tilde{q}(\tau') \right] \ , \cr
&& \cr
&& \cr
\sigma(\tau) &=& \tilde{q}''(\tau) \frac{\omega_{N}'\left[ \tilde{q}(\tau) \right]}{8\omega_{N}\left[ \tilde{q}(\tau) \right]^{3}}
- \tilde{q}''(0) \frac{\omega_{N}'\left[ \tilde{q}(0) \right]}{8\omega_{N}\left[ \tilde{q}(0) \right]^{3}} \ , \cr
&& \cr
&& \cr
\mathcal{D}(\tau) &=& \tilde{q}''(\tau) \Gamma_{mN}\left[ \tilde{q}(\tau) \right]\frac{\omega_{m}\left[ \tilde{q}(\tau) \right]^{2} + 3 \omega_{N}\left[ \tilde{q}(\tau) \right]^{2}}{\left\{ \omega_{m}\left[ \tilde{q}(\tau) \right]^{2} - \omega_{N}\left[ \tilde{q}(\tau) \right]^{2} \right\}^{2}} \ .  \cr
&& 
\end{eqnarray}
Recall that (\ref{Aprox3Terminos}) is accurate at least for the times given in (\ref{AccurateResult}). Moreover, during that time interval it is more accurate than both the first- and two-term approximations.

Observe that $t_{1N}(\tau)$ and $\alpha_{N}[\tilde{q}(\tau)]$ defined in (\ref{Aprox3TerminosCoef}) coincide with $t_{1N}(\tau)$ and $\alpha_{N}[\tilde{q}(\tau)]$ defined in (\ref{EscalaRapida}) and (\ref{Aprox1terminoCoef}), respectively. Moreover, $b_{N2}(0,\epsilon_{\mbox{\tiny pert}}^{2}\tau)$ is determined by solving the $\mathcal{O}(\epsilon_{\mbox{\tiny pert}}^{4})$ problem. Also, $b_{m2}(0,\epsilon_{\mbox{\tiny pert}}^{2}\tau)$ \ $(m\not= N)$ has two possibilities: (i) it is identically equal to zero if \ $\tilde{q}''(0)=0$ \ or \ $\Gamma_{mN}[\tilde{q}(0)] = 0$; (ii) it is a non-zero function determined by solving the $\mathcal{O}(\epsilon_{\mbox{\tiny pert}}^{4})$ problem if \ $\tilde{q}''(0) \not= 0$ \ and \ $\Gamma_{mN}[\tilde{q}(0)] \not= 0$. See Appendix C.

The three-term approximation $\tilde{A}_{0}^{(3)}(\xi , \tau)$ to $\tilde{A}_{0}(\xi , \tau)$ can be obtained by substituting (\ref{Aprox3Terminos}) in the expansion (\ref{ExpansionEnModos}) for $\tilde{A}_{0}(\xi , \tau)$.

Comparing (\ref{Aprox3Terminos}) with (\ref{Aprox2Terminos}), one finds the following facts:
\begin{enumerate}
\item[(a)] The three-term approximation incorporates information about the acceleration of the membrane.

This follows by observing from (\ref{Aprox3Terminos}) that $c_{n}^{(3)}(\tau)$ includes the terms $\sigma (\tau)$ and $\mathcal{D}(\tau)$ and by noticing from (\ref{Aprox3TerminosCoef}) that both $\sigma(\tau)$ and $\mathcal{D}(\tau)$ are proportional to the acceleration of the membrane.

\item[(b)] Modes \ $m\not= N$ \ start to evolve at their own frequency if \ $\tilde{q}''(0) \not= 0$ \ and \ $\Gamma_{mN}[\tilde{q}(0)] \not= 0$.

If \ $\tilde{q}''(0) \not= 0$ \ and \ $\Gamma_{mN}[\tilde{q}(0)] \not= 0$, then the term multiplied by \ $e^{-it_{1m}(\tau)}$ \ on the right-hand side of the expression for $c_{m}^{(3)}(\tau)$ $(m\not=N)$ in (\ref{Aprox3Terminos}) is non-zero and has an associated frequency \ $t_{1m}'(\tau) = \omega_{m}[\tilde{q}(\tau)]$, see Appendix C. Hence, each mode \ $m \not= N$ \ has a component that evolves at its associated frequency $\omega_{m}[\tilde{q}(\tau)]$. In fact, notice that the term multiplied by $e^{-it_{1m}(\tau)}$ is very similar to $c_{N}^{(1)}(\tau)$ in (\ref{Aprox1termino}). 

\item[(c)] The three-term approximation is only valid for three situations: (i) frequencies in a small band around the original mode's frequency $\omega_{N}\left[ \tilde{q}(\tau) \right]$, (ii) $\tilde{q}''(0) = 0$, or (iii) $\Gamma_{mN}[\tilde{q}(0)] = 0$ for all $m\not=N$.

Appendix B shows that the evolution of the electromagnetic field involves an infinite number of fast time-scales, namely one associated with each mode and given by $t_{1m}(\tau)$ in (\ref{Aprox3TerminosCoef}). In spite of this, we were able to reduce the difficulty of the problem to only one fast time-scale because we assumed that only mode $N$ was initially excited. This allowed us to deduce the one- and two-term approximations without any more assumptions. Nevertheless, if one requires an $n$-term approximation with \ $n \geq 3$, then it is shown in Appendix C that the problem of an infinite number of fast time-scales appears again and that one way to remedy this is to demand that at least one of the following three conditions is satisfied:
\begin{eqnarray}
\label{Validez3terminos}
\tilde{q}''(0) \ = \ 0 \ ,
\end{eqnarray}  
or
\begin{eqnarray}
\label{Validez3terminosN}
\Gamma_{mN}[\tilde{q}(0)] \ = \ 0 \ \qquad (m\not= N) ,
\end{eqnarray}  
or
\begin{eqnarray}
\label{Validez3terminos2}
\frac{d}{dt_{2}}\left\{ \ \frac{\omega_{m}\left[ \tilde{\tilde{q}}(t_{2})\right]}{\omega_{N}\left[ \tilde{\tilde{q}}(t_{2})\right]} \ \right\} &\simeq& 0 \ , \qquad (m\not= N).
\end{eqnarray}
In particular, (\ref{Validez3terminos}) is satisfied if an external agent moves the membrane from rest in such a way that its equation of motion is, for example, of the following forms: 
\begin{eqnarray}
\label{Ejemplod2q}
&(i)& \ \tilde{q}''(\tau) \ = \ -\mathcal{E}\frac{\tau^{n}}{\tau^{n} + 1}V_{1}'\left[ \tilde{q}(\tau) \right] \ , \cr
&& \cr 
&(ii)& \ \tilde{q}''(\tau) \ = \ \mathcal{E}\frac{\tau^{n}}{\tau^{n} + 1} - V_{2}'\left[ \tilde{q}(\tau) \right] \ ,
\end{eqnarray}
where \ $\mathcal{E}\not= 0$ \ is a (non-dimensional) constant, \ $n>0$, $V_{j}'\left[ \tilde{q}(\tau) \right]$ is the derivative of the (non-dimensional) potential $V_{j}( \tilde{q} )$ with respect to $\tilde{q}$ and evaluated at $\tilde{q}(\tau)$, and \ $V_{2}'\left[ \tilde{q}(0) \right] = 0$. Notice that the right-hand side of $(i)$ in (\ref{Ejemplod2q}) tends to \ $-\mathcal{E}V_{1}'\left[ \tilde{q}(\tau) \right]$ \ and that the right-hand side of $(ii)$ in (\ref{Ejemplod2q}) tends to \ $\mathcal{E} - V_{2}'\left[ \tilde{q}(\tau) \right] $ \ as \ $\tau \rightarrow + \infty$, that is, the right-hand side of the equations in (\ref{Ejemplod2q}) tend to a force that comes from a potential. 

On the other hand, \ $\Gamma_{mN}[\tilde{q}(0)] = 0$ \ $(m\not= N)$ could be satisfied for some electric susceptibility functions, while the condition in (\ref{Validez3terminos2}) is satisfied for frequencies $\omega_{m}\left[ \tilde{\tilde{q}}(t_{2})\right]$ in a small band around $\omega_{N}\left[ \tilde{\tilde{q}}(t_{2})\right]$. We note that one typically has  \ $\tilde{q}''(0) \not= 0$ \ and \ $\Gamma_{mN}[\tilde{q}(0)] \not= 0$ \ for some $m\not= N$. The former condition usually holds because the force on the membrane is, in general, not initially zero. Therefore, one has to restrict to modes with frequencies $\omega_{m}\left[ \tilde{q}(\tau)\right]$ in a small band around $\omega_{N}\left[ \tilde{q}(\tau)\right]$. Notice that modes with frequencies outside of this band can be neglected, since the two-term approximation tells us that they are negligibly excited.

If either one of (\ref{Validez3terminos}), (\ref{Validez3terminosN}), or (\ref{Validez3terminos2}) is not fulfilled, then one can only have a one- or two-term approximation for the coefficients $c_{n}(\tau)$ of the modes. Given that (\ref{Validez3terminos2}) is satisfied for frequencies $\omega_{m}\left[ \tilde{\tilde{q}}(t_{2})\right]$ in a small band around $\omega_{N}\left[ \tilde{\tilde{q}}(t_{2})\right]$, one may think that it would be proper to make the approximation \ $\omega_{m}\left[ \tilde{q}(\tau)\right] = \omega_{N}\left[ \tilde{q}(\tau)\right]$ \ for all frequencies in this band. We consider that this approximation is inappropriate because the frequencies are in fact different and this approximation would alter the system in a fundamental way. For example, the quasi-resonance factor  \ $\{ \omega_{m}[\tilde{q}(\tau)]^{2} - \omega_{N}[\tilde{q}(\tau)]^{2} \}^{-1}$  \ included in $c_{m}^{(2)}(\tau)$ $(m\not=N)$ in the two-term approximation, see (\ref{Aprox2Terminos}), would not appear if $\omega_{m}[\tilde{q}(\tau)] = \omega_{N}[\tilde{q}(\tau)]$ and it would be replaced by another term that is defined for $\omega_{m}[\tilde{q}(\tau)] = \omega_{N}[\tilde{q}(\tau)]$ in a similar way to a harmonic oscillator with resonant driving \cite{Ross}.

\end{enumerate}


\subsection{Physical interpretation}

In this subsection we add up all the information provided by the one-, two-, and three-term approximations to give a physical interpretation of how the electromagnetic field evolves in the presence of a mobile membrane. 

Suppose (as we have done in the deduction of all formulas up to now) that only mode $N$ is initially excited and that the membrane starts to move \textit{slowly} from rest. In a first approximation the electromagnetic field will follow mode $N$ with a phase and an amplitude that depend on the position of the membrane. In a more accurate approximation the phase and amplitude of mode $N$ include small corrections depending on the velocity and acceleration of the membrane. In particular, the phase includes a Doppler-like shift. Also, all the other modes become weakly excited. The reason for this is the following. As the membrane moves, the frequencies $\omega_{n}[\tilde{q}(\tau)]$ of each of the modes $G_{n}[\xi, \tilde{q}(\tau)]$ will vary. If $\omega_{n}[\tilde{q}(\tau)]$ is near $\omega_{N}[\tilde{q}(\tau)]$, then it is easy for light to \textit{jump} from mode $N$ to mode $n$ and, consequently, mode $n$ becomes excited. On the other hand, if $\omega_{n}[\tilde{q}(\tau)]$ is far from $\omega_{N}[\tilde{q}(\tau)]$, then it is difficult for light to \textit{jump} from mode $N$ to mode $n$ and, consequently, the excitation of mode $n$ is negligible. Therefore, only modes $n$ with frequencies $\omega_{n}[\tilde{q}(\tau)]$ in a small band around $\omega_{N}[\tilde{q}(\tau)]$ can have non-negligible excitation. The newly excited modes $n$ have a component that oscillates at the same frequency of the original mode $N$ but with a $\pm \pi/2$ phase shift. Also, a newly excited mode has a component that oscillates at its frequency $\omega_{n}[\tilde{q}(\tau)]$ if \ $\tilde{q}''(0) \not= 0$ \ and \ $\Gamma_{mN}[\tilde{q}(0)] \not= 0$.

Furthermore, the field can be described to good approximation by just the initially excited mode $N$ but with an amplitude and a frequency that depend on the position of the membrane if (\ref{ValidezAprox1Termino}) holds. These conditions essentially amount to requiring that the membrane \textit{moves sufficiently slowly} so that the speed of the membrane divided by the difference of the squares of the frequency of mode $N$ minus that of another mode $n$ is small. 

\subsection{Illustration of the results}

In this subsection we illustrate the results in the case where the electric susceptibility has the form
\begin{eqnarray}
\label{Susceptibilidad}
\tilde{\chi} \left[ \xi - \tilde{q}(\tau) \right] &=&
\left\{
\begin{array}{cc}
\chi_{0} & \mbox{if} \ \ \vert \xi - \tilde{q}(\tau) \vert \ < \ \frac{\tilde{\delta}_{0}}{2} \ , \cr
0 & \mbox{elsewhere.}
\end{array}
\right.
\end{eqnarray}
Demanding the continuity of $G_{n}(\xi ,\tilde{q}_{0})$ and its first (partial) derivative with respect to $\xi$ at the boundaries of the membrane $\xi = \tilde{q}_{0} \pm \tilde{\delta}_{0}/2$, the modes can be calculated explicitly in terms of elementary functions:
\begin{eqnarray}
\label{EigenfuncionesMesa}
G_{n}(\xi , \tilde{q}_{0}) &=& \left\{
\begin{array}{c}
A_{1n}\mbox{sin}(\omega_{n}\xi) \qquad \cr
\qquad\qquad \mbox{if} \ \ \ 0 \leq \xi \leq \tilde{q}_{0-} \ , \cr
\cr
A_{2n}\mbox{sin}\left[ \omega_{n0}\left( \xi - \tilde{q}_{0-} \right) + \phi_{2n}(\tilde{q}_{0}) \right] \cr
\qquad\qquad \mbox{if} \ \ \ \vert \xi - \tilde{q_{0}} \vert < \frac{\tilde{\delta}_{0}}{2} \ , \cr
\cr
A_{3n}\mbox{sin}\left[ \omega_{n}\left( \xi - \tilde{q}_{0+} \right) + \phi_{3n}(\tilde{q}_{0}) \right]  \cr
\qquad\qquad \mbox{if} \ \ \  \tilde{q}_{0+} \leq \xi \leq \xi_{L} \ .
\end{array}
\right. \cr
&& 
\end{eqnarray}
with
\begin{eqnarray}
\omega_{n0} &=& \omega_{n}\sqrt{1+4\pi \chi_{0}} \ , \cr
\tilde{q}_{0\pm} &=& \tilde{q}_{0} \pm \frac{\tilde{\delta_{0}}}{2} \ , \cr
\mbox{sen}\left[ \phi_{2n}(\tilde{q}_{0}) \right] &=& \frac{A_{1n}}{A_{2n}}\mbox{sin}\left( \omega_{n}\tilde{q}_{0-} \right) \ , \cr
\mbox{cos}\left[ \phi_{2n}(\tilde{q}_{0}) \right] &=& \frac{A_{1n}}{A_{2n}}\frac{ \mbox{cos}\left( \omega_{n}\tilde{q}_{0-} \right) }{\sqrt{1 + 4\pi \chi_{0}}} \ , \cr
\mbox{sen}\left[ \phi_{3n}(\tilde{q}_{0}) \right] &=& \frac{A_{1n}}{A_{3n}}\Big[ \mbox{cos}( \omega_{n0} \tilde{\delta}_{0} ) \mbox{sin}( \omega_{n} \tilde{q}_{0-} )  \cr
&&  + \frac{\mbox{sin}( \omega_{n0} \tilde{\delta}_{0}) \mbox{cos}( \omega_{n} \tilde{q}_{0-} )}{\sqrt{1 +4\pi\chi_{0}}} \Big] \ , \cr
\psi_{0n}(\tilde{q}_{0}) &=& \mbox{cos}( \omega_{n0} \tilde{\delta}_{0} ) \mbox{cos}( \omega_{n} \tilde{q}_{0-} ) \cr
&& - \sqrt{1 +4\pi\chi_{0}} \mbox{sin}( \omega_{n0} \tilde{\delta}_{0}) \mbox{sin}( \omega_{n} \tilde{q}_{0-} ) \ , \cr
\mbox{cos}\left[ \phi_{3n}(\tilde{q}_{0}) \right] &=& \frac{A_{1n}}{A_{3n}}\psi_{0n} (\tilde{q}_{0}) \ , \cr
\frac{A_{2n}}{A_{1n}} &=& \left[ 1 - \frac{4\pi \chi_{0}}{1 + 4\pi\chi_{0}}\mbox{cos}^{2}(\omega_{n}\tilde{q}_{0-}) \right]^{1/2} \cr
\frac{A_{3n}}{A_{2n}} &=& \left[ 1 + \frac{4\pi \chi_{0}}{1 + 4\pi\chi_{0}}\left( \frac{A_{1n}}{A_{2n}} \right)^{2} \psi_{0n}(\tilde{q}_{0})^{2} \right]^{1/2} \ . \cr
&&
\end{eqnarray}
Here $A_{1n}$ is obtained by requiring $G_{n}(\xi , \tilde{q}_{0})$ to be normalized, that is, by demanding that
\begin{eqnarray}
\label{OrtSE}
\int_{0}^{\xi_{L}}d\xi \ \tilde{\epsilon}(\xi - \tilde{q}_{0})  G_{n}(\xi , \tilde{q}_{0})^{2} &=& 1 \ .
\end{eqnarray}
An analytic expression for $A_{1n}$ can be given, but it is quite lengthy and can be calculated with ease using the expressions above and a symbolic programming language such as \textit{Mathematica}. For these reasons we have chosen not to present it.
 
Moreover, \ $\omega_{n} = \omega_{n}(\tilde{q}_{0})$ \ are functions of $\tilde{q}_{0}$. They are the solutions of the transcendental equation
\begin{eqnarray}
\label{EcTrascendente}
&& \mbox{sin}\left[ \omega_{n}(\xi_{L} -\tilde{q}_{0+}) \right] \times \cr
&& \times \left[ \mbox{sin}(\omega_{n}\tilde{q}_{0-})\mbox{sin}(\omega_{n0}\tilde{\delta}_{0})  - \frac{\mbox{cos}(
\omega_{n0}\tilde{\delta}_{0})}{\sqrt{1 + 4\pi\chi_{0}}} \mbox{cos}(\omega_{n}\tilde{q}_{0-}) \right] \cr
&=&  \frac{\mbox{cos}[ \omega_{n}(\xi_{L} - \tilde{q}_{0+}) ]}{\sqrt{1 + 4\pi \chi_{0}}}\times \cr
&& \times \left[ \mbox{sin}(\omega_{n}\tilde{q}_{0-})\mbox{cos}(\omega_{n0}\tilde{\delta}_{0})  + \frac{\mbox{sin}(\omega_{n0}\tilde{\delta}_{0})}{\sqrt{1 + 4\pi\chi_{0}}} \mbox{cos}(\omega_{n}\tilde{q}_{0-}) \right] \ .\cr
&&
\end{eqnarray} 

In the rest of this subsection we take the following parameters:
\begin{eqnarray}
\label{ParametrosF}
\xi_{L} &=& 20 \ , \ \ \ \tilde{\delta}_{0} \ =\ 1 \ , \ \ \ \epsilon_{\mbox{\tiny pert}} \ = \ 10^{-1} \ , \ \ \ \lambda_{0} \ = \ 1 \mbox{cm} , \cr
\chi_{0} &=& 10 \ , \ \ \ g_{0N} \ = \ 1 \ , \ \ \ \ \ g_{1N} \ = \ 0 \ .
\end{eqnarray}

We calculated the exact frequencies $\omega_{n}$ numerically using the parameters in (\ref{ParametrosF}) and equation (\ref{EcTrascendente}). Figure \ref{Figure2} illustrates the lowest $10$ (non-dimensional) angular frequencies $\omega_{n}$ of the cavity as a function of the position $\tilde{q}$ of the midpoint of the membrane. Notice that the $\omega_{n}$ are smooth functions of $\tilde{q}$ and that there are no crossings between frequencies.

\begin{figure}
\subfloat[$\tau = 0$]{\label{Figure3a} \includegraphics[width=8cm]{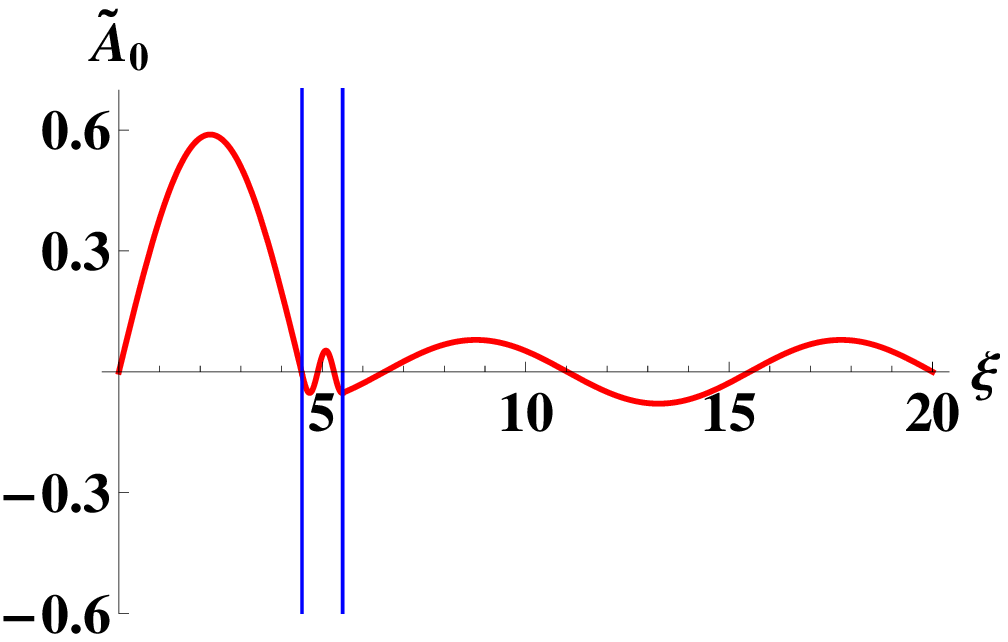}}\\
\subfloat[$\tau = 2.5$]{\label{Figure3b} \includegraphics[width=8cm]{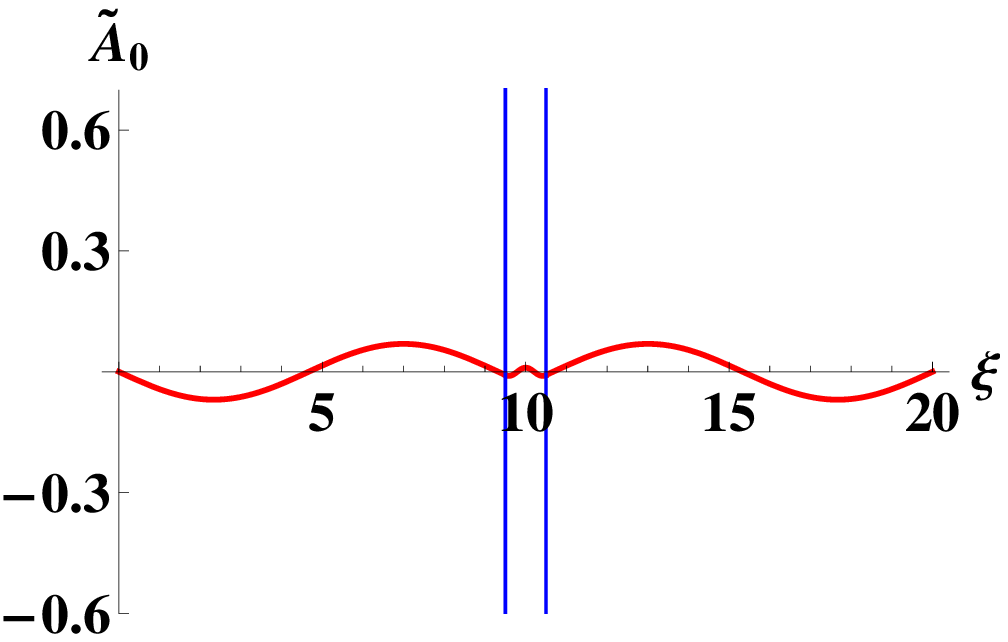}}\\
\subfloat[$\tau = 5$]{\label{Figure3c} \includegraphics[width=8cm]{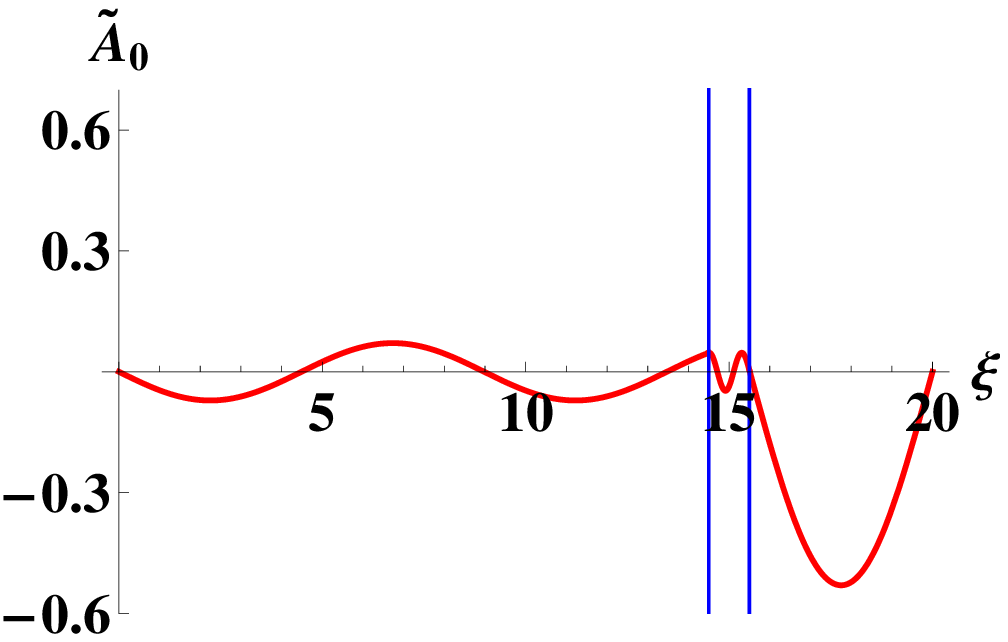}}
\caption{\label{Figure3} The figures show $\tilde{A}_{0}(\xi ,\tau)$ (red line) inside the cavity for several values of $\tau$ using the electric susceptibility in (\ref{Susceptibilidad}), the first-term approximation in (\ref{A01}), the parameters in (\ref{ParametrosF}), and $\tilde{q}(\tau)$ given in (\ref{IMotion}). The frequency of the excited instantaneous mode is $\omega_{7}$ (the $7$th lowest frequency) and it is illustrated in Figure \ref{Figure2}. The corresponding values of $\tau$ are indicated in each figure and the vertical blue lines indicate the boundaries of the membrane.}
\end{figure}

Finally, Figure \ref{Figure3} illustrates the evolution of $\tilde{A}_{0}(\xi ,\tau)$ using the first term approximation in (\ref{A01}) when the membrane follows the motion
\begin{eqnarray}
\label{IMotion}
\tilde{q}(\tau) &=& 10 -5\mbox{cos}(2\pi \epsilon_{\mbox{\tiny pert}} \tau ) \ .
\end{eqnarray}
Notice that the membrane follows a harmonic oscillation of large amplitude centered at the midpoint of the cavity.

Figure \ref{Figure3a} (\ref{Figure3c}) illustrates $\tilde{A}_{0}(\xi , \tau)$ when $\tilde{q}(\tau)$ is closest to the left (right) mirror, while Figure \ref{Figure3b} shows $\tilde{A}_{0}(\xi , \tau)$ when $\tilde{q}(\tau)$ is at the midpoint of the cavity. Notice that $\tilde{A}_{0}(\xi , \tau)$ is larger in the left (right) side of the cavity in Figure \ref{Figure3a} (\ref{Figure3c}). On may interpret this behavior as \textit{light being pushed from one side of the cavity to the other} by the motion of the membrane. A similar effect has been studied in quantum optomechanics \cite{PhotonShuttle}.

\section{The case of several modes}

In Sections IV and V we assumed that only one mode of the cavity was initially excited, namely mode $N$. In this section we consider the case where several modes can be initially excited.

We proceed in the same way as before, first expanding the potential $\tilde{A}_{0}(\xi , \tau)$ in terms of the modes, see (\ref{ExpansionEnModos}), and then obtaining the equations (\ref{10}) for the coefficients $c_{m}(\tau)$. The difference appears in the initial conditions (\ref{13}). They have to be replaced by 
\begin{eqnarray}
\label{MuchosModos}
c_{m}(0) &=& g_{0m} \ , \ \ c_{m}'(0) \ = \ g_{1m} \ ,
\end{eqnarray}
with $g_{0m}$ and $g_{1m}$ real numbers.

Now let $N$ be arbitrary but fixed. Consider the initial value problem with the differential equations in (\ref{10}) and the initial conditions
\begin{eqnarray}
\label{MuchosModos2}
c_{m}(0) &=& g_{0N}\delta_{mN} \ , \ \ c_{m}'(0) \ = \ g_{1N}\delta_{mN} \ .
\end{eqnarray}
This initial value problem was solved to good approximation in Sections IV and V. Let $c_{m,N}(\tau)$ denote its exact solution and $c_{m,N}^{(j)}(\tau)$ denote the $j$-term approximate solution. Since the differential equations in (\ref{10}) are linear, it follows that the exact solution to the initial value problem in (\ref{10}) and (\ref{MuchosModos}) is given by 
\begin{eqnarray}
\label{MuchosModos3}
c_{m}(\tau) &=& \sum_{N=1}^{+\infty} c_{m,N}(\tau) \ , 
\end{eqnarray} 
while a $j$-term approximate solution is given by 
\begin{eqnarray}
\label{MuchosModos4}
c_{m}(\tau) &\simeq& \sum_{N=1}^{+\infty} c_{m,N}^{(j)}(\tau) \ . 
\end{eqnarray} 

We now use the first-term approximation to $c_{m,N}(\tau)$. First observe from (\ref{Aprox1termino}) that
\begin{eqnarray}
\label{MuchosModos5}
c_{m,N}^{(1)}(\tau) &=& \left\{ \alpha_{N}\left[ \tilde{q}(\tau) \right] b_{N0} e^{-i[ t_{1N}(\tau) - \Theta_{N0}] } \ + \ c.c. \right\} \delta_{mN} \ , \cr
&&
\end{eqnarray} 
where $\alpha_{N}[ \tilde{q}(\tau)]$, $b_{N0}$, and $\Theta_{N0}$ are given in (\ref{Aprox1terminoCoef}) and $t_{1N}$ is defined in (\ref{Aprox3TerminosCoef}).

Substituting (\ref{MuchosModos5}) in the expansion (\ref{ExpansionEnModos}) for $\tilde{A}_{0}(\xi, \tau)$, we conclude that
the potential is approximately given by
\begin{eqnarray}
\label{MuchosModos6}
\tilde{A}_{0}(\xi , \tau)  &\simeq&  
\sum_{N=1}^{+\infty}\alpha_{N} \left[ \tilde{q}(\tau) \right] b_{N0} e^{-i[ t_{1N}(\tau) - \Theta_{N0}] } G_{N}[ \xi , \tilde{q}(\tau) ] \cr
&& \cr
&& + \ c.c. .
\end{eqnarray}
We end this section by noting that (\ref{MuchosModos6}) is an accurate approximation as long as (\ref{ValidezAprox1Termino}) holds for each $N$ with a non-zero coefficient involved in (\ref{MuchosModos6}).


\section{Conclusions}

In this article we considered a one-dimensional cavity composed of two perfect, fixed mirrors and a mobile membrane in between. Moreover, we assumed that the mirrors and the membrane are slabs of infinite length and width that are parallel to each other with vacuum between the membrane and the mirrors. We modelled the membrane as a linear, isotropic, non-magnetizable, non-conducting, and uncharged dielectric of thickness $\delta_{0}$ when it is at rest. Furthermore, the membrane can only move along the axis of the cavity, that is, along the line perpendicular to the mirrors and the membrane (the $x$-axis in the article). Also, we assumed that there is an electromagnetic field inside the cavity that can be deduced from a zero scalar potential and from a vector potential with direction along a line perpendicular to the axis of the cavity (the $z$-axis in the article). 

Since the membrane can move, the evolution of the vector potential is determined by a wave equation with time-dependent coefficients and modified by terms proportional to the velocity and acceleration of the membrane. Assuming that the membrane starts to move from rest and that the membrane moves appreciably in a time-scale much larger than the time-scale in which the field evolves appreciably, we were able to solve to good approximation the aforementioned equation for the vector potential using the method of multiple scales. This method allowed us to obtain simple analytic formulas that provide physical insight for the evolution of the electromagnetic field. We now describe the physical picture provided by our results. Suppose that a single-mode of the cavity is initially excited. In a first approximation the vector potential follows the initially excited mode with an amplitude and a phase that depend on the position of the membrane. In a more accurate approximation, the amplitude and phase of the original mode include small corrections due to the velocity and acceleration of the membrane. In particular, the phase includes a Doppler-like phase shift. Also, all the other modes are excited. The reason for this is the following. As the membrane moves, the frequencies of the other modes may approach or move away from that of the initially excited mode. If they are near, then it is easy for light to pass from the initially excited mode to the other mode and this other mode becomes excited. On the other hand, if they are far from each other, then it is difficult for light to pass from the initially excited mode to the other mode and this other mode has negligible excitation. Therefore, only those modes with frequency in a small band around the frequency of the initially excited mode can have non-negligible excitation. Furthermore, the newly excited modes have a component that evolves with the phase of the original mode with a $\pm \pi/2$ phase shift and, in general, they also have a component that evolves at their own frequency.

Also, we deduced the conditions under which the field can be described to good approximation by just the initially excited mode but with an amplitude and frequency that depend on the position of the membrane. These essentially amount to requiring that the membrane \textit{moves sufficiently slowly} so that the speed of the membrane divided by the difference of the squares of the frequency of the originally excited mode minus that of another mode is small. Furthermore, we generalized the results for the case where there are several initially excited modes. 

Finally, we emphasize that the approximate solutions obtained take into account both the velocity and the acceleration of the membrane and that they are valid for arbitrary displacements of the membrane, that is, they are not restricted to small deviations of the membrane around an equilibrium position. In these formulas describing the evolution of the field, the motion of the membrane can be determined by an external agent or, alternatively, they can be substituted in the self-consistent system of equations governing the dynamics of the membrane and the field deduced in \cite{PhysicaScripta}. In particular, the approximate expressions obtained allow one to investigate the position and time dependent correction to the mass of the membrane and the velocity dependent force affecting the membrane, effects that arise from the motion of the membrane and its coupling to the field \cite{PhysicaScripta}. This is the subject of future work. We emphasize that the only limitation of the results obtained is that the membrane must move \textit{sufficiently slowly} so that the assumption of the different time-scales of evolution of the membrane and the field mentioned above holds.

\begin{acknowledgments}
Luis Octavio Casta\~{n}os wishes to thank the IIMAS of the UNAM and the ITESM for support. Luis Octavio Casta\~{n}os and Ricardo Weder are fellows of the \textit{Sistema Nacional de Investigadores}. Research partially done while Luis Octavio Casta\~{n}os was affiliated to the IIMAS. Research partially supported by project PAPIIT-DGAPA UNAM IN102215.
\end{acknowledgments}


\appendix

\section{Modes}
Consider the Sturm-Liouville problem in (\ref{BVPmodos}).  Assume that $ \tilde{\chi}$ is Lebesgue measurable, nonnegative, and bounded  (this is true in particular if $\tilde{\chi}$ is nonnegative and piecewise continuous). Then, for some constant $c_1$ and any fixed $\tilde{q} \in [0,\xi_{L}]$, one has $ 1 \leq \tilde\varepsilon(\xi- \tilde{q}) \leq c_1$ for  almost every $ \xi \in [0,\xi_{L}]$. 

Denote by $L^{2}$ the standard space of Lebesgue measurable, square integrable functions on $(0,\xi_{L} )$, and by $L^2_{\tilde{q}}$ with $\tilde{q} \in [0,\xi_{L}]$,  the space $L^{2}$ endowed with the scalar product,
\begin{equation}
\label{DosN}
\left(\varphi, \psi\right)_{\tilde{q}} \ = \ \int_{0}^{\xi_{L}} \, \varphi(\xi) \, \psi(\xi)^{*}\, \tilde\varepsilon(\xi-\tilde{q})\,  d\xi.
\end{equation}
Note that the norms of $L^2$ and of $L^2_{\tilde{q}}$ are equivalent:
\begin{eqnarray}
\left\| \varphi  \right\|_{L^2} \ \leq \  \left\| \varphi  \right\|_{L^2_{\tilde{q}}} \ \leq \   c_1  \left\| \varphi  \right\|_{L^2}. 
\nonumber
\end{eqnarray}
  
Denote by $H_2$ the second Sobolev space on $(0,\xi_{L})$ and by $H_{1,0}$ the first Sobolev space of functions on $(0, \xi_{L})$ that vanish for $\xi=0, \xi_{L}$. For the definitions see \cite{Sobolev}. Also, define the following operator in $L^2_{\tilde{q}}$ 
\begin{equation}
\label{TresN}
\mathbb{A}_{\tilde{q}} \varphi \ = \ - \frac{1}{\tilde\varepsilon(\xi- \tilde{q})}\, \frac{d^2}{d \xi^2}\, \varphi,
\end{equation}
with domain $D(\mathbb{A}_{\tilde{q}})$ given by
\begin{equation}
\label{CuatroN}
D(\mathbb{A}_{\tilde{q}}) = H_2 \cap H_{1,0}.
\end{equation}
The operator   $\mathbb{A}_{\tilde{q}}$ is self-adjoint. Actually, it is the operator associated to the quadratic form
\begin{eqnarray}
h_{\tilde{q}}\left( \varphi, \psi \right) \ = \ \int_{0}^{\xi_{L}}\, d\xi \varphi' (\xi) \psi' (\xi)^{*}, \qquad \varphi, \psi \in H_{1,0}.
\nonumber
\end{eqnarray}
See Chapter 6 of \cite{Kato}. Moreover, by the Rellich compactness theorem \cite{Sobolev}, the imbedding of $ D(\mathbb{A}_{\tilde{q}})$ into  $L^2_{\tilde{q}}$ is compact, and it follows that the resolvent \ $R_{\tilde{q}}(z)= \left(\mathbb{A}_{\tilde{q}} -z\right)^{-1}$ \ is compact for $z$ in the resolvent set. From the results of Section 6 of Chapter 3 of \cite{Kato} it follows that the spectrum of $\mathbb{A}_{\tilde{q}}$ consists of isolated eigenvalues of finite multiplicity that can only accumulate at $\infty$.  Note that the eigenvalues of  $\mathbb{A}_{\tilde{q}}$ are precisely the squared eigenfrequencies \ $\omega_n(\tilde{q})^2$ \ $(n=1,2, ...)$ \ and, hence, they are of multiplicity one.

Denote by $J_{\tilde{q}}$ the identification operator from  $L^2_{\tilde{q}}$ onto $L^2$ given by,
\begin{eqnarray}
J_{\tilde{q}}\, \varphi(\xi) = \varphi(\xi), \qquad \varphi \in L^2_{\tilde{q}}. \nonumber
\end{eqnarray}
Clearly,  $J_{\tilde{q}}$ is invertible with bounded inverse. Let us denote by $\mathbb{B}_{\tilde{q}}$ the operator,
\begin{eqnarray}
\mathbb{B}_{\tilde{q}} =  J_{\tilde{q}} \mathbb{A}_{\tilde{q}} J_{\tilde{q}}^{-1}. 
\end{eqnarray}
The operator $\mathbb{B}_{\tilde{q}}$ can be understood as the operator $\mathbb{A}_{\tilde{q}}$ but seen as acting on the space $L^2$ whose norm is equivalent to the norm of $L^2_{\tilde{q}}$. Since $\mathbb{B}_{\tilde{q}}$ and  $\mathbb{A}_{\tilde{q}}$ are related by a similarity transformation, they have the same spectrum. Hence, the spectrum of $\mathbb{B}_{\tilde{q}}$ is composed of the squared eigenfrequencies $\omega_n(\tilde{q})^2$ \, $(n=1,2,...)$. 

Let us assume that $\tilde{\chi}$ is  continuous for all $ \xi \in [0,\xi_{L}] \setminus M$, where $M$ is a set of Lebesgue measure zero (of course, this will be true if  
$\tilde{\chi}$
 is piecewise continuous). Consequently,
\begin{equation}
\label{CincoN}
\lim_{\tilde{q} \rightarrow \tilde{q}_0} \tilde\varepsilon(\xi-\tilde{q})=  \tilde\varepsilon(\xi-\tilde{q}_0), \qquad \forall \xi \not \in \tilde{q}_0+M. 
\end{equation}
Moreover, by  (\ref{CincoN}),
\begin{eqnarray}
 \lim_{\tilde{q}\rightarrow \tilde{q}_0} \mathbb{B}_{\tilde{q}}\, \varphi =  \mathbb{B}_{\tilde{q}_0}\, \varphi, \qquad \forall \varphi \in D\left( \mathbb{B}_{\tilde{q}_0} \right)=
H_2 \cap H_{1,0},
\nonumber
\end{eqnarray}
where the limit is taken in the norm of $L^2$. It follows that the family of closed operators  $\mathbb{B}_{\tilde{q}}$ with $D\left(  \mathbb{B}_{\tilde{q}}  \right)= H_2 \cap H_{1,0}$ and $\tilde{q}  \in [0,\xi_{L}]$ is  strongly continuous. Let us denote by $ \hat{R}_{\tilde{q}}(z)= \left( \mathbb{B}_{\tilde{q}}-z\right)^{-1}$ the resolvent of $\mathbb{B}_q$.  By the second resolvent equation, for any $ z \in \mathbb{C} \setminus [0,\infty)$,
 $$
 \begin{array}{l}
 \hat{R}_{\tilde{q}}(z)- \hat{R}_{\tilde{q}_0}(z)= \\\\ - \left\{ \hat{R}_{\tilde{q}_0}(z)  \, \left[\frac{1}{\tilde\varepsilon(\xi-\tilde{q})}-\frac{1}{ \tilde\varepsilon(\xi-\tilde{q}_0)}\right] \right\} \frac{d^2}{d \xi^2} \hat{R}_{\tilde{q}}(z).
 \end{array}
 $$
 Since $D(\mathbb{B}_{\tilde{q}}) \subset H_2$, we have that,
 $$
 \left\| \frac{d^2}{d \xi^2} \hat{R}_{\tilde{q}}(z)\right\| \leq C,
 $$
  for a constant $C$ independent of $ \tilde{q} \in[0,\xi_{L}]$. Moreover, by (\ref{CincoN}) and since $ \hat{R}_{\tilde{q}_0}(z)$ is a compact operator,
  $$
  \lim_{\tilde{q} \rightarrow \tilde{q}_0} \left\|  \ \hat{R}_{\tilde{q}_0}(z)\, \left[ \frac{1}{\tilde\varepsilon(\xi-\tilde{q})}-\frac{1} {\tilde\varepsilon(\xi-\tilde{q}_0)}\right] \ \right\|=0.
  $$
  Hence, $\mathbb{B}_{\tilde{q}}$ converges to $\mathbb{B}_{\tilde{q}_0}$ in norm resolvent sense. Then, one can take the following results in \cite{Kato}: Chapter 2, Section 4.2 (in particular equation (4.18)); Chapter 4, Theorem 2.25 in Section 2.6 and Section 3.5; and Chapter 7, Section 1.3. It follows that we can apply to our problem the results in Sections 5.1 to 5.5 of Chapter 2 of \cite{Kato}. Note that in our case the transformation operator considered in Section 1.3 of Chapter 7 of \cite{Kato} is only required to be continuous and, moreover, that since $\mathbb{B}_{\tilde{q}}$ is similar to the selfadjoint operator $\mathbb{A}_{\tilde{q}}$, its eigenvalues are semisimple.     
  Then, by Theorem 5.1 in Section 5 of Chapter II of \cite{Kato}, the functions $\omega_n(\tilde{q})$ \ $(n= 1,2,...)$ are continuous and there are no crossings because the eigenvalues are of multiplicity $1$, that is, \ $\omega_n({\tilde{q}}) \neq   \omega_m({\tilde{q}})$ \ for all \ $\tilde{q} \in [0,\xi_{L}]$ \ and \ $n \neq  m$ \ $(m,n = 1, 2, ...)$. 
Furthermore, the eigenvectors, $G_n(\xi, \tilde{q})$ \  $(n=1,2,...)$ are continuous in $\tilde{q}$ (note that, since the eigenvalues are of multiplicity one, the projectors onto the $ \lambda$-groups are one dimensional).

Suppose, furthermore, that $\tilde\chi(\xi)$ is  piecewise absolutely continuous , with bounded derivative  (clearly, this will be true if  $\tilde\chi(\xi)$ is piecewise continuous with piecewise continuous derivative). In this case the operator family $\mathbb{B}_{\tilde{q}}$ is strongly continuously differentiable and continuously differentiable in norm resolvente sense. Then, by Theorem 5.4 in Chapter 2 of \cite{Kato}, the functions $\omega_n(\tilde{q})$ \ $(n= 1,2,...)$ are continuously differentiable, and, by Remark 5.10 of Chapter 2 of \cite{Kato}, the eigenvectors $G_n(\xi, \tilde{q})$ $(n=1,2,...)$ are continuously differentiable in $\tilde{q}$.

\section{Two time-scales}

The purpose of this appendix is to deduce a two term approximation of the solution of (\ref{12}) and (\ref{13}) using two time-scales. Throughout this appendix we use the mode-dependent quantities defined in (\ref{ModeDependent}).

Since the coefficients of $c_{m}(\tau)$ in (\ref{12}) are functions of $\epsilon_{\mbox{\tiny pert}}\tau$ and \ $0 < \epsilon_{\mbox{\tiny pert}} \ll 1$ \ (see item 1 in Section IV), it is clear that the slow time-scale is given by $\epsilon_{\mbox{\tiny pert}} \tau$, but the fast time-scale remains to be determined. Therefore, we consider the two time-scales
\begin{eqnarray}
\label{A2e15}
t_{1N} \ = \ f(\tau , \epsilon_{\mbox{\tiny pert}} ) \ , \ \ t_{2} \ = \ \epsilon_{\mbox{\tiny pert}} \tau \ ,
\end{eqnarray}
\noindent
where
\begin{enumerate}
\item $f(\tau , \epsilon_{\mbox{\tiny pert}} ) \geq 0$ for all $\tau \geq 0$.\\
      We demand this property because $\tau \geq 0$ and, therefore, we want $t_{1N} \geq 0$.
\item $(\partial f/\partial \tau )(\tau , \epsilon_{\mbox{\tiny pert}} ) > 0$ for all $\tau > 0$.\\
      We demand this property because the non-dimensional time $\tau$ is strictly increasing and we want the fast time-scale $t_{1N}$ to be strictly increasing. 
\item $f(0,\epsilon_{\mbox{\tiny pert}}) = 0$.\\
      We demand this property because we want the fast time-scale $t_{1N}$ to be zero when $\tau = 0$.
\item For fixed $\epsilon_{\mbox{\tiny pert}}$, $f(\tau , \epsilon_{\mbox{\tiny pert}} )$ is two times continuously differentiable as a function of $\tau$.\\
      We demand this property because two derivatives with respect to $\tau$ appear in the equations for $c_{m}(\tau)$ given in (\ref{12}).
\item For fixed $\tau$ we have $\epsilon_{\mbox{\tiny pert}}\tau \ll f(\tau , \epsilon_{\mbox{\tiny pert}} )$ as $\epsilon_{\mbox{\tiny pert}} \downarrow 0$.\\
      Here $\epsilon \downarrow 0$ means that $\epsilon_{\mbox{\tiny pert}}$ tends to zero through positive values. We demand this property because we want $t_{1N}$ to be the fast time-scale and $t_{2}$ to be the slow time-scale. 
\end{enumerate}

For each $m=1,2,...$ we define the function $d_{m}(t_{1N} , t_{2} )$ by 
\begin{eqnarray}
\label{A2e16}
d_{m}\left[ t_{1N}(\tau) , t_{2}(\tau) \right] \ = \ c_{m}(\tau) \qquad (\tau \geq 0).
\end{eqnarray}
We are now going to deduce the initial conditions and the system of differential equations that determine the evolution of the $d_{m}(t_{1N} , t_{2} )$. Substituting (\ref{A2e16}) in the initial conditions for $c_{m}(\tau)$ given in (\ref{13}), we obtain the corresponding initial conditions for $d_{m}(t_{1N} , t_{2} )$
\begin{eqnarray}
\label{A2e17} 
g_{1N}\delta_{mN} &=& \frac{\partial f}{\partial \tau}(0, \epsilon_{\mbox{\tiny pert}} ) \frac{\partial d_{m}}{\partial t_{1N}}(0, 0) + \epsilon_{\mbox{\tiny pert}} \frac{\partial d_{m}}{\partial t_{2}}(0, 0) \ , \cr
g_{0N}\delta_{mN} &=& d_{m}(0,0) \ . 
\end{eqnarray}

Substituting (\ref{A2e16}) in the system of differential equations for $c_{m}(\tau)$ given in (\ref{12}), we obtain the corresponding system of differential equations for $d_{m}(t_{1N} , t_{2} )$

\begin{widetext}
\begin{eqnarray}
\label{A2e18}
&& \left[ \frac{\partial f}{\partial \tau}(\tau , \epsilon_{\mbox{\tiny pert}}) \right]^{2}\frac{\partial^{2}d_{m}}{\partial t_{1N}^{2}}\left[ t_{1N}(\tau) , t_{2}(\tau) \right] \ + \ \left[ \frac{\partial^{2} f}{\partial \tau^{2}}(\tau , \epsilon_{\mbox{\tiny pert}}) \right]\frac{\partial d_{m}}{\partial t_{1N}}\left[ t_{1N}(\tau) , t_{2}(\tau) \right] \ + \ \omega_{\scriptscriptstyle{m}}\left[ \tilde{\tilde{q}}(\epsilon_{\mbox{\tiny pert}}\tau)\right]^{2}d_{m}\left[ t_{1N}(\tau) , t_{2}(\tau) \right] \cr
&& \cr
&& +2\epsilon_{\mbox{\tiny pert}} \left[ \frac{\partial f}{\partial \tau}(\tau , \epsilon_{\mbox{\tiny pert}}) \right]\frac{\partial^{2}d_{m}}{\partial t_{2} \partial t_{1N}}\left[ t_{1N}(\tau) , t_{2}(\tau) \right] \ + \ 2\epsilon_{\mbox{\tiny pert}}\tilde{\tilde{q}}'(\epsilon_{\mbox{\tiny pert}}\tau) \Gamma_{\scriptscriptstyle{mm}}\left[ \tilde{\tilde{q}}(\epsilon_{\mbox{\tiny pert}}\tau) \right] \left[ \frac{\partial f}{\partial \tau}(\tau , \epsilon_{\mbox{\tiny pert}})\right] \frac{\partial d_{m}}{\partial t_{1N}} \left[ t_{1N}(\tau) , t_{2}(\tau) \right] \cr
&& \cr
&& + \epsilon_{\mbox{\tiny pert}}^{2} \frac{\partial^{2}d_{m}}{\partial t_{2}^{2}}\left[ t_{1N}(\tau) , t_{2}(\tau) \right] \ + \ 2\epsilon_{\mbox{\tiny pert}}^{2}\tilde{\tilde{q}}'(\epsilon_{\mbox{\tiny pert}}\tau) \Gamma_{\scriptscriptstyle{mm}}\left[ \tilde{\tilde{q}}(\epsilon_{\mbox{\tiny pert}}\tau)\right] \frac{\partial d_{m}}{\partial t_{2}} \left[ t_{1N}(\tau) , t_{2}(\tau) \right] \cr
&&\cr
&& + \ \epsilon_{\mbox{\tiny pert}}^{2}\tilde{\tilde{q}}''(\epsilon_{\mbox{\tiny pert}}\tau) \Gamma_{\scriptscriptstyle{mm}}\left[ \tilde{\tilde{q}}(\epsilon_{\mbox{\tiny pert}}\tau)\right] d_{m}\left[ t_{1N}(\tau) , t_{2}(\tau) \right] \cr
&& \cr
&=& -\epsilon_{\mbox{\tiny pert}} \sum_{n=1 \atop n\not= m}^{+ \infty} \Gamma_{\scriptscriptstyle{mn}}\left[ \tilde{\tilde{q}}(\epsilon_{\mbox{\tiny pert}}\tau)\right] \left\{ \ 2\tilde{\tilde{q}}'(\epsilon_{\mbox{\tiny pert}}\tau) \left[ \frac{\partial f}{\partial \tau}(\tau , \epsilon_{\mbox{\tiny pert}})  \right] \frac{\partial d_{n}}{\partial t_{1N}} \left[ t_{1N}(\tau) , t_{2}(\tau) \right] \right. \cr
&& \cr
&& \left. \qquad\qquad\qquad\qquad + 2\epsilon_{\mbox{\tiny pert}}\tilde{\tilde{q}}'(\epsilon_{\mbox{\tiny pert}}\tau) \frac{\partial d_{n}}{\partial t_{2}} \left[ t_{1N}(\tau) , t_{2}(\tau) \right] \ + \ \epsilon_{\mbox{\tiny pert}} \tilde{\tilde{q}}''(\epsilon_{\mbox{\tiny pert}}\tau)d_{n} \left[ t_{1N}(\tau) , t_{2}(\tau) \right] \ \right\} \ .
\end{eqnarray}
\end{widetext}

We now obtain $f(\tau , \epsilon_{\mbox{\tiny pert}})$ from (\ref{A2e18}). First observe that in the case $\epsilon_{\mbox{\tiny pert}} \downarrow 0$ all the terms of (\ref{A2e18}) cancel out except for those in the first line. Now recall from (\ref{Adimensionales}) that $\epsilon_{\mbox{\tiny pert}} = \nu_{\mbox{\tiny osc}}/\nu_{0}$, that is, $\epsilon_{\mbox{\tiny pert}}$ is the time-scale $1/\nu_{0}$ in which the field evolves appreciably divided by the time-scale $1/\nu_{\mbox{\tiny osc}}$ in which the membrane evolves appreciably. Hence, $\epsilon_{\mbox{\tiny pert}} \downarrow 0$ describes the case where the field evolves infinitely faster than the membrane and, therefore, corresponds to the situation where the membrane is fixed. Moreover, we know that the coefficients $c_{m}(\tau) = d_{m}\left[ t_{1N}(\tau) , t_{2}(\tau) \right]$ of the modes must satisfy harmonic oscillator equations when the membrane is fixed (see (\ref{Pi2b}) in Section III). Hence, one must recover a harmonic oscillator equation (with angular frequency equal to $1$) for $d_{m}\left[ t_{1N}(\tau) , t_{2}(\tau) \right]$ in the fast time-scale $t_{1N}$. From (\ref{A2e18}) we see that this happens if and only if
\begin{eqnarray}
\label{A2e20b}
&& \left[ \frac{\partial f}{\partial \tau}(\tau , \epsilon_{\mbox{\tiny pert}}) \right]^{2} \ = \ \omega_{m}\left[ \tilde{\tilde{q}}(\epsilon_{\mbox{\tiny pert}}\tau ) \right]^{2} \ , \qquad (\tau \geq 0) \cr
&& \cr
&\Leftrightarrow& f(\tau , \epsilon_{\mbox{\tiny pert}}) \ = \ \int_{0}^{\tau}d\tau' \ \omega_{m}\left[ \tilde{\tilde{q}}(\epsilon_{\mbox{\tiny pert}}\tau' ) \right] \ , \qquad (\tau \geq 0). \cr
&&
\end{eqnarray}
Notice that to deduce the last line we used that both $(\partial f /\partial \tau )(\tau , \epsilon_{\mbox{\tiny pert}})$ and $\omega_{m}\left[ \tilde{\tilde{q}}(\epsilon_{\mbox{\tiny pert}}\tau ) \right]$ are positive, as well as the fact that $f(0,\epsilon_{\mbox{\tiny pert}}) = 0$, see items 2 and 3 above and item 1 in Section III. 

From (\ref{A2e20b}) we find that the fast time-scale is not uniquely defined. This is to be expected because every mode should have its own fast time-scale determined by its corresponding frequency $\omega_{m}\left[ \tilde{\tilde{q}}(\epsilon_{\mbox{\tiny pert}}\tau' ) \right]$. Since the initial conditions in (\ref{A2e17}) indicate that only mode $N$ is initially excited, from (\ref{A2e20b}) we find that the appropriate choice for the fast time-scale is
\begin{eqnarray}
\label{A2e20}
t_{1N} \ = \ f(\tau , \epsilon_{\mbox{\tiny pert}}) \ = \ \int_{0}^{\tau}d\tau' \ \omega_{N}\left[ \tilde{\tilde{q}}(\epsilon_{\mbox{\tiny pert}}\tau' ) \right] \ . \cr
&&
\end{eqnarray}

There are certain quantities that will be appearing quite frequently in the calculations that follow. Therefore, it is convenient to define them now and make reference to them as we proceed. We benefit from this because the results will have more succinct and transparent expressions. For each \ $m =1,2,...$ \ we define
\begin{eqnarray}
\label{A2eOp}
\mathcal{L}_{m0} &=& \omega_{\scriptscriptstyle{N}}\left[ \tilde{\tilde{q}}(t_{2})\right]^{2}\frac{\partial^{2}}{\partial t_{1N}^{2}} + \omega_{\scriptscriptstyle{m}}\left[ \tilde{\tilde{q}}(t_{2})\right]^{2} , \cr
&& \cr
\zeta_{m}(t_{2}) &=& \Gamma_{\scriptscriptstyle{mm}}\left[ \tilde{\tilde{q}}(t_{2}) \right]   +  \frac{\omega_{\scriptscriptstyle{N}}'\left[ \tilde{\tilde{q}}(t_{2})\right]}{2\omega_{\scriptscriptstyle{N}}\left[ \tilde{\tilde{q}}(t_{2})\right]} \ , \cr
&& \cr
\mathcal{L}_{m1} &=& 2\omega_{\scriptscriptstyle{N}}\left[ \tilde{\tilde{q}}(t_{2})\right]\left[ \frac{\partial^{2}}{\partial t_{2} \partial t_{1N}}
+  \tilde{\tilde{q}}'(t_{2})\zeta_{m}(t_{2})\frac{\partial }{\partial t_{1N}} \right] \cr
&& \cr
\mathcal{L}_{m2} &=& \frac{\partial^{2}}{\partial t_{2}^{2}} \ + \ 2\tilde{\tilde{q}}'(t_{2}) \Gamma_{\scriptscriptstyle{mm}}\left[ \tilde{\tilde{q}}(t_{2})\right] \frac{\partial }{\partial t_{2}} \cr 
&& + \ \tilde{\tilde{q}}''(t_{2}) \Gamma_{\scriptscriptstyle{mm}}\left[ \tilde{\tilde{q}}(t_{2})\right] , \cr
&& \cr
\mathcal{L}_{mn1} &=& 2\tilde{\tilde{q}}'(t_{2})\omega_{\scriptscriptstyle{N}}\left[ \tilde{\tilde{q}}(t_{2})\right] \Gamma_{\scriptscriptstyle{mn}}\left[ \tilde{\tilde{q}}(t_{2})\right] \frac{\partial }{\partial t_{1N}} \ , \cr
&& \cr
\mathcal{L}_{mn2} &=& 2\tilde{\tilde{q}}'(t_{2}) \Gamma_{\scriptscriptstyle{mn}}\left[ \tilde{\tilde{q}}(t_{2})\right] \frac{\partial }{\partial t_{2}} + \tilde{\tilde{q}}''(t_{2})\Gamma_{\scriptscriptstyle{mn}}\left[ \tilde{\tilde{q}}(t_{2})\right] , \cr
&& \cr
W_{m}(t_{2}) &=& \frac{\omega_{m}\left[ \tilde{\tilde{q}}(t_{2}) \right]}{\omega_{N}\left[ \tilde{\tilde{q}}(t_{2}) \right]} \ , \cr
&& \cr
T_{m}(t_{2}) &=& 2\tilde{\tilde{q}}'(t_{2})\frac{\Gamma_{\scriptscriptstyle{mN}}\left[ \tilde{\tilde{q}}(t_{2}) \right]}{\omega_{\scriptscriptstyle{N}}\left[ \tilde{\tilde{q}}(t_{2}) \right]} \ , \cr
&& \cr
\alpha_{m}\left[ \tilde{\tilde{q}}(t_{2}) \right] &=& \sqrt{\frac{\omega_{\scriptscriptstyle{N}}\left[ \tilde{\tilde{q}}(0)\right]}{\omega_{\scriptscriptstyle{N}}\left[ \tilde{\tilde{q}}(t_{2})\right]}} \times \cr
&& \times\mbox{exp}\left\{ \ - \int_{\tilde{\tilde{q}}(0)}^{{\tilde{\tilde{q}}(t_{2})}} dy \ \Gamma_{\scriptscriptstyle{mm}}(y) \ \right\} \ , \cr
&& \cr
\eta (t_{2}) &=& -i\frac{d}{dt_{2}}\left\{ \tilde{\tilde{q}}'(t_{2}) \frac{\omega_{N}'\left[ \tilde{\tilde{q}}(t_{2}) \right]}{4\omega_{N}\left[ \tilde{\tilde{q}}(t_{2}) \right]^{2}} \right\} \cr
&&  - i\tilde{\tilde{q}}'(t_{2})^{2}\phi \left[ \tilde{\tilde{q}}(t_{2}) \right] \ , \cr
&& \cr
\phi (y) &=& \frac{\Gamma_{NN}'(y)}{2\omega_{N}(y)} + \frac{\Gamma_{NN}(y)^{2}}{2\omega_{N}(y)} + \frac{\omega_{N}'(y)^{2}}{8\omega_{N}(y)^{3}} \cr
&& - 2\omega_{N}(y) \sum_{n=1 \atop n\not= N}^{+\infty} \frac{\Gamma_{Nn}(y)\Gamma_{nN}(y)}{\omega_{n}(y)^{2} - \omega_{N}(y)^{2}} \ .
\end{eqnarray}

Substituting (\ref{A2e20}) in (\ref{A2e17}) and (\ref{A2e18}), using the definitions in (\ref{A2eOp}),and expressing all quantities in terms of $t_{1N}$ and $t_{2}$ we obtain the following system of differential equations
\begin{eqnarray}
\label{A2e18b}
&& \left( \mathcal{L}_{m0} + \epsilon_{\mbox{\tiny pert}} \mathcal{L}_{m1} + \epsilon_{\mbox{\tiny pert}}^{2} \mathcal{L}_{m2} \right) d_{m}( t_{1N} , t_{2} ) \cr
&& \cr
&=& - \sum_{n=1 \atop n\not= m}^{+ \infty} \left( \epsilon_{\mbox{\tiny pert}} \mathcal{L}_{mn1} + \epsilon_{\mbox{\tiny pert}}^{2}\mathcal{L}_{mn2} \right) d_{n}(t_{1N},t_{2}) \ , 
\end{eqnarray}
with the initial conditions
\begin{eqnarray}
\label{A2e23}
g_{1N}\delta_{\scriptscriptstyle{mN}} &=& \omega_{\scriptscriptstyle{N}}\left[ \tilde{\tilde{q}}(0)\right]  \frac{\partial d_{m}}{\partial t_{1N}}(0, 0) + \epsilon_{\mbox{\tiny pert}} \frac{\partial d_{m}}{\partial t_{2}}(0, 0) \ , \cr
g_{0N}\delta_{\scriptscriptstyle{mN}} &=& d_{m}(0,0) \ . 
\end{eqnarray}
We now solve (\ref{A2e18b}) and (\ref{A2e23}) to good approximation. 

In the following we assume that each $d_{m}(t_{1N}, t_{2})$ has the following asymptotic expansion
\begin{eqnarray}
\label{A2e24}
d_{m} &\sim& Y_{\scriptscriptstyle{m0}} + \epsilon_{\mbox{\tiny pert}}Y_{\scriptscriptstyle{m1}} + \epsilon_{\mbox{\tiny pert}}^{2} Y_{\scriptscriptstyle{m2}} + ... \ .
\end{eqnarray}
If one takes \ $d_{m} \simeq Y_{\scriptscriptstyle{m0}}$ for each $m$, then one has a first term approximation. Similarly, if one takes \ $d_{m} \simeq Y_{\scriptscriptstyle{m0}} + \epsilon_{\mbox{\tiny pert}}Y_{\scriptscriptstyle{m1}}$ \ for each $m$, then one has a two term approximation. In general, one has an $n$-term approximation if one takes \ $d_{m} \simeq Y_{\scriptscriptstyle{m0}} + \epsilon_{\mbox{\tiny pert}}Y_{\scriptscriptstyle{m1}} + \epsilon_{\mbox{\tiny pert}}^{2} Y_{\scriptscriptstyle{m2}} + ... + \epsilon_{\mbox{\tiny pert}}^{n-1} Y_{\scriptscriptstyle{m,(n-1)}}$ \ for each $m$. We are now going to see how these are determined.

Substituting (\ref{A2e24}) in (\ref{A2e18b}) and (\ref{A2e23}) and equating equal powers of $\epsilon_{\mbox{\tiny pert}}$ we arrive at the following first three initial value problems for each $m =1,2,...$:
\\
\\
$\mathcal{O}(1)$:\\
\begin{eqnarray}
\label{ProblemaO1}
\mathcal{L}_{m0}Y_{\scriptscriptstyle{m0}}(t_{1N},t_{2}) &=& 0 \ , \nonumber
\end{eqnarray}
\begin{eqnarray}
\label{A2e25}
Y_{\scriptscriptstyle{m0}}(0,0) &=& g_{0N}\delta_{\scriptscriptstyle{mN}} \ , \cr
&& \cr
\omega_{\scriptscriptstyle{N}}\left[ \tilde{\tilde{q}}(0)\right]  \frac{\partial Y_{\scriptscriptstyle{m0}}}{\partial t_{1N}}(0, 0) &=& g_{1N}\delta_{\scriptscriptstyle{mN}}  \ .
\end{eqnarray}

\noindent
$\mathcal{O}(\epsilon_{\mbox{\tiny pert}})$:
\begin{eqnarray}
\mathcal{L}_{m0}Y_{\scriptscriptstyle{m1}}(t_{1N},t_{2}) &=&  -\mathcal{L}_{m1}Y_{\scriptscriptstyle{m0}}(t_{1N},t_{2}) \cr
&&  - \sum_{n=1 \atop n\not= m}^{+ \infty} \mathcal{L}_{mn1}Y_{\scriptscriptstyle{n0}}(t_{1N},t_{2}) \ ,\nonumber
\end{eqnarray}
\begin{eqnarray}
\label{A2e26}
Y_{\scriptscriptstyle{m1}}(0,0) &=& 0 \ , \cr
&& \cr
\omega_{\scriptscriptstyle{N}}\left[ \tilde{\tilde{q}}(0)\right]  \frac{\partial Y_{\scriptscriptstyle{m1}}}{\partial t_{1N}}(0, 0) + \frac{\partial Y_{\scriptscriptstyle{m0}}}{\partial t_{2}}(0, 0) &=& 0  \ .
\end{eqnarray}

\noindent
$\mathcal{O}(\epsilon_{\mbox{\tiny pert}}^{2})$: 
\begin{eqnarray}
\label{A2e26b}
&& \mathcal{L}_{m0} Y_{m2}(t_{1N},t_{2}) \cr
&=&  -\mathcal{L}_{m1}Y_{m1}(t_{1N},t_{2}) -\mathcal{L}_{m2}Y_{m0}(t_{1N},t_{2}) \cr
&& -\sum_{n=1 \atop n \not= m}^{+ \infty} \mathcal{L}_{mn1}Y_{n1}(t_{1N},t_{2}) -\sum_{n=1 \atop n \not= m}^{+ \infty} \mathcal{L}_{mn2}Y_{n0}(t_{1N},t_{2}) \ ,  \nonumber
\end{eqnarray}
\begin{eqnarray}
\label{A2e27}
Y_{\scriptscriptstyle{m2}}(0,0) &=& 0 \ , \cr
&& \cr
\omega_{\scriptscriptstyle{N}}\left[ \tilde{\tilde{q}}(0)\right]  \frac{\partial Y_{\scriptscriptstyle{m2}}}{\partial t_{1N}}(0, 0) + \frac{\partial Y_{\scriptscriptstyle{m1}}}{\partial t_{2}}(0, 0) &=& 0  \ .
\end{eqnarray}

Notice that the $\mathcal{O}(1)$ problem corresponds to uncoupled harmonic oscillator equations in the fast time-scale $t_{1N}$, a result that was expected because the modes satisfy uncoupled harmonic oscillator equations when the membrane is fixed. Moreover, the $\mathcal{O}(\epsilon_{\mbox{\tiny pert}}^{n})$ ($n\geq 1$) problems correspond to uncoupled, driven harmonic oscillator equations in the fast time-scale $t_{1N}$. Observe that the coupling between the modes is going to be introduced through the driving because it depends on the previous values $Y_{m0}$, $Y_{m1}$, ..., $Y_{m(n-1)}$. 

We first solve the $\mathcal{O}(1)$ problem. If $m = N$, one immediately finds that
\begin{eqnarray}
\label{A2e33}
Y_{\scriptscriptstyle{N0}}(t_{1N}, t_{2}) &=& a_{N0}(t_{2})e^{it_{1N}} + b_{N0}(t_{2})e^{-it_{1N}} \ , 
\end{eqnarray}
where
\begin{eqnarray}
\label{A2e34}
b_{N0}(0) &=& \frac{g_{0N}}{2} + i\frac{g_{1N}}{2\omega_{N}\left[ \tilde{\tilde{q}}(0) \right]}\ , \cr
a_{N0}(0) &=& b_{N0}(0)^{*} \ . 
\end{eqnarray}
To obtain (\ref{A2e34}) we used that $g_{0N}$ and $g_{1N}$ are real numbers, see Section III.

Notice that if $m \not= N$, then 
\begin{eqnarray}
\label{A2e31}
Y_{\scriptscriptstyle{m0}}(t_{1N}, t_{2}) &=& 0 \ ,
\end{eqnarray}
is a solution of the $\mathcal{O}(1)$ problem. In fact, if one solves the $\mathcal{O}(1)$ problem for $m\not= N$ and subsequently solves the $\mathcal{O}(\epsilon_{\mbox{\tiny pert}})$ problem, then one finds that (\ref{A2e31}) is indeed the unique solution. In order to lighten the burden of the calculations we take (\ref{A2e31}) from the outset. Moreover, this result can also be motivated physically. One expects that the field will approximately follow the instantaneous mode if the membrane moves \textit{sufficiently slowly}. This physical intuition is expressed mathematically by requiring that the first term $Y_{m0}(t_{1N},t_{2})$ of the asymptotic expansion of $d_{m}(t_{1N},t_{2})$ is zero except for the mode $N$ that was initially excited.

Notice that the coefficients $a_{N0}(t_{2})$ and $b_{N0}(t_{2})$ in (\ref{A2e33}) are not known for $t_{2}\not=0$. They are determined by solving the $\mathcal{O}(\epsilon_{\mbox{\tiny pert}})$ problem. This is what we do now.

Substituting (\ref{A2e33}) and (\ref{A2e31}) in the $\mathcal{O}(\epsilon_{\mbox{\tiny pert}})$ differential equation and solving the resulting driven harmonic oscillator equations, one finds that
\begin{eqnarray}
\label{A2e38}
Y_{\scriptscriptstyle{m1}}(t_{1N},t_{2}) 
&=& a_{m1}(t_{2})e^{iW_{m}(t_{2})t_{1N}} + b_{m1}(t_{2})e^{-iW_{m}(t_{2})t_{1N}} \cr
&& \cr
&& + \frac{iT_{m}(t_{2})}{W_{m}(t_{2})^{2} -1}\Big[ b_{N0}(t_{2})e^{-it_{1N}} \cr
&& \qquad\qquad\qquad - a_{N0}(t_{2})e^{it_{1N}} \Big]  \ \ \ (m\not= N), \cr
&& \cr
Y_{\scriptscriptstyle{N1}}(t_{1N},t_{2}) 
&=& a_{N1}(t_{2})e^{it_{1N}} + b_{N1}(t_{2})e^{-it_{1N}} \cr
&& \cr
&& + \frac{(1+2it_{1N})B_{\scriptscriptstyle{N0}}(t_{2})}{4\omega_{N}\left[ \tilde{\tilde{q}}(t_{2}) \right]^{2}}e^{-it_{1N}} \cr
&& \cr
&& - \frac{(1-2it_{1N})A_{\scriptscriptstyle{N0}}(t_{2})}{4\omega_{N}\left[ \tilde{\tilde{q}}(t_{2}) \right]^{2}}e^{it_{1N}} , 
\end{eqnarray}
where
\begin{eqnarray}
\label{A2e44}
\frac{A_{\scriptscriptstyle{N0}}(t_{2})}{i2\omega_{\scriptscriptstyle{N}}\left[ \tilde{\tilde{q}}(t_{2}) \right]} &=& a_{\scriptscriptstyle{N0}}'(t_{2})  \ + \ \tilde{\tilde{q}}'(t_{2}) \zeta_{N}(t_{2}) a_{\scriptscriptstyle{N0}}(t_{2}) \ , \cr
&& \cr
&& \cr
\frac{B_{\scriptscriptstyle{N0}}(t_{2})}{i2\omega_{\scriptscriptstyle{N}}\left[ \tilde{\tilde{q}}(t_{2}) \right]} &=& b_{\scriptscriptstyle{N0}}'(t_{2}) \ + \ \tilde{\tilde{q}}'(t_{2}) \zeta_{N}(t_{2}) b_{\scriptscriptstyle{N0}}(t_{2})  \ , \cr
&& 
\end{eqnarray}
Notice that one obtains $B_{\scriptscriptstyle{N0}}(t_{2})$ from $A_{\scriptscriptstyle{N0}}(t_{2})$ by changing $a_{\scriptscriptstyle{N0}}(t_{2})$ to $b_{\scriptscriptstyle{N0}}(t_{2})$. We remark that the initial conditions for $Y_{\scriptscriptstyle{m1}}(t_{1N},t_{2})$ ($m\not= N$) and $Y_{\scriptscriptstyle{N1}}(t_{1N},t_{2})$ have not yet been imposed. The reason for this will become clear when we solve the $\mathcal{O}(\epsilon_{\mbox{\tiny pert}}^{2})$ problem further below.

Before proceeding one must eliminate secular terms from (\ref{A2e38}), that is, one must eliminate terms that destroy de order of the asymptotic expansion in (\ref{A2e24}). In our case, these correspond to terms in (\ref{A2e38}) that become unbounded as $t_{1N}$ increases while $t_{2}$ is fixed. We observe that the secular terms in (\ref{A2e38}) are those that have $t_{1N}$ as a factor. Then, secular terms disappear from $Y_{N1}$  if and only if
\begin{eqnarray}
B_{\scriptscriptstyle{N0}}(t_{2}) \ = \ A_{\scriptscriptstyle{N0}}(t_{2}) \ = \ 0 \ .
\end{eqnarray}
This condition is equivalent to the following differential equation

\begin{eqnarray}
\label{A2e43}
&& z'(t_{2})  + \tilde{\tilde{q}}'(t_{2})\zeta_{N}(t_{2}) z(t_{2})  \ = \ 0 \ , \cr
\cr
&& \mbox{with} \ \ z(t_{2})  = \ a_{\scriptscriptstyle{N0}}(t_{2}) \ , \ \  b_{\scriptscriptstyle{N0}}(t_{2})\ .
\end{eqnarray} 
The differential equations in (\ref{A2e43}) are linear and of order $1$. Hence, they can be solved analytically \cite{Ross} to give 
\begin{eqnarray}
\label{A2e46}
&& z(t_{2}) \ = \ z(0)\alpha_{N}\left[ \tilde{\tilde{q}}(t_{2}) \right] \ , \cr
&& \cr
&& \mbox{with} \ \ z(t_{2}) \ = \ a_{\scriptscriptstyle{N0}}(t_{2}) \ , \ \ b_{\scriptscriptstyle{N0}}(t_{2}) \ .
\end{eqnarray}
Recall that the values of $a_{\scriptscriptstyle{N0}}(0)$ and $b_{\scriptscriptstyle{N0}}(0)$ had already been obtained in (\ref{A2e34}) and that $\alpha_{N}\left[ \tilde{\tilde{q}}(t_{2}) \right]$ was defined in (\ref{A2eOp}).

Having eliminated the secular terms (\ref{A2e38}) takes the form
\begin{eqnarray}
Y_{\scriptscriptstyle{N1}}(t_{1N},t_{2})
&=& a_{N1}(t_{2})e^{it_{1N}} + b_{N1}(t_{2})e^{-it_{1N}} \ , \nonumber
\end{eqnarray}
\noindent
and
\begin{eqnarray}
\label{A2e49}
&& Y_{\scriptscriptstyle{m1}}(t_{1N},t_{2}) \cr
&=& a_{m1}(t_{2})e^{iW_{m}(t_{2})t_{1N}} + b_{m1}(t_{2})e^{-iW_{m}(t_{2})t_{1N}} \cr
&& \cr
&& + \frac{iT_{m}(t_{2})}{W_{m}(t_{2})^{2}-1}\left[ b_{N0}(t_{2})e^{-it_{1N}} - a_{N0}(t_{2})e^{it_{1N}} \right] \ , \cr
&& (m\not= N). 
\end{eqnarray}
Notice that the coefficients $a_{m1}(t_{2})$, $b_{m1}(t_{2})$, $a_{N1}(t_{2})$, and $b_{N1}(t_{2})$ are not known. They are determined by solving the $\mathcal{O}(\epsilon_{\mbox{\tiny pert}}^{2})$ problem. 

Also, from (\ref{A2e33})-(\ref{A2e31}) and (\ref{A2e46}) we find that the first term approximation to $d_{m}(t_{1N}, t_{2})$ is now completely specified. It is given by 
\begin{eqnarray}
\label{A2e47}
Y_{\scriptscriptstyle{N0}}(t_{1N},t_{2}) 
&=& \alpha_{N}\left[ \tilde{\tilde{q}}(t_{2}) \right] \left[ b_{N0}(0)^{*}e^{it_{1N}} + b_{N0}(0)e^{-it_{1N}}  \right]  , \cr
&& \cr
Y_{\scriptscriptstyle{m0}}(t_{1N},t_{2}) &=& 0 \ \ (m \not= N),
\end{eqnarray}
where $b_{N0}(0)$ is given in (\ref{A2e34}) and $\alpha_{N}\left[ \tilde{\tilde{q}}(t_{2}) \right]$ is defined in (\ref{A2eOp}).

We now solve the $\mathcal{O}(\epsilon_{\mbox{\tiny pert}}^{2})$ problem. As mentioned before, not only will this give us an expression for $Y_{\scriptscriptstyle{m2}}(t_{1N},t_{2})$, but it will also allow us to fully specify $Y_{\scriptscriptstyle{m1}}(t_{1N},t_{2})$, that is, it will allow us to determine the unknown coefficients $a_{m1}(t_{2})$, $b_{m1}(t_{2})$, $a_{N1}(t_{2})$, and $b_{N1}(t_{2})$ in (\ref{A2e49}).

First assume that $m\not= N$. Substituting (\ref{A2e49}) and (\ref{A2e47}) in the differential equation of the $\mathcal{O}(\epsilon_{\mbox{\tiny pert}}^{2})$ problem in (\ref{A2e26b}), one again obtains a driven harmonic oscillator equation that can be solved analytically \cite{Ross}. One obtains

\begin{widetext}
\begin{eqnarray}
\label{A2e63}
Y_{\scriptscriptstyle{m2}}(t_{1N}, t_{2}) &=& a_{\scriptscriptstyle{m2}}(t_{2})e^{iW_{m}(t_{2})t_{1N}} + b_{\scriptscriptstyle{m2}}(t_{2})e^{-iW_{m}(t_{2})t_{1N}} \cr
&& \cr
&& + e^{iW_{m}(t_{2})t_{1N}}\left\{ -\frac{A_{m1}(t_{2})}{4W_{m}(t_{2})^{2}} + i\frac{A_{m1}(t_{2})t_{1N}}{2W_{m}(t_{2})} \ + \ i\frac{D_{m1}(t_{2})}{8W_{m}(t_{2})^{3}} + \frac{D_{m1}(t_{2})t_{1N}}{4W_{m}(t_{2})^{2}}  - \ i\frac{D_{m1}(t_{2})t_{1N}^{2}}{4W_{m}(t_{2})} \right\}  \cr
&& \cr
&& + e^{-iW_{m}(t_{2})t_{1N}}\left\{ \frac{B_{m1}(t_{2})}{4W_{m}(t_{2})^{2}} + i\frac{B_{m1}(t_{2})t_{1N}}{2W_{m}(t_{2})} \ - \ i\frac{F_{m1}(t_{2})}{8W_{m}(t_{2})^{3}} + \frac{F_{m1}(t_{2})t_{1N}}{4W_{m}(t_{2})^{2}} \ + \ i\frac{F_{m1}(t_{2})t_{1N}^{2}}{4W_{m}(t_{2})} \right\}  \cr
&& \cr
&& + e^{it_{1N}} \frac{H_{mN}^{(3)}(t_{2}) +R_{m}^{(1)}(t_{2})}{W_{m}(t_{2})^{2}-1} \ + \ e^{-it_{1N}} \frac{H_{mN}^{(4)}(t_{2}) +R_{m}^{(2)}(t_{2})}{W_{m}(t_{2})^{2}-1} \ , \cr
&& \cr
&& + \sum_{n=1 \atop n\not= m,N}^{+\infty} \left[ \ e^{iW_{n}(t_{2})t_{1N}} \frac{H_{mn}^{(1)}(t_{2})}{W_{m}(t_{2})^{2}-W_{n}(t_{2})^{2}} \ + \ e^{-iW_{n}(t_{2})t_{1N}} \frac{H_{mn}^{(2)}(t_{2})}{W_{m}(t_{2})^{2}-W_{n}(t_{2})^{2}} \right]  \ , \cr
&&
\end{eqnarray}
\end{widetext}

where we have introduced the quantities
\begin{eqnarray}
\label{A2e63b}
A_{\scriptscriptstyle{m1}}(t_{2}) &=& \frac{2i}{\omega_{\scriptscriptstyle{N}}\left[ \tilde{\tilde{q}}(t_{2}) \right]}\frac{d}{dt_{2}}\left[ W_{m}(t_{2}) a_{\scriptscriptstyle{m1}}(t_{2})  \right] \cr
&& \cr
&& + \frac{i2\tilde{\tilde{q}}'(t_{2})}{\omega_{\scriptscriptstyle{N}}\left[ \tilde{\tilde{q}}(t_{2}) \right]} W_{m}(t_{2})a_{\scriptscriptstyle{m1}}(t_{2}) \zeta_{m}(t_{2}) \ , \cr
&&\cr
&& \cr
B_{\scriptscriptstyle{m1}}(t_{2}) &=& \frac{2i}{\omega_{\scriptscriptstyle{N}}\left[ \tilde{\tilde{q}}(t_{2}) \right]}\frac{d}{dt_{2}}\left[ W_{m}(t_{2}) b_{\scriptscriptstyle{m1}}(t_{2})  \right] \cr
&& \cr
&& + \frac{i2\tilde{\tilde{q}}'(t_{2})}{\omega_{\scriptscriptstyle{N}}\left[ \tilde{\tilde{q}}(t_{2}) \right]} W_{m}(t_{2})b_{\scriptscriptstyle{m1}}(t_{2}) \zeta_{m}(t_{2}) \ , \cr
&&\cr
&& \cr
D_{\scriptscriptstyle{m1}}(t_{2}) &=& 2\frac{W_{m}(t_{2})W_{m}'(t_{2})}{\omega_{\scriptscriptstyle{N}}\left[ \tilde{\tilde{q}}(t_{2}) \right]}a_{m1}(t_{2}) \ , \cr
&& \cr
&& \cr
F_{\scriptscriptstyle{m1}}(t_{2}) &=& 2\frac{W_{m}(t_{2})W_{m}'(t_{2})}{\omega_{\scriptscriptstyle{N}}\left[ \tilde{\tilde{q}}(t_{2}) \right]}b_{m1}(t_{2}) \ , \cr
&& \cr
&& \cr
H_{\scriptscriptstyle{mn}}^{(1)}(t_{2}) &=& -i2\tilde{\tilde{q}}'(t_{2})W_{n}(t_{2})\frac{\Gamma_{\scriptscriptstyle{mn}}\left[ \tilde{\tilde{q}}(t_{2}) \right]}{\omega_{\scriptscriptstyle{N}}\left[ \tilde{\tilde{q}}(t_{2}) \right]}a_{n1}(t_{2}) \ , \cr
&& \cr
&& \cr
H_{\scriptscriptstyle{mn}}^{(2)}(t_{2}) &=& i2\tilde{\tilde{q}}'(t_{2})W_{n}(t_{2})\frac{\Gamma_{\scriptscriptstyle{mn}}\left[ \tilde{\tilde{q}}(t_{2}) \right]}{\omega_{\scriptscriptstyle{N}}\left[ \tilde{\tilde{q}}(t_{2}) \right]}b_{n1}(t_{2}) \ , \cr
&& \cr
&& \cr
H_{\scriptscriptstyle{mN}}^{(3)}(t_{2}) &=& H_{\scriptscriptstyle{mN}}^{(1)}(t_{2}) -\frac{\Gamma_{\scriptscriptstyle{mN}}\left[ \tilde{\tilde{q}}(t_{2}) \right]}{\omega_{\scriptscriptstyle{N}}\left[ \tilde{\tilde{q}}(t_{2}) \right]^{2}} \times \cr
&& \times \left[ 2\tilde{\tilde{q}}'(t_{2})a_{N0}'(t_{2}) + \tilde{\tilde{q}}''(t_{2})a_{N0}(t_{2})  \right] \ , \cr 
&& \cr
&& \cr
H_{\scriptscriptstyle{mN}}^{(4)}(t_{2}) &=& H_{\scriptscriptstyle{mN}}^{(2)}(t_{2}) -\frac{\Gamma_{\scriptscriptstyle{mN}}\left[ \tilde{\tilde{q}}(t_{2}) \right]}{\omega_{\scriptscriptstyle{N}}\left[ \tilde{\tilde{q}}(t_{2}) \right]^{2}} \times \cr
&& \left[ 2\tilde{\tilde{q}}'(t_{2})b_{N0}'(t_{2}) + \tilde{\tilde{q}}''(t_{2})b_{N0}(t_{2})  \right] \ , \cr 
&& \cr
&& \cr
\alpha_{\scriptscriptstyle{m0}}(t_{2}) &=& \frac{iT_{m}(t_{2})a_{N0}(t_{2})}{W_{m}(t_{2})^{2} -1} \ , \cr
&& \cr
&& \cr
\beta_{\scriptscriptstyle{m0}}(t_{2}) &=& \frac{iT_{m}(t_{2})b_{N0}(t_{2})}{W_{m}(t_{2})^{2} -1} \ , \cr
&& \cr
&& \cr
R_{m}^{(1)}(t_{2}) &=& i\frac{2\alpha_{m0}'(t_{2})}{\omega_{\scriptscriptstyle{N}}\left[ \tilde{\tilde{q}}(t_{2}) \right]} \cr
&& \cr
&& + i\frac{2\tilde{\tilde{q}}'(t_{2}) \zeta_{m}(t_{2})}{\omega_{\scriptscriptstyle{N}}\left[ \tilde{\tilde{q}}(t_{2}) \right]} \alpha_{m0}(t_{2}) \cr
&& \cr
&& + \frac{i 2 \tilde{\tilde{q}}'(t_{2})}{\omega_{\scriptscriptstyle{N}}\left[ \tilde{\tilde{q}}(t_{2}) \right]} \sum_{n=1 \atop n\not= m,N}^{+\infty} \Gamma_{mn}\left[ \tilde{\tilde{q}}(t_{2}) \right] \alpha_{n0}(t_{2}) \ , \cr
&& \cr
&& \cr
R_{m}^{(2)}(t_{2}) &=& i\frac{2\beta_{m0}'(t_{2})}{\omega_{\scriptscriptstyle{N}}\left[ \tilde{\tilde{q}}(t_{2}) \right]} \cr
&& \cr
&& + i\frac{2\tilde{\tilde{q}}'(t_{2}) \zeta_{m}(t_{2})}{\omega_{\scriptscriptstyle{N}}\left[ \tilde{\tilde{q}}(t_{2}) \right]} \beta_{m0}(t_{2}) \cr
&& \cr
&& + \frac{i 2 \tilde{\tilde{q}}'(t_{2})}{\omega_{\scriptscriptstyle{N}}\left[ \tilde{\tilde{q}}(t_{2}) \right]} \sum_{n=1 \atop n\not= m,N}^{+\infty} \Gamma_{mn}\left[ \tilde{\tilde{q}}(t_{2}) \right] \beta_{n0}(t_{2}) \ . \cr
&&
\end{eqnarray}
Observe that in each consecutive pair of quantities introduced in (\ref{A2e63b}), one is obtained from the other by changing $a$ by $b$, $D$ by $F$, $\alpha$ by $\beta$, and $(1)$ by $(2)$ and by adjusting some signs.

We now have to eliminate secular terms from $Y_{\scriptscriptstyle{m2}}(t_{1N}, t_{2})$. Notice from (\ref{A2e63}) that, for fixed $t_{2}$, the terms multiplied by $t_{1N}$ and $t_{1N}^{2}$ grow without bound and the terms multiplied by $t_{1N}^{2}$ eventually dominate. Hence, secular terms disappear from $Y_{\scriptscriptstyle{m2}}(t_{1N}, t_{2})$ if and only if
\begin{eqnarray}
\label{A2e64}
&D_{m1}(t_{2})& = A_{m1}(t_{2}) = F_{m1}(t_{2}) = B_{m1}(t_{2}) = 0 \ , \ (t_{2} \geq 0)   \cr
&& \cr
&\Leftrightarrow& 
\left\{
\begin{array}{cc}
a_{m1}(t_{2})W_{m}'(t_{2}) = 0 \ , & A_{m1}(t_{2}) = 0 \ , \cr
b_{m1}(t_{2})W_{m}'(t_{2}) = 0 \ , & B_{m1}(t_{2}) = 0 \ , \cr
\mbox{for} \ \ t_{2} \geq 0 \ . 
\end{array}
\right. \cr
&& \cr
&\Leftarrow& a_{m1}(t_{2}) = b_{m1}(t_{2}) = 0 \ , \ \ (t_{2} \geq 0 ).
\end{eqnarray}
In the second step in (\ref{A2e64}) we used the definitions of $D_{m1}(t_{2})$ and $F_{m1}(t_{2})$ in (\ref{A2e63b}) along with the fact that \ $W_{m}(t_{2}) \not= 0$ \ and \ $\omega_{N}[\tilde{\tilde{q}}(t_{2})] \not= 0$ \ for all \ $t_{2}$, see the definition of $W_{m}(t_{2})$ in (\ref{A2eOp}) and item 1 in Section III. Also, in the last step in (\ref{A2e64}) we used that \ $A_{m1}(t_{2}) = B_{m1}(t_{2}) = 0$ \ for all $t_{2} \geq 0$ if $a_{m1}(t_{2}) = b_{m1}(t_{2}) = 0$ \ for all \ $t_{2} \geq 0$, see the definitions of $A_{m1}(t_{2})$ and $B_{m1}(t_{2})$ in (\ref{A2e63b}). Moreover, we note that, except for a discrete set of values of $t_{2}$, one has $W_{m}'(t_{2}) \not= 0$ because \ $W_{m}(t_{2}) = \omega_{m}[\tilde{\tilde{q}}(t_{2})]/\omega_{N}[\tilde{\tilde{q}}(t_{2})]$ \ is, in general, not constant, see Figure \ref{Figure2} for an example of how the $\omega_{n}\left[ \tilde{\tilde{q}}(t_{2}) \right]$ vary with respect to $\tilde{\tilde{q}}(t_{2})$. Therefore, the last line in (\ref{A2e64}) is, in general, also a necessary condition for the disappearance of secular terms from $Y_{\scriptscriptstyle{m2}}(t_{1N}, t_{2})$. 

Taking \ $a_{m1}(t_{2}) = b_{m1}(t_{2}) = 0$ \ for all \ $t_{2} \geq 0$ \ and \ $m\not= N$, it follows that from (\ref{A2e63b}) that $H_{mn}^{(1)}(t_{2}) = H_{mn}^{(2)}(t_{2}) = 0$ \ for all \ $t_{2} \geq 0$ \ and \ $n\not= N$, so that $Y_{\scriptscriptstyle{m1}}(t_{1N}, t_{2})$ in (\ref{A2e38}) and $Y_{\scriptscriptstyle{m2}}(t_{1N}, t_{2})$ in (\ref{A2e63}) reduce to
\begin{eqnarray}
\label{A2e38bb}
&& Y_{\scriptscriptstyle{m1}}(t_{1N},t_{2}) \cr
&=& \frac{iT_{m}(t_{2})}{W_{m}(t_{2})^{2}-1}\Big[ b_{N0}(t_{2})e^{-it_{1N}} - a_{N0}(t_{2})e^{it_{1N}} \Big] \ , \cr
&& (m\not= N), 
\end{eqnarray}
and
\begin{eqnarray}
\label{A2e63bb}
Y_{\scriptscriptstyle{m2}}(t_{1N}, t_{2}) &=& a_{\scriptscriptstyle{m2}}(t_{2})e^{iW_{m}(t_{2})t_{1N}} + b_{\scriptscriptstyle{m2}}(t_{2})e^{-iW_{m}(t_{2})t_{1N}} \cr
&& \cr
&& + e^{it_{1N}} \frac{H_{mN}^{(3)}(t_{2}) +R_{m}^{(1)}(t_{2})}{W_{m}(t_{2})^{2}-1} \ , \cr
&& \cr
&& + e^{-it_{1N}} \frac{H_{mN}^{(4)}(t_{2}) +R_{m}^{(2)}(t_{2})}{W_{m}(t_{2})^{2}-1} \ , \cr
&& (m\not= N), 
\end{eqnarray}
with the quantities defined in (\ref{A2e63b}).

Notice that $Y_{\scriptscriptstyle{m1}}(t_{1N},t_{2})$ is completely specified in (\ref{A2e38bb}). Nevertheless, we still have to verify that it does indeed satisfy the initial conditions of the $\mathcal{O}(\epsilon_{\mbox{\tiny pert}})$ problem in (\ref{A2e26}). Recall that we did not impose them when we solved the $\mathcal{O}(\epsilon_{\mbox{\tiny pert}})$ in (\ref{A2e38}). We postponed this in order to use the simplified expression in (\ref{A2e38bb}). Using the initial conditions in (\ref{A2e26}), the values of $a_{N0}(0)$ and $b_{N0}(0)$ in (\ref{A2e34}), the definitions of $T_{m}(t_{2})$ and $W_{m}(t_{2})$ in (\ref{A2eOp}), and the expression for $Y_{\scriptscriptstyle{m1}}(t_{1N},t_{2})$ in (\ref{A2e38bb}) it follows that
\begin{eqnarray}
\label{A2e65}
&& Y_{m1}(0,0) = 0 \ \Leftrightarrow \ - \frac{g_{1N}T_{m}(0)}{[ W_{m}(0)^{2}-1]\omega_{N}[\tilde{\tilde{q}}(0)]} = 0 \cr
&& \cr
&\Leftrightarrow&  g_{1N} = 0 \ \ \ \mbox{or} \ \ \ \Gamma_{mN}[\tilde{\tilde{q}}(0)] = 0  \ \ \ \mbox{or} \ \ \ \tilde{\tilde{q}}'(0) = 0 \ , \cr
&&
\end{eqnarray}
and that
\begin{eqnarray}
\label{A2e65b}
&& \frac{ \partial Y_{m1}}{\partial t_{1N}}(0,0) = 0 \ \Leftrightarrow \ \frac{g_{0N}T_{m}(0)}{W_{m}(0)^{2}-1} = 0 \cr
&\Leftrightarrow&  g_{0N} = 0 \ \ \ \mbox{or} \ \ \ \Gamma_{mN}[\tilde{\tilde{q}}(0)] = 0  \ \ \ \mbox{or} \ \ \ \tilde{\tilde{q}}'(0) = 0 \ . \cr
&&
\end{eqnarray}
This is the first time we use the assumption $\tilde{q}'(0) = \epsilon_{\mbox{\tiny pert}}\tilde{\tilde{q}}'(0) = 0$ given in item 3 of Section IV. Using this assumption in (\ref{A2e65}) and (\ref{A2e65b}) we find that $Y_{\scriptscriptstyle{m1}}(t_{1N},t_{2})$ in (\ref{A2e38bb}) does indeed satisfy the initial conditions of the $\mathcal{O}(\epsilon_{\mbox{\tiny pert}})$ problem in (\ref{A2e38}) for the case $m\not= N$.

We are now going to solve the  $\mathcal{O}(\epsilon_{\mbox{\tiny pert}}^{2})$ problem in (\ref{A2e27}) for the case $m = N$. Substituting the expressions for $Y_{n1}$ and $Y_{n0}$ in (\ref{A2e49}), (\ref{A2e47}), and (\ref{A2e38bb}) into the $\mathcal{O}(\epsilon_{\mbox{\tiny pert}}^{2})$ differential equation in (\ref{A2e27}) with $m = N$, one obtains a driven harmonic oscillator equation that can be solved analytically \cite{Ross} to yield
\begin{eqnarray}
\label{A2e75}
Y_{\scriptscriptstyle{N2}}(t_{1N}, t_{2}) &=& a_{\scriptscriptstyle{N2}}(t_{2})e^{it_{1N}} + b_{\scriptscriptstyle{N2}}(t_{2})e^{-it_{1N}} \cr
&& -\frac{e^{it_{1N}}}{4}(1 -i2t_{1N})A_{\scriptscriptstyle{N2}}(t_{2}) \cr
&& +\frac{e^{-it_{1N}}}{4}(1 +i2t_{1N})B_{\scriptscriptstyle{N2}}(t_{2}) \ ,
\end{eqnarray}
where we have introduced the quantities

\begin{widetext}
\begin{eqnarray}
\label{A2e72}
A_{\scriptscriptstyle{N2}}(t_{2}) &=& \frac{1}{\omega_{\scriptscriptstyle{N}}\left[ \tilde{\tilde{q}}(t_{2})\right]^{2}}
\left\{ 
a_{\scriptscriptstyle{N0}}''(t_{2}) + 2\Gamma_{\scriptscriptstyle{NN}}\left[ \tilde{\tilde{q}}(t_{2}) \right] \tilde{\tilde{q}}'(t_{2})a_{\scriptscriptstyle{N0}}'(t_{2}) + \Gamma_{\scriptscriptstyle{NN}}\left[ \tilde{\tilde{q}}(t_{2}) \right] \tilde{\tilde{q}}''(t_{2})a_{\scriptscriptstyle{N0}}(t_{2})
\right\} \cr
&& \cr
&& +\frac{2i}{\omega_{\scriptscriptstyle{N}}\left[ \tilde{\tilde{q}}(t_{2})\right]}
\left[
a_{\scriptscriptstyle{N1}}'(t_{2}) + \tilde{\tilde{q}}'(t_{2}) \zeta_{N}(t_{2}) a_{\scriptscriptstyle{N1}}(t_{2}) \right] -i2\tilde{\tilde{q}}'(t_{2})\sum_{n=1 \atop n\not= N}^{+\infty} \frac{\Gamma_{\scriptscriptstyle{Nn}}\left[ \tilde{\tilde{q}}(t_{2}) \right]}{\omega_{\scriptscriptstyle{N}}\left[ \tilde{\tilde{q}}(t_{2}) \right]}\alpha_{\scriptscriptstyle{n0}}(t_{2}) \ , \cr
&& \cr
&& \cr
B_{\scriptscriptstyle{N2}}(t_{2}) &=& -\frac{1}{\omega_{\scriptscriptstyle{N}}\left[ \tilde{\tilde{q}}(t_{2})\right]^{2}}
\left\{ 
b_{\scriptscriptstyle{N0}}''(t_{2}) + 2\Gamma_{\scriptscriptstyle{NN}}\left[ \tilde{\tilde{q}}(t_{2}) \right] \tilde{\tilde{q}}'(t_{2})b_{\scriptscriptstyle{N0}}'(t_{2}) + \Gamma_{\scriptscriptstyle{NN}}\left[ \tilde{\tilde{q}}(t_{2}) \right] \tilde{\tilde{q}}''(t_{2})b_{\scriptscriptstyle{N0}}(t_{2})
\right\} \cr
&& \cr
&& +\frac{2i}{\omega_{\scriptscriptstyle{N}}\left[ \tilde{\tilde{q}}(t_{2})\right]}
\left[
b_{\scriptscriptstyle{N1}}'(t_{2}) + \tilde{\tilde{q}}'(t_{2}) \zeta_{N}(t_{2}) b_{\scriptscriptstyle{N1}}(t_{2}) \right] +i2\tilde{\tilde{q}}'(t_{2})\sum_{n=1 \atop n\not= N}^{+\infty} \frac{\Gamma_{\scriptscriptstyle{Nn}}\left[ \tilde{\tilde{q}}(t_{2}) \right]}{\omega_{\scriptscriptstyle{N}}\left[ \tilde{\tilde{q}}(t_{2}) \right]}\beta_{\scriptscriptstyle{n0}}(t_{2}) \ .
\end{eqnarray}
\end{widetext}

From (\ref{A2e75}) we observe that secular terms disappear from $Y_{\scriptscriptstyle{N2}}(t_{1N}, t_{2})$ if and only if
\begin{eqnarray}
A_{\scriptscriptstyle{N2}}(t_{2}) = 0 \ , \ \ B_{\scriptscriptstyle{N2}}(t_{2}) = 0 \ \ (t_{2} \geq 0) \ .
\end{eqnarray}
Using the definitions of $A_{N2}(t_{2})$ and $B_{N2}(t_{2})$ in (\ref{A2e72}), this condition is equivalent to the differential equation

\newpage
\begin{eqnarray}
&& z'(t_{2}) + \tilde{\tilde{q}}'(t_{2}) \zeta_{N}(t_{2}) z(t_{2}) \cr
&& = \eta (t_{2})z_{0}(t_{2}) \ , 
\end{eqnarray}
with
\begin{eqnarray}
z(t_{2}) = \left\{
\begin{array}{c}
z(t_{2}) = a_{\scriptscriptstyle{N1}}(t_{2}) \ , \ \ z_{0}(t_{2}) = a_{\scriptscriptstyle{N0}}(t_{2}) \cr
\mbox{or} \cr
z(t_{2}) = b_{\scriptscriptstyle{N1}}(t_{2}) \ , \ \ z_{0}(t_{2}) = -b_{\scriptscriptstyle{N0}}(t_{2}) \cr
\end{array}
\right. \cr
\end{eqnarray}
Solving this differential equation \cite{Ross} one concludes that secular terms disappear from $Y_{\scriptscriptstyle{N2}}(t_{1N},t_{2})$ if and only if
\begin{eqnarray}
\label{A2e76}
z(t_{2}) &=& \alpha_{N}\left[ \tilde{\tilde{q}}(t_{2}) \right] \left[ z(0) + z_{0}(0) \int_{0}^{t_{2}}dt_{2}' \ \eta (t_{2}') \right] \ , \cr
&&
\end{eqnarray}
with
\begin{eqnarray}
z(t_{2}) &=& \left\{
\begin{array}{c}
z(t_{2}) = a_{\scriptscriptstyle{N1}}(t_{2}) \ , \ \ z_{0}(t_{2}) = a_{\scriptscriptstyle{N0}}(t_{2}) \cr
\mbox{or} \cr
z(t_{2}) = b_{\scriptscriptstyle{N1}}(t_{2}) \ , \ \ z_{0}(t_{2}) = -b_{\scriptscriptstyle{N0}}(t_{2}) \cr
\end{array}
\right. \cr
&&
\end{eqnarray}
Notice that the values $a_{\scriptscriptstyle{N1}}(0)$ and $b_{\scriptscriptstyle{N1}}(0)$ remain to be found. These are obtained by applying the $\mathcal{O}(\epsilon_{\mbox{\tiny pert}})$ initial conditions in (\ref{A2e26}) to $Y_{\scriptscriptstyle{N1}}(t_{2})$. One finds that
\begin{eqnarray}
Y_{\scriptscriptstyle{N1}}(0,0) = 0 \ \ &\Leftrightarrow& \ \ a_{\scriptscriptstyle{N1}}(0) +b_{\scriptscriptstyle{N1}}(0) = 0 \ , \nonumber
\end{eqnarray}
and
\begin{eqnarray}
&& \omega_{\scriptscriptstyle{N}}\left[ \tilde{\tilde{q}}(0) \right]\frac{\partial Y_{\scriptscriptstyle{N1}}}{\partial t_{1N}}(0,0) + \frac{\partial Y_{\scriptscriptstyle{N0}}}{\partial t_{2}}(0,0) = 0 \cr
&& \cr
&\Leftrightarrow& \ \ a_{\scriptscriptstyle{N1}}(0) - b_{\scriptscriptstyle{N1}}(0) = 0 \ . \nonumber
\end{eqnarray}
From these two results it is straightforward to conclude that
\begin{eqnarray}
\label{A2e84}
a_{\scriptscriptstyle{N1}}(0) \ = \ b_{\scriptscriptstyle{N1}}(0) \ = \ 0 \ .
\end{eqnarray}

Substituting the values of $a_{\scriptscriptstyle{N1}}(0)$ and $b_{\scriptscriptstyle{N1}}(0)$ given in (\ref{A2e84}) into the expressions for $a_{\scriptscriptstyle{N1}}(t_{2})$ and $b_{\scriptscriptstyle{N1}}(t_{2})$ established in (\ref{A2e76}), and then substituting the result in the expression for $Y_{\scriptscriptstyle{N1}}(t_{1N},t_{2})$ in (\ref{A2e49}), one obtains that
\begin{eqnarray}
\label{A2e86}
Y_{\scriptscriptstyle{N1}}(t_{1N},t_{2}) &=& \alpha_{N}\left[ \tilde{\tilde{q}}(t_{2}) \right] \int_{0}^{t_{2}}dt_{2}' \ \eta (t_{2}') \ \times \cr
&& \times \left[ b_{\scriptscriptstyle{N0}}(0)^{*}e^{it_{1N}} -b_{\scriptscriptstyle{N0}}(0)e^{-it_{1N}} \right]   \ . \cr
&&
\end{eqnarray}
Recall that $\eta (t_{2})$ is given in (\ref{A2eOp}), while $b_{\scriptscriptstyle{N0}}(0)$ was established in (\ref{A2e34}).

Moreover, without secular terms $Y_{\scriptscriptstyle{N2}}(t_{2})$  in  (\ref{A2e75}) takes the form
\begin{eqnarray}
\label{A2e76b}
Y_{\scriptscriptstyle{N2}}(t_{1N},t_{2}) \ = \  a_{\scriptscriptstyle{N2}}(t_{2})e^{it_{1N}} + b_{\scriptscriptstyle{N2}}(t_{2})e^{-it_{1N}} \ .
\end{eqnarray}

We are now in a position to write a two-term approximation of $d_{n}(t_{1N},t_{2})$ using the definition of the time-scales in (\ref{A2e15}) and (\ref{A2e20}), and the asymptotic expansion in (\ref{A2e24}). First, using the expressions for $Y_{\scriptscriptstyle{N0}}(t_{1N},t_{2})$ and $Y_{\scriptscriptstyle{N1}}(t_{1N},t_{2})$ established in (\ref{A2e47}) and (\ref{A2e86}) and the expressions for $a_{N0}(t_{2})$ and $b_{N0}(t_{2})$ in (\ref{A2e46}), we conclude that a two term approximation for $c_{\scriptscriptstyle{N}}(\tau)$ is given by
\begin{eqnarray}
\label{A2e86bb}
&& d_{\scriptscriptstyle{N}}(t_{1N} ,t_{2}) \cr
&& \cr
&\simeq& Y_{\scriptscriptstyle{N0}}(t_{1N} ,t_{2}) + \epsilon_{\mbox{\tiny pert}} Y_{\scriptscriptstyle{N1}}( t_{1N},t_{2}) \ , \cr
&& \cr
&=&\alpha_{N}\left[ \tilde{\tilde{q}}(t_{2}) \right] \times \cr
&& \cr
&& \times \left\{ \ b_{\scriptscriptstyle{N0}}(0)^{*}e^{it_{1N}}\left[ 1 + \epsilon_{\mbox{\tiny pert}}\int_{0}^{t_{2}} dt_{2}' \ \eta (t_{2}') \right] \right. \cr
&& \cr
&& \ \ \left. +b_{\scriptscriptstyle{N0}}(0)e^{-it_{1N}}\left[ 1 - \epsilon_{\mbox{\tiny pert}}\int_{0}^{t_{2}}dt_{2}' \ \eta (t_{2}') \right] \ \right\}  . \cr
&&
\end{eqnarray} 
Recall that $\eta (t_{2})$ is given in (\ref{A2eOp}), while $b_{\scriptscriptstyle{N0}}(0)$ was established in (\ref{A2e34}).

Using the expressions for $Y_{\scriptscriptstyle{m0}}(t_{1N},t_{2})$ and $Y_{\scriptscriptstyle{m1}}(t_{1N},t_{2})$ in (\ref{A2e47}) and (\ref{A2e38bb}) and the expressions for $a_{N0}(t_{2})$ and $b_{N0}(t_{2})$ in (\ref{A2e46}), we conclude that a two term approximation of $d_{m}(t_{1N},t_{2})$ with $m\not= N$ is given by
\begin{eqnarray}
\label{A2e68}
&& d_{m}( t_{1N}, t_{2}) \ , \cr
&& \cr
&\simeq& Y_{m0}( t_{1N}, t_{2}) + \epsilon_{\mbox{\tiny pert}}Y_{m1}(t_{1N}, t_{2}) \ , \cr
&& \cr
&=&  \frac{i \epsilon_{\mbox{\tiny pert}} T_{m}(t_{2})}{W_{m}(t_{2})^{2}-1} \alpha_{N}\left[ \tilde{\tilde{q}}(t_{2}) \right] \times \cr
&& \times \left[ b_{N0}(0)e^{-it_{1N}} - b_{N0}(0)^{*} e^{it_{1N}} \right] \ . \cr
&&
\end{eqnarray}
where $b_{N0}(0)$ is given in (\ref{A2e34}).

Finally, we comment on the accuracy of the two time-scales approximation \cite{Holmes}. The two time-scales approximation is an accurate approximation to $c_{\scriptscriptstyle{N}}(\tau)$ for all $\tau$ such that \ $0 \leq t_{2}(\tau) = \epsilon_{\mbox{\tiny pert}} \tau \leq \mathcal{O}(1)$. Since \ $\epsilon_{\mbox{\tiny pert}} \tau = \nu_{\mbox{\tiny osc}}t$ (see (\ref{XTadimensional}) and (\ref{Adimensionales})), it follows that the latter condition is equivalent to
\begin{eqnarray}
\label{A2e88}
0 \ \leq \ t \ \leq \ \mathcal{O}\left( \nu_{\mbox{\tiny osc}}^{-1} \right) \ .
\end{eqnarray}
Therefore, the two time-scale approximation will be accurate up until $t = \mathcal{D} \nu_{\mbox{\tiny osc}}^{-1}$ for some $\mathcal{D} > 0$. For example, if the membrane is oscillating, then the two time-scale approximation will be accurate during some oscillation periods of the membrane. Moreover, during the time interval in (\ref{A2e88}) the two term approximation will be more accurate than the one term approximation.


\section{Three time-scales}

In Appendix B we obtained an approximation for the coefficients of the modes of the field that is accurate during an interval of the form \ $0\leq t \leq \mathcal{O}( \nu_{\mbox{\tiny osc}}^{-1} )$ \ where $\nu_{\mbox{\tiny osc}}^{-1}$ is the time-scale in which the membrane evolves appreciably, see (\ref{A2e88}). Hence, the approximation obtained might not be accurate for very long times. The purpose of this appendix is to remedy this by solving the equations for the coefficients using the method of multiple scales with three time-scales. 

From what we learned in Appendix B we know that the fast time-scale is given by 
\begin{eqnarray}
\label{A3e3p1p1}
t_{1N} &=& \int_{0}^{\tau} d\tau ' \omega_{N}\left[ \tilde{\tilde{q}}(\epsilon_{\mbox{\tiny pert}}\tau ') \right] \ ,
\end{eqnarray}
while a slow time-scale is given by
\begin{eqnarray}
\label{A3e3p1p2}
t_{2} &=& \epsilon_{\mbox{\tiny pert}}\tau \ .
\end{eqnarray}
Now we introduce a slower time-scale given by
\begin{eqnarray}
\label{A3e3p1p3}
t_{3} &=& \epsilon_{\mbox{\tiny pert}}^{2} \tau \ ,
\end{eqnarray}
These three time-scales allow one to obtain an approximate solution of the initial value problem in (\ref{12}) and (\ref{13}) that is accurate for long times. In fact, the theory of multiple scales \cite{Holmes} indicates that the approximate solution of (\ref{12}) and (\ref{13}) is accurate for times \ $\tau = \nu_{0}t$ \ such that
\begin{eqnarray}
\label{AccurateResultbisbis}
0 \ \leq \ \epsilon_{\mbox{\tiny pert}}^{2} \tau \ \leq \ \mathcal{O}\left( 1 \right) 
&\Leftrightarrow& 0 \ \leq \ t \ \leq \ \mathcal{O}\left(  \frac{1}{\epsilon_{\mbox{\tiny pert}} \nu_{\mbox{\tiny osc}}} \right) . \ \ 
\end{eqnarray}

In order to use the three time-scales introduced above, define for \ $m=1,2,...$
\begin{eqnarray}
\label{A3e3p2}
d_{m}[ t_{1N}(\tau) , t_{2}(\tau ), t_{3}(\tau) ] &=& c_{m} (\tau ) \ .
\end{eqnarray}

Throughout the appendix we make use of the quantities defined in (\ref{A2eOp}) in Appendix B. Now, we also introduce the following quantities for each \ $m= 1$, $2$, ...:
\begin{eqnarray}
\label{A3e3p4}
\mathcal{L}_{m2}' &=& \mathcal{L}_{m2} \ + \ 2\omega_{N} \left[ \tilde{\tilde{q}}(t_{2}) \right] \frac{\partial^{2}}{\partial t_{3} \partial t_{1N}} \ , \cr
&& \cr
&& \cr
\mathcal{L}_{m3} &=& 2\frac{\partial^{2}}{\partial t_{3} \partial t_{2}} \ + \ 2\tilde{\tilde{q}}'(t_{2}) \Gamma_{mm}\left[ \tilde{\tilde{q}}(t_{2})\right] \frac{\partial }{\partial t_{3}}  \ , \cr
&& \cr
\mathcal{L}_{mn3} &=& 2\tilde{\tilde{q}}'(t_{2}) \Gamma_{mn}\left[ \tilde{\tilde{q}}(t_{2})\right] \frac{\partial }{\partial t_{3}} \ .
\end{eqnarray}
Notice that $\mathcal{L}_{m2}$ is defined in (\ref{A2eOp}). 

Substituting $d_{m}[t_{1N}(\tau), t_{2}(\tau), t_{3}(\tau) ]$ given in (\ref{A3e3p2}) into the equation for $c_{m}(\tau)$ in (\ref{12}) and using the differential operators defined in (\ref{A2eOp}) and (\ref{A3e3p4}), one concludes that $d_{m}(t_{1N},t_{2},t_{3})$ must satisfy the equations
\begin{eqnarray}
\label{A3e3p7}
&& \Big[ \ \mathcal{L}_{m0}  + \epsilon_{\mbox{\tiny pert}} \mathcal{L}_{m1} + \epsilon_{\mbox{\tiny pert}}^{2} \mathcal{L}_{m2}  + \epsilon_{\mbox{\tiny pert}}^{3} \mathcal{L}_{m3} \cr
&& \qquad\qquad\qquad \ + \ \epsilon_{\mbox{\tiny pert}}^{4}\frac{\partial^{2}}{\partial t_{3}^{2}} \ \Big] d_{m}(t_{1N},t_{2},t_{3}) \cr
&=& - \sum_{n=1 \atop n \not= m}^{+\infty} \left[ \epsilon_{\mbox{\tiny pert}} \mathcal{L}_{mn1} + \epsilon_{\mbox{\tiny pert}}^{2} \mathcal{L}_{mn2}  + \epsilon_{\mbox{\tiny pert}}^{3} \mathcal{L}_{mn3} \right] \times \cr
&& \qquad \qquad \qquad \times  d_{n}(t_{1N},t_{2},t_{3}) \ .
\end{eqnarray}
Moreover, the initial conditions for $d_{m}(t_{1N},t_{2},t_{3})$ are obtained by substituting the definition of  $d_{m}[t_{1N}(\tau),t_{2}(\tau),t_{3}(\tau)]$ in (\ref{A3e3p2}) into the initial conditions (\ref{13}) for $c_{m}(\tau)$. One obtains 
\begin{eqnarray}
d_{m}(\mathbf{0}) &=& g_{0N}\delta_{mN} \ ,
\nonumber
\end{eqnarray}
and
\begin{eqnarray}
\label{A3e3p8}
&& \omega_{N}\left[ \tilde{\tilde{q}}(0) \right] \frac{\partial d_{m}}{\partial t_{1N}}(\mathbf{0}) + \epsilon_{\mbox{\tiny pert}} \frac{\partial d_{m}}{\partial t_{2}}(\mathbf{0}) + \epsilon_{\mbox{\tiny pert}}^{2} \frac{\partial d_{m}}{\partial t_{3}}(\mathbf{0}) \cr
&=& g_{1N}\delta_{mN} \ .
\end{eqnarray}
Here and in the following \ $\mathbf{0} = (0,0,0)$. Also, recall that $g_{0N}$ and $g_{1N}$ are real quantities, see Section III.

Now assume that each $d_{m}(t_{1N},t_{2},t_{3})$ has an asymptotic expansion of the form
\begin{eqnarray}
\label{A3e3p9}
d_{m} &\sim& Y_{\scriptscriptstyle{m0}} + \epsilon_{\mbox{\tiny pert}}Y_{\scriptscriptstyle{m1}} + \epsilon_{\mbox{\tiny pert}}^{2} Y_{\scriptscriptstyle{m2}} + ... \ .
\end{eqnarray}

Substituting the asymptotic expansion given in (\ref{A3e3p9}) into the differential equation and initial conditions for $d_{m}(t_{1N},t_{2},t_{3})$ in (\ref{A3e3p7}) and (\ref{A3e3p8}) and then equating equal powers of $\epsilon_{\mbox{\tiny pert}}$, one arrives at the following four problems for each \ $m =1,2,...$: 
\\
\\
$\mathcal{O}(1)$:
\begin{eqnarray}
\mathcal{L}_{m0} Y_{m0} &=& 0 \ , \nonumber
\end{eqnarray}
with
\begin{eqnarray}
\label{A3e3p12}
Y_{m0}(\mathbf{0}) &=& g_{0N}\delta_{mN} \ , \cr
\omega_{N}\left[ \tilde{\tilde{q}}(0) \right] \frac{\partial Y_{m0}}{\partial t_{1N}}(\mathbf{0}) &=& g_{1N}\delta_{mN} \ .
\end{eqnarray}

\noindent
$\mathcal{O}(\epsilon_{\mbox{\tiny pert}} )$:
\begin{eqnarray}
\mathcal{L}_{m0} Y_{m1} &=& -\mathcal{L}_{m1} Y_{m0} - \sum_{n=1 \atop n \not= m}^{+\infty} \mathcal{L}_{mn1}Y_{n0} \ , \nonumber
\end{eqnarray}
with
\begin{eqnarray}
\label{A3e3p13}
Y_{m1}(\mathbf{0}) &=& 0 \ , \cr
\omega_{N}\left[ \tilde{\tilde{q}}(0) \right] \frac{\partial Y_{m1}}{\partial t_{1N}}(\mathbf{0}) + \frac{\partial Y_{m0}}{\partial t_{2}}(\mathbf{0}) &=& 0 \ .
\end{eqnarray}

\noindent
$\mathcal{O}(\epsilon_{\mbox{\tiny pert}}^{2} )$:
\begin{eqnarray}
\mathcal{L}_{m0} Y_{m2} &=& -\mathcal{L}_{m2}' Y_{m0} -\mathcal{L}_{m1} Y_{m1}  \cr
&& - \sum_{n=1 \atop n \not= m}^{+\infty} \left( \mathcal{L}_{mn2}Y_{n0} +  \mathcal{L}_{mn1}Y_{n1} \right) \ , \nonumber
\end{eqnarray}
with
\begin{eqnarray}
\label{A3e3p14}
Y_{m2}(\mathbf{0}) &=& 0 \ , \cr
\omega_{N}\left[ \tilde{\tilde{q}}(0) \right] \frac{\partial Y_{m2}}{\partial t_{1N}}(\mathbf{0}) + \frac{\partial Y_{m1}}{\partial t_{2}}(\mathbf{0}) + \frac{\partial Y_{m0}}{\partial t_{3}}(\mathbf{0}) &=& 0 \ . \cr
&&
\end{eqnarray}

\noindent
$\mathcal{O}(\epsilon_{\mbox{\tiny pert}}^{3} )$:
\begin{eqnarray}
&& \mathcal{L}_{m0} Y_{m3} \cr
&& \cr
&=& -\mathcal{L}_{m3} Y_{m0} -\mathcal{L}_{m2}' Y_{m1} -\mathcal{L}_{m1} Y_{m2}  \cr
&& - \sum_{n=1 \atop n \not= m}^{+\infty} \left( \mathcal{L}_{mn3}Y_{n0} +  \mathcal{L}_{mn2}Y_{n1} +  \mathcal{L}_{mn1}Y_{n2} \right) \ , \nonumber
\end{eqnarray}
with
\begin{eqnarray}
\label{A3e3p15}
Y_{m3}(\mathbf{0}) &=& 0 \ , \cr
\omega_{N}\left[ \tilde{\tilde{q}}(0) \right] \frac{\partial Y_{m3}}{\partial t_{1N}}(\mathbf{0}) + \frac{\partial Y_{m2}}{\partial t_{2}}(\mathbf{0}) + \frac{\partial Y_{m1}}{\partial t_{3}}(\mathbf{0}) &=& 0 \ . \cr
&&
\end{eqnarray}

\noindent
In the initial value problems above, observe how the subindexes add up to give a constant number in each of the problems. Moreover, it is important to note that the $\mathcal{O}(1)$ and $\mathcal{O}(\epsilon_{\mbox{\tiny pert}})$ problems above are exactly the same as the $\mathcal{O}(1)$ and $\mathcal{O}(\epsilon_{\mbox{\tiny pert}})$ problems using two time-scales, see (\ref{A2e25}) and (\ref{A2e26}) in Appendix B. Also, notice that the differential equation for the $\mathcal{O}(1)$ problem in (\ref{A3e3p12}) corresponds to a harmonic oscillator in the fast time-scale $t_{1N}$, see the definition of $\mathcal{L}_{m0}$ in (\ref{A2eOp}). Hence, we obtain that all $Y_{m0}(t_{1N},t_{2},t_{3})$ satisfy uncoupled harmonic oscillator equations in the fast time-scale. This reflects the fact that the coefficients of the modes satisfy uncoupled harmonic oscillator equations when the membrane is fixed, see Section III. Finally, observe that the differential equations in the higher order problems $\mathcal{O}(\epsilon_{\mbox{\tiny pert}}^{n} )$ $(n\geq 1)$ correspond to driven harmonic oscillator equations.

First, we solve the $\mathcal{O}(1)$ problem. Since the differential equation in (\ref{A3e3p12}) corresponds to a harmonic oscillator in the fast time-scale $t_{1N}$, one immediately finds that
\begin{eqnarray}
\label{A3e104}
Y_{N0}(t_{1N},t_{2},t_{3}) &=& a_{N0}(t_{2},t_{3})e^{it_{1N}} + b_{N0}(t_{2},t_{3})e^{-it_{1N}} \ , \cr
&& \cr
Y_{m0}(t_{1N},t_{2},t_{3}) &=& 0 \qquad (m\not= N) \ .
\end{eqnarray}
Notice that we took $Y_{m0}$ equal to zero when $m \not= N$ because of what we learned in Appendix B, namely that in the first term approximation only the coefficient of the mode that is initially excited should be different from zero. Physically this amounts to demanding that (to lowest order) the field should follow mode $N$ as the membrane moves if only mode $N$ is initially excited. Also, observe that $Y_{m0} = 0$ with $m\not= N$ does indeed satisfy both the differential equation and the initial conditions of the $\mathcal{O}(1)$ problem.

Applying the $\mathcal{O}(1)$ initial conditions in (\ref{A3e3p12}) to $Y_{N0}$ in (\ref{A3e104}), one finds that
\begin{eqnarray}
\label{A3e105}
b_{N0}(0,0) &=& \frac{g_{0N}}{2} + \frac{ig_{1N}}{2\omega_{N}\left[ \tilde{\tilde{q}}(0) \right]} \ , \cr
a_{N0}(0,0) &=& b_{N0}(0,0)^{*} \ .
\end{eqnarray}   
To obtain the second line of (\ref{A3e105}) we used that both $g_{0N}$ and $g_{1N}$ are real.

We now solve the $\mathcal{O}(\epsilon_{\mbox{\tiny pert}})$ problem. Substituting $Y_{N0}$ and $Y_{m0}$ given in (\ref{A3e104}) into the right-hand side of the differential equation of the $\mathcal{O}(\epsilon_{\mbox{\tiny pert}})$ problem in (\ref{A3e3p13}) and solving the resulting harmonic oscillator equations with driving, one obtains 
\begin{eqnarray}
Y_{N1}(t_{1N},t_{2},t_{3}) 
&=& a_{N1}(t_{2},t_{3})e^{it_{1N}}  + b_{N1}(t_{2},t_{3})e^{-it_{1N}} \cr
&& + e^{it_{1N}}\frac{A(t_{2},t_{3})}{4}(1 -i2t_{1N}) \cr
&& + e^{-it_{1N}}\frac{B(t_{2},t_{3})}{4}(1 +i2t_{1N}) \ , \nonumber
\end{eqnarray}
and
\begin{eqnarray}
\label{A3e108}
&& Y_{m1}(t_{1N},t_{2},t_{3}) \cr
&=& \frac{i2\tilde{\tilde{q}}'(t_{2})\Gamma_{mN}\left[ \tilde{\tilde{q}}(t_{2}) \right]}{\omega_{m}\left[ \tilde{\tilde{q}}(t_{2}) \right]^{2} - \omega_{N}\left[ \tilde{\tilde{q}}(t_{2}) \right]^{2}} \omega_{N}\left[ \tilde{\tilde{q}}(t_{2}) \right] \times \cr
&& \times \left[ b_{N0}(t_{2},t_{3})e^{-it_{1N}} - a_{N0}(t_{2},t_{3})e^{it_{1N}} \right] \ , \cr
&& \cr
&& (m\not= N) \ ,
\end{eqnarray}
where we have introduced the quantities $A(t_{2},t_{3})$ and $B(t_{2},t_{3})$. The former is defined by
\begin{eqnarray}
\label{A3e112}
A(t_{2},t_{3}) &=& \frac{-2i}{\omega_{N}\left[ \tilde{\tilde{q}}(t_{2}) \right]}\left\{ \ \frac{\partial a_{N0}}{\partial t_{2}}(t_{2},t_{3}) \ + \ a_{N0}(t_{2},t_{3}) \times \right. \cr
&& \left. \times \tilde{\tilde{q}}'(t_{2})\zeta_{N}(t_{2}) \ \right\} \ ,
\end{eqnarray}
while $B(t_{2},t_{3})$ is obtained from $A(t_{2},t_{3})$ by changing $i$ to $-i$ and $a_{N0}$ to $b_{N0}$. Also, $\zeta_{N}(t_{2})$ is defined in (\ref{A2eOp}).

It is important to note that we have taken the complementary function associated to $Y_{m1}$ $(m\not= N)$ in (\ref{A3e108}) equal to zero, that is, we have taken the general solution of the homogeneous differential equation associated to  $Y_{m1}$ $(m\not= N)$ equal to zero. The reason for this is that we obtained that it is equal to zero in the two time-scales approach in Appendix B. Moreover, observe that $Y_{m1}$ $(m\not= N)$ in (\ref{A3e108}) satisfies the $\mathcal{O}(\epsilon_{\mbox{\tiny pert}})$ initial conditions in (\ref{A3e3p13}) because $\tilde{\tilde{q}}'(0)=0$. Recall that \ $\tilde{\tilde{q}}'(0)=0$ \ means that the membrane starts from rest and that we made this assumption in item 3 of Section IV.

Now we must eliminate secular terms from $Y_{N1}$, that is, we must demand that terms that become unbounded in $Y_{N1}$ for $t_{2}$ and $t_{3}$ fixed become zero. From (\ref{A3e108}) we observe that secular terms in $Y_{N1}$ disappear if and only if
\begin{eqnarray}
A(t_{2},t_{3}) = B(t_{2},t_{3}) = 0 \ , \ \ t_{2},t_{3} \geq 0 \ . \nonumber
\end{eqnarray}
Using the definitions of $A(t_{2},t_{3})$ and $B(t_{2},t_{3})$ in (\ref{A3e112}), it follows that this condition is equivalent to the differential equation
\begin{eqnarray}
&& \frac{\partial z}{\partial t_{2}}(t_{2},t_{3}) \ = \ -  \tilde{\tilde{q}}'(t_{2}) \zeta_{N} (t_{2})z(t_{2},t_{3}) \ , \cr
&& \mbox{with} \ \ z = a_{N0}, \ b_{N0} \ .
\end{eqnarray}
Recall that $\zeta_{N}(t_{2})$ is defined in (\ref{A2eOp}).

Solving this differential equation one concludes that secular terms in $Y_{N1}$ disappear if and only if
\begin{eqnarray}
\label{A3ea0b0}
&& z(t_{2},t_{3}) = z(0,t_{3}) \alpha_{N}\left[ \tilde{\tilde{q}}(t_{2}) \right]  \ ,  \cr
&& \mbox{with} \ z = a_{N0}, \ b_{N0}.
\end{eqnarray}
Here $\alpha_{N}\left[ \tilde{\tilde{q}}(t_{2}) \right]$ is defined in (\ref{A2eOp}). Notice that both $a_{N0}(t_{2},t_{3})$ and $b_{N0}(t_{2},t_{3})$ are not completely determined yet, since $a_{N0}(0,t_{3})$ and $b_{N0}(0,t_{3})$ have not been specified. One needs to solve the $\mathcal{O}(\epsilon_{\mbox{\tiny pert}}^{2})$ problem to do this.

Without secular terms $Y_{N1}$ in (\ref{A3e108}) takes the form
\begin{eqnarray}
\label{A3e114}
Y_{N1}(t_{1N},t_{2},t_{3}) 
&=& a_{N1}(t_{2},t_{3})e^{it_{1N}}  + b_{N1}(t_{2},t_{3})e^{-it_{1N}}  \ . \cr
&&
\end{eqnarray}
We still have to impose the $\mathcal{O}(\epsilon_{\mbox{\tiny pert}})$ initial conditions on $Y_{N1}$. Substituting into the aforementioned initial conditions the values of $Y_{N0}$, $Y_{N1}$, $a_{N0}$, and $b_{N0}$ given in (\ref{A3e104}), (\ref{A3ea0b0}), and (\ref{A3e114}), and using that $\tilde{\tilde{q}}'(0) = 0$ one finds that
\begin{eqnarray}
\label{A3e127}
a_{N1}(0,0) = b_{N1}(0,0) = 0 \ .
\end{eqnarray}
Recall that the assumption $\tilde{\tilde{q}}'(0) = 0$ was made in item 3 of Section IV.

We now solve the $\mathcal{O}(\epsilon_{\mbox{\tiny pert}}^{2})$ problem in (\ref{A3e3p14}). In the following assume that $m \not= N$. Substituting into the right-hand side of the differential equation of the $\mathcal{O}(\epsilon_{\mbox{\tiny pert}}^{2})$ problem in (\ref{A3e3p14}) the values of $Y_{N0}$ and $Y_{m0}$ given in (\ref{A3e104}), $Y_{m1}$ given in (\ref{A3e108}), $a_{N0}$ and $b_{N0}$ given in (\ref{A3ea0b0}), and $Y_{N1}$ given in (\ref{A3e114}), one obtains a driven harmonic oscillator equation for $Y_{N2}$ in the fast time-scale $t_{1N}$. The general solution of this equation is
\begin{eqnarray}
\label{A3e116}
Y_{N2}(t_{1N},t_{2},t_{3}) &=& a_{N2}(t_{2},t_{3})e^{it_{1N}} + b_{N2}(t_{2},t_{3})e^{-it_{1N}} \cr
&& + e^{it_{1N}}\frac{A_{1}(t_{2},t_{3})}{4}(1-i2t_{1N}) \cr
&& + e^{-it_{1N}}\frac{B_{1}(t_{2},t_{3})}{4}(1+i2t_{1N}) \ ,
\end{eqnarray} 
where we have introduced the quantities $A_{1}(t_{2},t_{3})$ and $B_{1}(t_{2},t_{3})$. The former is defined by
\begin{widetext}
\begin{eqnarray}
A_{1}(t_{2},t_{3}) &=& \frac{-2i}{\omega_{N}[ \tilde{\tilde{q}}(t_{2}) ]}\left\{ \frac{\partial a_{N1}}{\partial t_{2}}(t_{2},t_{3})+\frac{\partial a_{N0}}{\partial t_{3}}(t_{2},t_{3}) + \tilde{\tilde{q}}'(t_{2})\Gamma_{NN}\left[ \tilde{\tilde{q}}(t_{2}) \right] a_{N1}(t_{2},t_{3}) \right\} \cr
&& \cr
&& \cr
&& - \frac{\Gamma_{NN}\left[ \tilde{\tilde{q}}(t_{2}) \right] }{\omega_{N}\left[ \tilde{\tilde{q}}(t_{2}) \right]^{2}} \left[ 2\tilde{\tilde{q}}'(t_{2})\frac{\partial a_{N0}}{\partial t_{2}}(t_{2},t_{3}) + \tilde{\tilde{q}}''(t_{2})a_{N0}(t_{2},t_{3}) \right] \cr
&& \cr
&& \cr
&& - \frac{1}{\omega_{N}\left[ \tilde{\tilde{q}}(t_{2}) \right]^{2}} \left\{  \frac{\partial^{2} a_{N0}}{\partial t_{2}^{2}}(t_{2},t_{3}) + i \tilde{\tilde{q}}'(t_{2})\omega_{N}'\left[ \tilde{\tilde{q}}(t_{2}) \right] a_{N1}(t_{2},t_{3}) \right\} \cr
&& \cr
&& \cr
&& - 4 \tilde{\tilde{q}}'(t_{2})^{2} a_{N0}(t_{2},t_{3}) \sum_{n=1 \atop n\not= N}^{+\infty} \frac{\Gamma_{Nn}\left[ \tilde{\tilde{q}}(t_{2}) \right] \Gamma_{nN}\left[ \tilde{\tilde{q}}(t_{2}) \right] }{\omega_{n}\left[ \tilde{\tilde{q}}(t_{2})\right]^{2} - \omega_{N}\left[ \tilde{\tilde{q}}(t_{2})\right]^{2}} \ , 
\end{eqnarray}
and $B_{1}(t_{2},t_{3})$ is obtained from $A_{1}(t_{2},t_{3})$ by changing $i$ to $-i$, $a_{N1}$ to $b_{N1}$, and $a_{N0}$ to $b_{N0}$.
\end{widetext}

We now eliminate secular terms from $Y_{N2}$ given in (\ref{A3e116}). Using the expressions of $a_{N0}(t_{2},t_{3})$ and $b_{N0}(t_{2},t_{3})$ in (\ref{A3ea0b0}) it follows that secular terms in $Y_{N2}$ disappear if and only if
\begin{eqnarray}
\label{A3e116b}
&& A_{1}(t_{2},t_{3}) = B_{1}(t_{2},t_{3}) = 0 \ , \cr
&& \cr
&& \cr
&\Leftrightarrow& 
z(t_{2},t_{3}) = \alpha_{N}\left[ \tilde{\tilde{q}}(t_{2}) \right] \left[ z(0,t_{3}) -t_{2}\frac{\partial z_{1}}{\partial t_{3}}(0,t_{3}) \right. \cr
&& \qquad\qquad\qquad\qquad\qquad \left. - z_{2}(0,t_{3}) \int_{0}^{t_{2}}dt_{2}' \eta (t_{2}') \right] \ , \cr
&& \cr
&& \cr
&&
\mbox{with} \ \left\{
\begin{array}{c}
z = b_{N1} , \ z_{1} = b_{N0}, \ z_{2} = b_{N0} \ , \cr
\mbox{or} \cr
z=a_{N1}, \ z_{1} = a_{N0}, \ z_{2} = -a_{N0} \ .
\end{array}
\right.
\end{eqnarray}
Now one must also eliminate secular terms from $a_{N1}(t_{2},t_{3})$ and $b_{N1}(t_{2},t_{3})$, that is, one must eliminate terms that become unbounded as $t_{2}$ grows with $t_{3}$ fixed. From (\ref{A3e116b}) one finds that secular terms disappear from $a_{N1}(t_{2},t_{3})$ and $b_{N1}(t_{2},t_{3})$ if and only if
\begin{eqnarray}
\label{A3e121}
&& \frac{\partial a_{N0}}{\partial t_{3}}(0,t_{3}) = \frac{\partial b_{N0}}{\partial t_{3}}(0,t_{3}) = 0 \ , \ \ (t_{3} \geq 0) \ , \cr
&& \cr
&\Leftrightarrow& 
b_{N0}(0,t_{3}) = b_{N0}(0,0) = \frac{g_{0N}}{2} + \frac{ig_{1N}}{2\omega_{N}\left[ \tilde{\tilde{q}}(0) \right]} \ , \cr
&& a_{N0}(0,t_{3}) = a_{N0}(0,0) = b_{N0}(0,0)^{*} \ . \cr
&&
\end{eqnarray}
In deducing (\ref{A3e121}) we used the fact that the factor of $z_{2}$ in (\ref{A3e116b}) is not a multiple of $t_{2}$. This is seen explicitly further below in (\ref{etaF2}). Also, notice that we used the initial conditions in (\ref{A3e105}) in the last line of (\ref{A3e121}).

Substituting the values of $a_{N0}(0,t_{3})$ and $b_{N0}(0,t_{3})$ given in  (\ref{A3e121}) into the expressions for $a_{N0}(t_{2},t_{3})$ and $b_{N0}(t_{2},t_{3})$ in (\ref{A3ea0b0}), one concludes that
\begin{eqnarray}
\label{A3ea0b0F}
&& z(t_{2},t_{3}) = z(0,0)\alpha_{N}\left[ \tilde{\tilde{q}}(t_{2}) \right] \ ,
 \cr
&& \mbox{with} \ \ z = a_{N0}, \ b_{N0} \ .
\end{eqnarray} 
Therefore, $a_{N0}(t_{2},t_{3})$ and $b_{N0}(t_{2},t_{3})$ are now completely specified and it turns out that they do not depend on $t_{3}$. Also, notice from (\ref{A3e105}) and (\ref{A3ea0b0F}) that
\begin{eqnarray}
\label{ElConjugado}
a_{N0}(t_{2},t_{3}) &=& b_{N0}(t_{2},t_{3})^{*} \ .
\end{eqnarray}

Eliminating secular terms from $Y_{N2}$, $a_{N1}$, and $b_{N1}$ according to (\ref{A3e116b}) and (\ref{A3e121}), one obtains that
\begin{eqnarray}
\label{A3e125}
Y_{N2}(t_{1N},t_{2},t_{3}) &=& a_{N2}(t_{2},t_{3})e^{it_{1N}} + b_{N2}(t_{2},t_{3})e^{-it_{1N}} \ , \cr
&& \cr
&& \cr
a_{N1}(t_{2},t_{3}) &=& \alpha_{N}\left[ \tilde{\tilde{q}}(t_{2}) \right] \Big\{  \ a_{N1}(0,t_{3}) \cr
&& \qquad + b_{N0}(0,0)^{*} \int_{0}^{t_{2}}dt_{2}' \eta (t_{2}') \Big\} \ , \cr
&& \cr
&& \cr
b_{N1}(t_{2},t_{3}) &=& \alpha_{N}\left[ \tilde{\tilde{q}}(t_{2}) \right] \Big\{ b_{N1}(0,t_{3}) \cr 
&& \qquad - b_{N0}(0,0) \int_{0}^{t_{2}}dt_{2}' \eta (t_{2}') \Big\} \ . \cr
&&
\end{eqnarray}
Notice that $a_{N1}(t_{2},t_{3})$ and $b_{N1}(t_{2},t_{3})$ are not yet completely specified because $a_{N1}(0,t_{3})$ and $b_{N1}(0,t_{3})$ remain to be determined. This is done by solving the $\mathcal{O}(\epsilon_{\mbox{\tiny pert}}^{3})$ problem. Also, the initial conditions for $a_{N2}(t_{2},t_{3})$ and $b_{N2}(t_{2},t_{3})$ are obtained from the $\mathcal{O}(\epsilon_{\mbox{\tiny pert}}^{2})$ initial conditions in (\ref{A3e3p14}) by substituting the expressions for $Y_{N0}$, $a_{N0}(0,0)$, $b_{N0}(0,0)$, $Y_{N1}$, $a_{N1}(0,0)$, $b_{N1}(0,0)$, and $Y_{N2}$  given in (\ref{A3e104}), (\ref{A3e105}), (\ref{A3e114}), (\ref{A3e127}), and (\ref{A3e125}). One obtains
\begin{eqnarray}
\label{NNN}
b_{N2}(0,0) &=& i\tilde{\tilde{q}}''(0)\frac{g_{1N}\omega_{N}'\left[ \tilde{\tilde{q}}(0) \right]}{8 \omega_{N}\left[ \tilde{\tilde{q}}(0) \right]^{4}} \ ,  \qquad \cr
a_{N2}(0,0) &=& b_{N2}(0,0)^{*} \ . 
\end{eqnarray}

We now solve the $\mathcal{O}(\epsilon_{\mbox{\tiny pert}}^{2})$ for $Y_{m2}$ with $m\not= N$. Substituting in the $\mathcal{O}(\epsilon_{\mbox{\tiny pert}}^{2})$ differential equation given in (\ref{A3e3p14}) the expressions for $Y_{N0}$ and $Y_{m0}$ in (\ref{A3e104}), $Y_{m1}$ in (\ref{A3e108}), $Y_{N1}$ in (\ref{A3e114}), $a_{N0}$ and $b_{N0}$ in (\ref{A3ea0b0F}), and  $a_{N1}$ and $b_{N1}$ in (\ref{A3e125}), and solving the resulting harmonic oscillator equations with driving, one obtains that
\begin{eqnarray}
\label{A3e3p23}
&& Y_{m2}(t_{1N},t_{2},t_{3}) \cr
&=& a_{m2}(t_{2},t_{3})e^{iW_{m}(t_{2})t_{1N}} + b_{m2}(t_{2},t_{3})e^{-iW_{m}(t_{2})t_{1N}} \cr
&& \cr
&& + \frac{e^{it_{1N}}D(t_{2},t_{3}) + e^{-it_{1N}}F(t_{2},t_{3})}{W_{m}(t_{2})^{2} -1} \ , \cr
&& (m \not= N),
\end{eqnarray}
where we have introduced the quantities 

\begin{widetext}
\begin{eqnarray}
\label{A3e3p20}
h_{9}(t_{2}) &=& -i\tilde{\tilde{q}}'(t_{2})\frac{2 \Gamma_{mN}\left[ \tilde{\tilde{q}}(t_{2}) \right] }{\omega_{N}\left[ \tilde{\tilde{q}}(t_{2}) \right]} \alpha_{N}\left[ \tilde{\tilde{q}}(t_{2}) \right] a_{N0}(0,0)\int_{0}^{t_{2}} dt_{2}' \eta (t_{2}') \cr  
&& - \frac{2 a_{N0}(t_{2},t_{3}) }{\omega_{N}\left[ \tilde{\tilde{q}}(t_{2}) \right]} \left\{ \ \tilde{\tilde{q}}''(t_{2})\frac{\Gamma_{mN}\left[ \tilde{\tilde{q}}(t_{2}) \right]}{2 \omega_{N}\left[ \tilde{\tilde{q}}(t_{2}) \right]} \ +  \frac{\partial}{\partial t_{2}}\left[ \frac{T_{m}(t_{2})}{W_{m}(t_{2})^{2} -1} \right] \right. \cr
&& \left.  -\tilde{\tilde{q}}'(t_{2})\frac{T_{m} (t_{2}) \Gamma_{NN}\left[ \tilde{\tilde{q}}(t_{2}) \right] }{W_{m}(t_{2})^{2} -1}   \ 
+ \tilde{\tilde{q}}'(t_{2})\sum_{n=1 \atop n\not= N}^{+\infty} \frac{T_{n}(t_{2}) \Gamma_{mn} \left[ \tilde{\tilde{q}}(t_{2}) \right]}{W_{n}(t_{2})^{2}-1} -\tilde{\tilde{q}}'(t_{2})^{2}\frac{\Gamma_{mN}\left[ \tilde{\tilde{q}}(t_{2}) \right]}{\omega_{N}\left[ \tilde{\tilde{q}}(t_{2}) \right]}\zeta_{N}(t_{2}) \ \right\}  \ , \cr
&& \cr
h_{10}(t_{2}) &=& -i\tilde{\tilde{q}}'(t_{2})\frac{2 \Gamma_{mN}\left[ \tilde{\tilde{q}}(t_{2}) \right] }{\omega_{N}\left[ \tilde{\tilde{q}}(t_{2}) \right]} \alpha_{N}\left[ \tilde{\tilde{q}}(t_{2}) \right] \ ,  \cr
&& \cr
D(t_{2},t_{3}) &=& h_{9}(t_{2}) + h_{10}(t_{2})a_{N1}(0,t_{3}) \ , \ \qquad\qquad\qquad F(t_{2},t_{3}) \ = \ h_{9}(t_{2})^{*} +h_{10}(t_{2})^{*}b_{N1}(0,t_{3}) \ .
\end{eqnarray}
Notice that we used that $a_{N0}(t_{2},t_{3})$ and $b_{N0}(t_{2},t_{3})$ do not depend on $t_{3}$, see (\ref{A3ea0b0F}).
\end{widetext}

Applying the $\mathcal{O}(\epsilon_{\mbox{\tiny pert}}^{2})$ initial conditions given in (\ref{A3e3p14}) to $Y_{m2}$ given in (\ref{A3e3p23}) and using that $\tilde{\tilde{q}}'(0) = 0$ (see item 3 in Section IV), it is straightforward to show that
\begin{eqnarray}
\label{A3e3p26}
a_{m2}(0,0) + b_{m2}(0,0) &=& g_{0N}\tilde{\tilde{q}}''(0)\Gamma_{mN}\left[ \tilde{\tilde{q}}(0) \right] \times \cr
&& \times \frac{\omega_{m}\left[ \tilde{\tilde{q}}(0) \right]^{2} + 3 \omega_{N}\left[ \tilde{\tilde{q}}(0) \right]^{2}}{\left\{ \omega_{m}\left[ \tilde{\tilde{q}}(0) \right]^{2} - \omega_{N}\left[ \tilde{\tilde{q}}(0) \right]^{2} \right\}^{2}} \ , \cr
&& \cr
&& \cr
a_{m2}(0,0) - b_{m2}(0,0) &=& -ig_{1N}\tilde{\tilde{q}}''(0)\frac{\Gamma_{mN}\left[ \tilde{\tilde{q}}(0) \right]}{\omega_{m}\left[ \tilde{\tilde{q}}(0) \right]} \times \cr
&& \times \frac{3 \omega_{m}\left[ \tilde{\tilde{q}}(0) \right]^{2} + \omega_{N}\left[ \tilde{\tilde{q}}(0) \right]^{2}}{\left\{ \omega_{m}\left[ \tilde{\tilde{q}}(0) \right]^{2} - \omega_{N}\left[ \tilde{\tilde{q}}(0) \right]^{2} \right\}^{2}} \ . \cr
&&
\end{eqnarray} 
Observe that both $a_{m2}(0,0)$ and $b_{m2}(0,0)$ are, in general, different from zero because $g_{0N}$ and $g_{1N}$ cannot both be zero (there would be no field in that case) and the rest of the quantities involved in the right-hand sides of (\ref{A3e3p26}) are, in general, different from zero. As a consequence, the terms that vary as $e^{\pm iW_{m}(t_{2})t_{1N}}$ in (\ref{A3e3p23}) will not disappear, in general, from $Y_{m2}$.

We now solve the $\mathcal{O}(\epsilon_{\mbox{\tiny pert}}^{3})$ for $Y_{N3}$. This will allow us to completely specify $a_{N1}(t_{2},t_{3})$ and $b_{N1}(t_{2},t_{3})$. 

Substituting into the $\mathcal{O}(\epsilon_{\mbox{\tiny pert}}^{3})$ equation in (\ref{A3e3p15}) the expressions for $Y_{N0}$ and $Y_{m0}$ in (\ref{A3e104}), $Y_{m1}$ in (\ref{A3e108}), $Y_{N1}$ in (\ref{A3e114}), $a_{N0}$ and $b_{N0}$ in (\ref{A3ea0b0F}), $Y_{N2}$, $a_{N1}$, and $b_{N1}$ in (\ref{A3e125}), and $Y_{m2}$ in (\ref{A3e3p23}), and solving the resulting harmonic oscillator equations with driving, one obtains that
\begin{eqnarray}
\label{A3e3p28}
Y_{N3}(t_{1N},t_{2},t_{3}) &=& a_{N3}(t_{2},t_{3})e^{it_{1N}} + b_{N3}(t_{2},t_{3})e^{-it_{1N}} \cr
&& \cr
&& + e^{it_{1N}}\frac{( 1-i2t_{1N} )}{4\omega_{N}\left[ \tilde{\tilde{q}}(t_{2}) \right]^{2}}R(t_{2},t_{3}) \cr
&& + e^{-it_{1N}}\frac{ (1+i2t_{1N}) }{4\omega_{N}\left[ \tilde{\tilde{q}}(t_{2}) \right]^{2}}S(t_{2},t_{3}) \cr
&& \cr
&& + \sum_{n=1 \atop n\not= N}^{+ \infty} \frac{R_{n}(t_{2},t_{3})e^{iW_{n}(t_{2})t_{1N}}}{\omega_{N}\left[ \tilde{\tilde{q}}(t_{2}) \right]^{2} - \omega_{n}\left[ \tilde{\tilde{q}}(t_{2}) \right]^{2}} \cr
&& + \sum_{n=1 \atop n\not= N}^{+ \infty} \frac{S_{n}(t_{2},t_{3})e^{-iW_{n}(t_{2})t_{1N}}}{\omega_{N}\left[ \tilde{\tilde{q}}(t_{2}) \right]^{2} - \omega_{n}\left[ \tilde{\tilde{q}}(t_{2}) \right]^{2}} \ , \cr
&&
\end{eqnarray}
where
\begin{eqnarray}
\label{A3e3p33}
\frac{i R(t_{2},t_{3})}{2\omega_{N}\left[ \tilde{\tilde{q}}(t_{2}) \right]} &=& \frac{\partial a_{N2}}{\partial t_{2}}(t_{2},t_{3}) + \tilde{\tilde{q}}'(t_{2})\zeta_{N}(t_{2}) a_{N2}(t_{2},t_{3})  \cr
&&  + a_{N0}(t_{2},t_{3}) \left( i\frac{f_{9}(t_{2}) -f_{11}(t_{2})}{2\omega_{N}\left[ \tilde{\tilde{q}}(t_{2}) \right]} \right. \cr
&&  + i\frac{f_{10}(t_{2}) -f_{12}(t_{2})}{2 \omega_{N}\left[ \tilde{\tilde{q}}(t_{2}) \right]}a_{N1}(0,t_{3}) \cr
&& \left. +\frac{1}{a_{N0}(0,0)}\frac{\partial a_{N1}}{\partial t_{3}}(0,t_{3}) \right) \ , \cr
&& \cr
&& \cr
\frac{S(t_{2},t_{3})}{i2\omega_{N}\left[ \tilde{\tilde{q}}(t_{2}) \right]} &=& \frac{\partial b_{N2}}{\partial t_{2}}(t_{2},t_{3}) + \tilde{\tilde{q}}'(t_{2})\zeta_{N}(t_{2}) b_{N2}(t_{2},t_{3})  \cr
&& \cr
&& - b_{N0}(t_{2},t_{3}) \left( i\frac{f_{16}(t_{2}) -f_{14}(t_{2})}{2\omega_{N}\left[ \tilde{\tilde{q}}(t_{2}) \right]} \right. \cr
&& \cr
&& + i\frac{f_{17}(t_{2}) -f_{15}(t_{2})}{2\omega_{N}\left[ \tilde{\tilde{q}}(t_{2}) \right]}b_{N1}(0,t_{3}) \cr
&& \cr
&& \left. -\frac{1}{b_{N0}(0,0)}\frac{\partial b_{N1}}{\partial t_{3}}(0,t_{3}) \right)  \ , \cr
&& \cr
\frac{i R_{n}(t_{2},t_{3})}{2\omega_{N}\left[ \tilde{\tilde{q}}(t_{2}) \right]} &=& \tilde{\tilde{q}}'(t_{2})\Gamma_{Nn}\left[ \tilde{\tilde{q}}(t_{2}) \right] W_{n}(t_{2}) a_{n2}(t_{2},t_{3}) \ , \cr
&& \cr
&& \cr
\frac{S_{n}(t_{2},t_{3})}{i2\omega_{N}\left[ \tilde{\tilde{q}}(t_{2}) \right]} &=&  \tilde{\tilde{q}}'(t_{2})\Gamma_{Nn}\left[ \tilde{\tilde{q}}(t_{2}) \right] W_{n}(t_{2}) b_{n2}(t_{2},t_{3}) \ . \cr
&&
\end{eqnarray}
Here we had to introduce the following quantities which also make use of $D(t_{2},t_{3})$ and $F(t_{2},t_{3})$ in (\ref{A3e3p20}), the quantities defined in (\ref{A2eOp}), and the result that $a_{N0}(t_{2},t_{3})$ and $b_{N0}(t_{2},t_{3})$ do not depend on $t_{3}$ (see (\ref{A3ea0b0F})):
\begin{eqnarray}
\label{A3e3p30}
\frac{h_{5}(t_{2})}{a_{N0}(t_{2},t_{3})} &=& -\frac{h_{7}(t_{2})}{b_{N0}(t_{2},t_{3})} \ = \ \int_{0}^{t_{2}} dt_{2}' \eta (t_{2}') \ , \cr
&& \cr
&& \cr
h_{6}(t_{2}) &=& \frac{a_{N0}(t_{2},t_{3})}{a_{N0}(0,0)} \ , \qquad h_{8}(t_{2}) \ = \ \frac{b_{N0}(t_{2},t_{3})}{b_{N0}(0,0)} \ , \cr
&& \cr
&& \cr
f_{1}(t_{2}) &=& \eta'(t_{2}) -2\tilde{\tilde{q}}'(t_{2}) \eta(t_{2}) \zeta_{N}(t_{2}) \cr
&&\cr
&& -\tilde{\tilde{q}}''(t_{2}) \zeta_{N}(t_{2}) \int_{0}^{t_{2}} dt_{2}' \ \eta (t_{2}') \cr
&& \cr
&& -\tilde{\tilde{q}}'(t_{2}) \left[ \int_{0}^{t_{2}} dt_{2}' \ \eta (t_{2}') \right] \frac{d}{dt_{2}}\zeta_{N}(t_{2}) \cr
&& \cr
&& + \tilde{\tilde{q}}'(t_{2})^{2}\zeta_{N}(t_{2})^{2} \int_{0}^{t_{2}} dt_{2}' \ \eta (t_{2}') \cr
&& \cr
&& \cr
f_{2}(t_{2}) &=& -\frac{\tilde{\tilde{q}}''(t_{2})}{a_{N0}(0,0)} \zeta_{N}(t_{2}) -\frac{\tilde{\tilde{q}}'(t_{2})}{a_{N0}(0,0)} \frac{d}{dt_{2}} \zeta_{N}(t_{2}) \cr
&&\cr
&& + \frac{\tilde{\tilde{q}}'(t_{2})^{2}}{a_{N0}(0,0)}\zeta_{N}(t_{2})^{2} \ , \cr
&& \cr
&& \cr
f_{4}(t_{2}) &=& -\frac{a_{N0}(0,0)}{b_{N0}(0,0)}f_{2}(t_{2}) \ , \cr
&& \cr
&& \cr
f_{5}(t_{2}) &=& \eta (t_{2}) - \tilde{\tilde{q}}'(t_{2})\zeta_{N}(t_{2}) \int_{0}^{t_{2}}dt_{2}' \ \eta (t_{2}') \ , \cr
&& \cr
&& \cr
f_{6}(t_{2}) &=& - \frac{\tilde{\tilde{q}}'(t_{2})\zeta_{N}(t_{2})}{a_{N0}(0,0)} \ , \ \ \ f_{8}(t_{2}) \ = \ \frac{\tilde{\tilde{q}}'(t_{2}) \zeta_{N}(t_{2})}{b_{N0}(0,0)}  \ , \cr
&& \cr
&& \cr
f_{9}(t_{2}) &=& i2\tilde{\tilde{q}}'(t_{2}) \sum_{n=1 \atop n\not= N}^{+\infty} \Gamma_{Nn}\left[ \tilde{\tilde{q}}(t_{2}) \right] \times \cr
&& \times\left\{ \frac{\partial}{\partial t_{2}} \left[ \frac{T_{n}(t_{2})}{W_{n}(t_{2})^{2} -1} \right] \right. \cr
&& \cr
&& \left. -\frac{\omega_{N}\left[ \tilde{\tilde{q}}(t_{2}) \right] h_{9}(t_{2})}{a_{N0}(t_{2},t_{3})\left[ W_{n}(t_{2})^{2} -1 \right]} \right\} \cr
&& \cr
&& \cr
&& +i \left[ \tilde{\tilde{q}}''(t_{2}) -2\tilde{\tilde{q}}'(t_{2})^{2} \zeta_{N}(t_{2}) \right] \times \cr
&& \times \sum_{n=1 \atop n\not= N}^{+\infty}  \frac{ T_{n}(t_{2}) \Gamma_{Nn}\left[ \tilde{\tilde{q}}(t_{2}) \right] }{W_{n}(t_{2})^{2} -1} \cr
&& \cr
&& \cr
&& \cr
f_{10}(t_{2}) &=& -i2\tilde{\tilde{q}}'(t_{2}) \omega_{N}\left[ \tilde{\tilde{q}}(t_{2}) \right] \frac{h_{10}(t_{2})}{a_{N0}(t_{2},t_{3})}  \times \cr
&& \times \sum_{n=1 \atop n\not= N}^{+\infty} \frac{\Gamma_{Nn}\left[ \tilde{\tilde{q}}(t_{2}) \right] }{W_{n}(t_{2})^{2} -1} \ , \cr
&& \cr
&& \cr
f_{11}(t_{2}) &=& f_{1}(t_{2}) + 2\tilde{\tilde{q}}'(t_{2})\Gamma_{NN}\left[ \tilde{\tilde{q}}(t_{2}) \right] f_{5}(t_{2}) \cr
&& + \tilde{\tilde{q}}''(t_{2}) \Gamma_{NN}\left[ \tilde{\tilde{q}}(t_{2}) \right] \frac{h_{5}(t_{2})}{a_{N0}(t_{2},t_{3})} \ ,
\end{eqnarray}
and
\begin{enumerate}
\item[(i)] $f_{12}(t_{2})$ is obtained from $f_{11}(t_{2})$ by changing $f_{1}(t_{2})$ to $f_{2}(t_{2})$,  $f_{5}(t_{2})$ to $f_{6}(t_{2})$, and $h_{5}(t_{2})$ to $h_{6}(t_{2})$.

\item[(ii)] $f_{16}(t_{2})$ is obtained from $f_{11}(t_{2})$ by changing $h_{5}(t_{2})$ to $h_{7}(t_{2})$ and $a_{N0}(t_{2},t_{3})$ to $-b_{N0}(t_{2},t_{3})$.

\item[(iii)] $f_{17}(t_{2})$ is obtained from $f_{11}(t_{2})$ by changing $f_{1}(t_{2})$ to $f_{4}(t_{2})$, $f_{5}(t_{2})$ to $f_{8}(t_{2})$, $h_{5}(t_{2})$ to $h_{8}(t_{2})$, and $a_{N0}(t_{2},t_{3})$ to $-b_{N0}(t_{2},t_{3})$.

\item[(iv)] $f_{15}(t_{2})$ is obtained from $f_{10}(t_{2})$ by changing $h_{10}(t_{2})/a_{N0}(t_{2},t_{3})$ to $h_{10}(t_{2})^{*}/b_{N0}(t_{2},t_{3})$.

\item[(v)] $f_{14}(t_{2})$ is obtained from $f_{9}(t_{2})$ by changing $h_{9}(t_{2})/a_{N0}(t_{2},t_{3})$ to $h_{9}(t_{2})^{*}/b_{N0}(t_{2},t_{3})$.
\end{enumerate}

From (\ref{A3e3p28}) it follows that secular terms disappear from $Y_{N3}$ if and only if
\begin{eqnarray}
&& R(t_{2},t_{3}) = 0 \ , \ \ S(t_{2},t_{3}) = 0 \ . 
\end{eqnarray}
Using the definitions of $S(t_{2},t_{3})$ and $R(t_{2},t_{3})$ in (\ref{A3e3p33}), one finds that this is equivalent to
\begin{eqnarray}
\label{A3e3p29b}
a_{N2}(t_{2},t_{3}) &=& \alpha_{N}\left[ \tilde{\tilde{q}}(t_{2}) \right] \Big\{ a_{N2}(0,t_{3}) - t_{2}\frac{\partial a_{N1}}{\partial t_{3}}(0,t_{3}) \cr
&& + a_{N0}(0,0)\left[ \psi_{1}(t_{2}) + \psi_{2}(t_{2})a_{N1}(0,t_{3}) \right]  \Big\}  \cr
&& \cr
&& \cr
b_{N2}(t_{2},t_{3}) &=& \alpha_{N}\left[ \tilde{\tilde{q}}(t_{2}) \right] \Big\{ b_{N2}(0,t_{3}) - t_{2}\frac{\partial b_{N1}}{\partial t_{3}}(0,t_{3}) \cr 
&& + b_{N0}(0,0)\left[ \psi_{3}(t_{2}) + \psi_{4}(t_{2})b_{N1}(0,t_{3}) \right]  \Big\} \ , \cr
&&
\end{eqnarray}
where we have introduced the quatities
\begin{eqnarray}
\label{A2e2p29b}
\psi_{1}(t_{2}) &=& \int_{0}^{t_{2}} dt_{2}' \  i \frac{f_{11}(t_{2}') - f_{9}(t_{2}')}{2\omega_{N}\left[ \tilde{\tilde{q}}(t_{2}') \right]} \ , \cr
&& \cr
&& \cr
\psi_{2}(t_{2}) &=& \int_{0}^{t_{2}} dt_{2}' \  i \frac{f_{12}(t_{2}') - f_{10}(t_{2}')}{2\omega_{N}\left[ \tilde{\tilde{q}}(t_{2}') \right]} \ , \cr
&& \cr
&& \cr
\psi_{3}(t_{2}) &=& \int_{0}^{t_{2}} dt_{2}' \  i \frac{f_{16}(t_{2}') - f_{14}(t_{2}')}{2\omega_{N}\left[ \tilde{\tilde{q}}(t_{2}') \right]} \ , \cr
&& \cr
&& \cr
\psi_{4}(t_{2}) &=& \int_{0}^{t_{2}} dt_{2}' \  i \frac{f_{17}(t_{2}') - f_{15}(t_{2}')}{2\omega_{N}\left[ \tilde{\tilde{q}}(t_{2}') \right]} \ . 
\end{eqnarray}

We now use the expressions for $a_{N2}(t_{2},t_{3})$ and $b_{N2}(t_{2},t_{3})$ given in (\ref{A3e3p29b}) to completely specify $a_{N1}(t_{2},t_{3})$ and $b_{N1}(t_{2},t_{3})$ (recall that $a_{N1}(0,t_{3})$ and $b_{N1}(0,t_{3})$ remain to be determined). This is done by eliminating secular terms from $a_{N2}(t_{2},t_{3})$ and $b_{N2}(t_{2},t_{3})$. From (\ref{A3e3p29b}) we find that secular terms disappear from $a_{N2}(t_{2},t_{3})$ and $b_{N2}(t_{2},t_{3})$ if and only if
\begin{eqnarray}
\label{Nuevob}
\frac{\partial a_{N1}}{\partial t_{3}}(0,t_{3}) &=& \frac{\partial b_{N1}}{\partial t_{3}}(0,t_{3}) = 0 \ , \ \ (t_{3} \geq 0) \ .
\end{eqnarray}
We note that in deducing (\ref{Nuevob}) one uses that $\psi_{j}(t_{2})$ ($j=1$, $2$, $3$, $4$) does not give rise to terms proportional to $t_{2}$. This is seen explicitly further below in (\ref{etaF}).

Integrating (\ref{Nuevob}) and using the initial conditions in (\ref{A3e127}) one finds that this is equivalent to
\begin{eqnarray}
\label{A3e3p35}
a_{N1}(0,t_{3}) = b_{N1}(0,t_{3}) = 0 \ \ \ (t_{3} \geq 0) .
\end{eqnarray}

Substituting the values of $a_{N1}(0,t_{3})$ and $b_{N1}(0,t_{3})$ given in (\ref{A3e3p35}) into the expressions for $a_{N1}(t_{2},t_{3})$ and $b_{N1}(t_{2},t_{3})$ given in (\ref{A3e125}), one concludes that
\begin{eqnarray}
\label{A3e125b}
a_{N1}(t_{2},t_{3}) &=& \alpha_{N}\left[ \tilde{\tilde{q}}(t_{2}) \right] b_{N0}(0,0)^{*} \int_{0}^{t_{2}}dt_{2}' \eta (t_{2}')  \ , \cr
&& \cr
b_{N1}(t_{2},t_{3}) &=& - \alpha_{N}\left[ \tilde{\tilde{q}}(t_{2}) \right] b_{N0}(0,0) \int_{0}^{t_{2}}dt_{2}' \eta (t_{2}')  \ . \cr
&&
\end{eqnarray}
Also, substituting (\ref{A3e3p35}) into $ a_{N2}(t_{2},t_{3})$ and $ b_{N2}(t_{2},t_{3})$ given in (\ref{A3e3p29b}) and using $a_{N0}(0,0)=b_{N0}(0,0)^{*}$ given in (\ref{ElConjugado}), one obtains that
\begin{eqnarray}
\label{A3e3p29}
a_{N2}(t_{2},t_{3}) &=& \alpha_{N}\left[ \tilde{\tilde{q}}(t_{2}) \right] \left[ a_{N2}(0,t_{3}) + b_{N0}(0,0)^{*}\psi_{1}(t_{2}) \right] \ , \cr
&& \cr
&& \cr
b_{N2}(t_{2},t_{3}) &=& \alpha_{N}\left[ \tilde{\tilde{q}}(t_{2}) \right] \left[ b_{N2}(0,t_{3}) + b_{N0}(0,0)\psi_{3}(t_{2})  \right] \ . \cr
&&
\end{eqnarray}

If one only requires a one- or two-term approximation for $d_{n}$, then one could stop here because all the necessary quantities have already been determined. Nevertheless, we are now going to solve the $\mathcal{O}(\epsilon_{\mbox{\tiny pert}}^{3})$ problem in (\ref{A3e3p15}) for $Y_{m3}$ with $m\not= N$. This will serve to point out the difficulties one encounters when one requires an $n$-term approximation for $d_{n}$ with $n\geq 3$.

First substitute into the $\mathcal{O}(\epsilon_{\mbox{\tiny pert}}^{3})$ differential equation the expressions for $Y_{m0}$ and $Y_{N0}$ in (\ref{A3e104}), $Y_{m1}$ and $Y_{N1}$ in (\ref{A3e108}) and (\ref{A3e114}), and $Y_{N2}$ and $Y_{m2}$ in (\ref{A3e125}) and (\ref{A3e3p23}). Solving the resulting harmonic oscillator equations with driving, we obtain

\begin{widetext}
\begin{eqnarray}
\label{A3e3p43}
&& Y_{m3}(t_{1N},t_{2},t_{3})  \cr
&=& \frac{e^{it_{1N}}J(t_{2},t_{3})}{\omega_{m}\left[ \tilde{\tilde{q}}(t_{2}) \right]^{2} -\omega_{N}\left[ \tilde{\tilde{q}}(t_{2}) \right]^{2}} \ + \frac{ie^{iW_{m}(t_{2})t_{1N}}}{4W_{m}(t_{2})}\left\{ \ \left[ \frac{H_{10}(t_{2},t_{3})}{2W_{m}(t_{2})^{2}} - i\frac{L_{0}(t_{2},t_{3})}{W_{m}(t_{2})} \right]\left[ 1 -2it_{1N}W_{m}(t_{2})\right] - H_{10}(t_{2},t_{3})t_{1N}^{2} \ \right\} \cr
&& \cr
&& + \frac{e^{-it_{1N}}K(t_{2},t_{3})}{\omega_{m}\left[ \tilde{\tilde{q}}(t_{2}) \right]^{2} -\omega_{N}\left[ \tilde{\tilde{q}}(t_{2}) \right]^{2}} \ - \frac{ie^{-iW_{m}(t_{2})t_{1N}}}{4W_{m}(t_{2})}\left\{ \ \left[ \frac{H_{20}(t_{2},t_{3})}{2W_{m}(t_{2})^{2}} + i\frac{P_{0}(t_{2},t_{3})}{W_{m}(t_{2})} \right] \left[ 1 + 2it_{1N}W_{m}(t_{2})  \right] - H_{20}(t_{2},t_{3})t_{1N}^{2} \ \right\} \cr
&& \cr
&& +a_{m3}(t_{2},t_{3})e^{iW_{m}(t_{2})t_{1N}} \ + b_{m3}(t_{2},t_{3})e^{-iW_{m}(t_{2})t_{1N}} \ + \sum_{n=1 \atop n\not= m,N}^{+\infty}\frac{e^{iW_{n}(t_{2})t_{1N}}R_{mn0}(t_{2},t_{3}) + e^{-iW_{n}(t_{2})t_{1N}}S_{mn0}(t_{2},t_{3})}{W_{m}(t_{2})^{2} - W_{n}(t_{2})^{2}} \ , \cr
&&
\end{eqnarray}
with

\begin{eqnarray}
\label{JJ}
J(t_{2},t_{3}) 
&=& 
\tilde{\tilde{q}}''(t_{2}) \Gamma_{mm}\left[ \tilde{\tilde{q}}(t_{2}) \right] \frac{i T_{m}(t_{2})}{W_{m}(t_{2})^{2} -1} a_{N0}(t_{2},t_{3}) \ +2 \tilde{\tilde{q}}'(t_{2})\Gamma_{mm}\left[ \tilde{\tilde{q}}(t_{2}) \right] a_{N0}(t_{2},t_{3})\frac{\partial}{\partial t_{2}}\left[ \frac{iT_{m}(t_{2})}{W_{m}(t_{2})^{2}-1} \right] \cr
&& +2\tilde{\tilde{q}}'(t_{2}) \Gamma_{mm}\left[ \tilde{\tilde{q}}(t_{2}) \right]\frac{i T_{m}(t_{2})}{W_{m}(t_{2})^{2} -1} \frac{\partial a_{N0}}{\partial t_{2}}(t_{2},t_{3}) \ + a_{N0}(t_{2},t_{3})\frac{\partial^{2}}{\partial t_{2}^{2}}\left[ \frac{i T_{m}(t_{2})}{W_{m}(t_{2})^{2} -1} \right] \cr
&& + 2\frac{\partial a_{N0}}{\partial t_{2}}(t_{2},t_{3}) \frac{\partial}{\partial t_{2}}\left[ \frac{i T_{m}(t_{2})}{W_{m}(t_{2})^{2} -1} \right] \ + \frac{i T_{m}(t_{2})}{W_{m}(t_{2})^{2} -1}\frac{\partial^{2}a_{N0}}{\partial t_{2}^{2}}(t_{2},t_{3}) \ - \tilde{\tilde{q}}''(t_{2})\Gamma_{mN}\left[ \tilde{\tilde{q}}(t_{2}) \right]a_{N1}(t_{2},t_{3}) \cr
&& -2\omega_{N}\left[ \tilde{\tilde{q}}(t_{2}) \right]\frac{\partial}{\partial t_{2}}\left[ \frac{iD(t_{2},t_{3})}{W_{m}(t_{2})^{2} -1}  \right] \ -\tilde{\tilde{q}}'(t_{2})\Big\{ \omega_{N}'\left[ \tilde{\tilde{q}}(t_{2}) \right] + 2\Gamma_{mm}\left[ \tilde{\tilde{q}}(t_{2}) \right] \omega_{N}\left[ \tilde{\tilde{q}}(t_{2}) \right] \Big\} \frac{iD(t_{2},t_{3})}{W_{m}(t_{2})^{2} -1} \cr
&& -2\tilde{\tilde{q}}'(t_{2})\Gamma_{mN}\left[ \tilde{\tilde{q}}(t_{2}) \right]\frac{\partial a_{N1}}{\partial t_{2}}(t_{2},t_{3}) \ -i2\tilde{\tilde{q}}'(t_{2})\omega_{N}\left[ \tilde{\tilde{q}}(t_{2}) \right] \Gamma_{mN}\left[ \tilde{\tilde{q}}(t_{2}) \right]a_{N2}(t_{2},t_{3}) \cr
&& + \tilde{\tilde{q}}''(t_{2}) a_{N0}(t_{2},t_{3}) \sum_{n=1 \atop n \not= m, N}^{+\infty} iT_{n}(t_{2})\frac{\Gamma_{mn}\left[ \tilde{\tilde{q}}(t_{2}) \right]}{W_{n}(t_{2})^{2} -1} \ -2i \tilde{\tilde{q}}'(t_{2}) \omega_{N}\left[ \tilde{\tilde{q}}(t_{2}) \right] D(t_{2},t_{3}) \sum_{n=1 \atop n \not= m, N}^{+\infty} \frac{\Gamma_{mn}\left[ \tilde{\tilde{q}}(t_{2}) \right]}{W_{n}(t_{2})^{2} -1} \cr
&& + 2\tilde{\tilde{q}}'(t_{2}) \sum_{n=1 \atop n\not= m,N}^{+ \infty} \Gamma_{mn}\left[ \tilde{\tilde{q}}(t_{2}) \right] \frac{\partial}{\partial t_{2}}\left\{ \ a_{N0}(t_{2},t_{3})\frac{iT_{n}(t_{2})}{W_{n}(t_{2})^{2} -1} \ \right\} \ , \cr
&& \cr
&& \cr
-\frac{R_{mn0}(t_{2},t_{3})}{ ia_{n2}(t_{2},t_{3})} &=& \frac{S_{mn0}(t_{2},t_{3})}{ib_{n2}(t_{2},t_{3})} \ = \ 2\tilde{\tilde{q}}'(t_{2})\frac{ \Gamma_{mn} \left[ \tilde{\tilde{q}}(t_{2}) \right] }{\omega_{N}\left[ \tilde{\tilde{q}}(t_{2}) \right]} W_{n}(t_{2}) \ , \ \ \frac{H_{10}(t_{2},t_{3})}{a_{m2}(t_{2},t_{3})} \ = \ \frac{H_{20}(t_{2},t_{3})}{b_{m2}(t_{2},t_{3})} \ = \ \frac{2 W_{m}(t_{2})}{\omega_{N}\left[ \tilde{\tilde{q}}(t_{2}) \right]} W_{m}'(t_{2})  \ , \cr
&& \cr
&& \cr
L_{0}(t_{2},t_{3}) &=& -\tilde{\tilde{q}}'(t_{2}) \frac{2i \zeta_{m}(t_{2}) W_{m}(t_{2})}{\omega_{N}\left[ \tilde{\tilde{q}}(t_{2}) \right]}a_{m2}(t_{2},t_{3}) \ -\frac{2i W_{m}'(t_{2})}{\omega_{N}\left[ \tilde{\tilde{q}}(t_{2}) \right]} a_{m2}(t_{2},t_{3}) \ - \frac{2 i W_{m}(t_{2})}{\omega_{N}\left[ \tilde{\tilde{q}}(t_{2}) \right]} \frac{\partial a_{m2}}{\partial t_{2}}(t_{2},t_{3}) \ , \cr
&& \cr
&& \cr
P_{0}(t_{2},t_{3}) &=& \tilde{\tilde{q}}'(t_{2}) \frac{2 i \zeta_{m}(t_{2}) W_{m}(t_{2})}{\omega_{N}\left[ \tilde{\tilde{q}}(t_{2}) \right]}b_{m2}(t_{2},t_{3}) \ + \frac{2 i W_{m}'(t_{2})}{\omega_{N}\left[ \tilde{\tilde{q}}(t_{2}) \right]} b_{m2}(t_{2},t_{3}) \ + \frac{2 i W_{m}(t_{2})}{\omega_{N}\left[ \tilde{\tilde{q}}(t_{2}) \right]} \frac{\partial b_{m2}}{\partial t_{2}}(t_{2},t_{3}) \ , 
\end{eqnarray}
and $K(t_{2},t_{3})$ is obtained from $J(t_{2},t_{3})$ by changing $i$ to $-i$, $a_{N0}$ to $b_{N0}$, $D$ to $F$, $a_{N1}$ to $b_{N1}$, $a_{N2}$ to $b_{N2}$. Notice that we made use of some quantities defined in (\ref{A2eOp}) and (\ref{A3e3p20}).
\end{widetext}

From (\ref{A3e3p43}) it follows that secular terms disappear from $Y_{m3}(t_{1N},t_{2},t_{3})$ for all \ $t_{1N},t_{2},t_{3} \geq 0$ \ if and only if 
\begin{eqnarray}
\label{A3e3p44b}
&& H_{10}(t_{2},t_{3}) = H_{20}(t_{2},t_{3}) = L_{0}(t_{2},t_{3}) = P_{0}(t_{2},t_{3}) = 0 \ , \cr
&& (t_{2},t_{3} \geq 0).
\end{eqnarray}
Using the definitions of $H_{10}$, $H_{20}$, $L_{0}$, and $P_{0}$ in (\ref{JJ}) and the fact that $\omega_{N}[\tilde{\tilde{q}}(t_{2})]$ and $W_{m}(t_{2})$ are positive (see (\ref{A2eOp}) and item 1 of Section III), one finds that (\ref{A3e3p44b}) is equivalent to
\begin{eqnarray}
\label{A3e3p44}
&& a_{m2}(t_{2},t_{3})W_{m}'(t_{2}) = b_{m2}(t_{2},t_{3})W_{m}'(t_{2}) = 0 \ , \cr
&& L_{0}(t_{2},t_{3}) = P_{0}(t_{2},t_{3}) = 0 \ , \ \ (t_{2},t_{3} \geq 0) \ .
\end{eqnarray}
From (\ref{A3e3p26}) and (\ref{A2eOp}) one has, in general,
\begin{eqnarray}
\label{A3e3p45}
a_{m2}(t_{2},t_{3}) &\not=& 0 \ , \ \ b_{m2}(t_{2},t_{3}) \not= 0 \ , \ \ W_{m}'(t_{2}) \not= 0 \ . \qquad
\end{eqnarray}
From (\ref{A3e3p45}) one observes that the first two equalities in (\ref{A3e3p44}) involving $a_{m2}(t_{2},t_{3})W_{m}'(t_{2})$ and $b_{m2}(t_{2},t_{3})W_{m}'(t_{2})$ cannot be satisfied. The origin of this difficulty is that there are actually many fast time-scales involved in the problem, one associated to each (instantaneous) mode. In fact, in Appendix B it was shown that the fast time-scale $t_{1N}(\tau)$ should be chosen to be
\begin{eqnarray}
\label{rapidab}
\int_{0}^{\tau}d\tau' \ \omega_{m}\left[ \tilde{\tilde{q}}(\epsilon_{\mbox{\tiny pert}}\tau' ) \right] \ .
\end{eqnarray}
In other words, each (instantaneous) mode has its own fast time-scale given by (\ref{rapidab}). This is physically reasonable because each mode has its own angular frequency $\omega_{m}[ \tilde{\tilde{q}}(\epsilon_{\mbox{\tiny pert}}\tau' ) ]$. We chose the fast time-scale in both Appendix B and this appendix to be (\ref{A3e3p1p1}) because we considered initial conditions such that only (instantaneous) mode $N$ is initially excited. This choice allowed us to obtain a two-term approximation in Appendix B and in this appendix. If one wants a three-term or higher approximation in Appendix B or in this appendix, then one encounters the problem stated in (\ref{A3e3p44}) and (\ref{A3e3p45}). Now, one encounters this problem only for (instantaneous) modes $m \not= N$ and for the aforementioned higher order approximations because these other modes are not initially excited and in these higher order approximations these other modes begin to oscillate at their respective frequencies. 

One can remedy this difficulty in three ways:
\begin{enumerate}
\item[(i)] $\tilde{q}''(0) = 0$ or, equivalently $\tilde{\tilde{q}}''(0) = 0$ (see (\ref{4})):

One can consider the case where the membrane is moved by an external agent in such a way that $\tilde{\tilde{q}}''(0) = 0$. For example, see (\ref{Ejemplod2q}) for possible trajectories of the membrane that satisfy this condition. Substituting $\tilde{\tilde{q}}''(0) = 0$ in (\ref{A3e3p26}) one concludes that
\begin{eqnarray}
\label{Nuevo1}
a_{m2}(0,0) = b_{m2}(0,0) = 0 \ \ (m\not= N).
\end{eqnarray}
Hence, one can take 
\begin{eqnarray}
\label{Nuevo2}
a_{m2}(t_{2},t_{3}) = b_{m2}(t_{2},t_{3}) = 0 \ \ (m\not= N).
\end{eqnarray}
From (\ref{Nuevo2}) and the definitions of $L_{0}$ and $P_{0}$ in (\ref{JJ}) one concludes that (\ref{A3e3p44}) is satisfied and, consequently, secular terms are eliminated from $Y_{m3}$ $(m\not= N)$.

\item[(ii)] $\Gamma_{mN}\left[ \tilde{q}(0) \right] = 0$ for all \ $m\not=N$ or, equivalently, $\Gamma_{mN}\left[ \tilde{\tilde{q}}(0) \right] = 0$ \ for all \ $m\not=N$ (see (\ref{4})):

One could consider special electric susceptibilities $\tilde{\chi}[\xi - \tilde{q}(\tau)]$ such that \ $\Gamma_{mN}\left[ \tilde{\tilde{q}}(0) \right] = 0$ \ for all \ $m\not=N$. Substituting this into (\ref{A3e3p26}) leads to (\ref{Nuevo1}) and, consequently, one can take (\ref{Nuevo2}). Using the same argument as in item (i) above, it follows that secular terms are eliminated from $Y_{m3}$ $(m\not= N)$.

\item[(iii)] $W_{m}'(t_{2}) \simeq 0$:

From (\ref{A3e104}) and (\ref{A3e108}) one can observe that $Y_{m0}(t_{1N},t_{2},t_{3}) = 0$ and that $Y_{m1}(t_{1N},t_{2},t_{3})$ is proportional to $\{ \omega_{m}[\tilde{\tilde{q}}(t_{2})]^{2} - \omega_{N}[\tilde{\tilde{q}}(t_{2})]^{2} \}^{-1}$ $(m\not=N)$. In item 1 Section III it is stated that \ $\omega_{m}[\tilde{\tilde{q}}(t_{2})] \rightarrow + \infty$ \ if \ $n \rightarrow + \infty$. Hence, $Y_{m1}(t_{1N},t_{2},t_{3})$ will be negligible except for $m\not=N$ such that $\omega_{m}[\tilde{\tilde{q}}(t_{2})]$ is in a small band around $\omega_{N}[\tilde{\tilde{q}}(t_{2})]$. Given this observation, we can assume that the membrane interacts only with a narrow range of frequencies around $\omega_{N}[ \tilde{\tilde{q}}(t_{2}) ]$. It then follows that
\begin{eqnarray}
\label{ApproxProblema}
W_{m}(t_{2}) \ = \ \frac{\omega_{m}\left[ \tilde{\tilde{q}}(t_{2}) \right]}{\omega_{N}\left[ \tilde{\tilde{q}}(t_{2}) \right]}  \ \simeq \ 1 \qquad (t_{2} \geq 0) ,
\end{eqnarray}
so that
\begin{eqnarray}
\label{A3e3p46}
W_{m}'(t_{2}) \ \simeq \ 0 \qquad (t_{2} \geq 0) .
\end{eqnarray}
We mention that we prefer not to make the approximation that all frequencies in the narrow band are equal, that is, the approximation that
\begin{eqnarray}
\label{ApproxProblemab}
W_{m}(t_{2}) \ = \ \frac{\omega_{m}\left[ \tilde{\tilde{q}}(t_{2}) \right]}{\omega_{N}\left[ \tilde{\tilde{q}}(t_{2}) \right]} \ = \ 1 \qquad (t_{2} \geq 0) ,
\end{eqnarray}
for all frequencies in the narrow band. The reason for this is that we actually do know that the frequencies are different (see Section III and item (e) in Section VB) and we consider that assuming that they are equal would alter the system in a fundamental way. For example, the quasi-resonance factor  \ $\{ \omega_{m}[\tilde{\tilde{q}}(t_{2})]^{2} - \omega_{N}[\tilde{\tilde{q}}(t_{2})]^{2} \}^{-1}$  \ included in $Y_{m1}$ $(m\not= N)$ in (\ref{A3e108}) would not appear if $\omega_{m}[\tilde{\tilde{q}}(t_{2})] = \omega_{N}[\tilde{\tilde{q}}(t_{2})]$ and it would be replaced by another term that is defined for $\omega_{m}[\tilde{\tilde{q}}(t_{2})] = \omega_{N}[\tilde{\tilde{q}}(t_{2})]$ in a similar way to a harmonic oscillator with resonant driving \cite{Ross}.

With the approximation in (\ref{A3e3p46}) it follows from (\ref{A3e3p44}) that secular terms disappear from $Y_{m3}(t_{1N},t_{2},t_{3})$ $(m\not= N)$ for all \ $t_{1N},t_{2},t_{3} \geq 0$ \ if and only if 
\begin{eqnarray}
\label{Nuevo3}
&& L_{0}(t_{2},t_{3}) = P_{0}(t_{2},t_{3}) = 0 \ , \ \ (t_{2},t_{3} \geq 0) \ .
\end{eqnarray}
Solving the resulting equations for $a_{m2}$ and $b_{m2}$ one concludes that
\begin{eqnarray}
\label{Nuevo4}
&& z(t_{2},t_{3}) \ = \
z(0,t_{3})
\alpha_{m}\left[ \tilde{\tilde{q}}(t_{2}) \right] \ , \ \ \ (t_{2},t_{3} \geq 0) , \cr
&& \mbox{with} \ \ z=a_{m2},b_{m2}.
\end{eqnarray}
Notice that $\alpha_{m}\left[ \tilde{\tilde{q}}(t_{2}) \right]$ is defined in (\ref{A2eOp}).

\end{enumerate}

\noindent
Before proceeding, we emphasize that choosing $t_{1N}(\tau)$ given in (\ref{A3e3p1p1}) at least allows one to obtain a one- and two-term approximation using three time-scales without making any further assumptions or approximations. The problem appears when one requires an $n$-term approximation with $n\geq 3$.

We are now going to give a summary of the results obtained above. We use quantities defined in (\ref{A2eOp}).
\\
\\ 
(a) \ A first-term approximation to $d_{n}(t_{1N},t_{2},t_{3})$ is given by
\begin{eqnarray}
\label{Uno1}
&& d_{N}(t_{1N},t_{2},t_{3}) \ \simeq \ Y_{N0}(t_{1N},t_{2},t_{3})  \ , \cr
&& \cr
&& = \alpha_{N}\left[ \tilde{\tilde{q}}(t_{2}) \right] \Big[ b_{N0}(0,0)^{*}e^{it_{1N}} + b_{N0}(0,0)e^{-it_{1N}} \Big]\ , \cr
&& \cr
&& \cr
&& d_{m}(t_{1N},t_{2},t_{3}) \ \simeq \ Y_{m0}(t_{1N},t_{2},t_{3}) \ = \ 0  \ , \ \  (m\not= N), \cr
&&
\end{eqnarray}
with
\begin{eqnarray}
\label{Uno2}
b_{N0}(0,0) &=& \frac{g_{0N}}{2} + \frac{ig_{1N}}{2\omega_{N}\left[ \tilde{\tilde{q}}(0) \right]} \ .
\end{eqnarray}
This follows from (\ref{A3e3p9}) and the results in (\ref{A3e104}), (\ref{A3e105}), and (\ref{A3ea0b0F}).
\\
\\
(b) \ A two-term approximation to $d_{n}(t_{1N},t_{2},t_{3})$ is given by
\begin{eqnarray}
\label{Dos1}
&& d_{N}(t_{1N},t_{2},t_{3}) \ \simeq \ Y_{N0}\ + \ \epsilon_{\mbox{\tiny pert}} Y_{N1} \ , \cr
&& \cr
&& = \ \alpha_{N}\left[ \tilde{\tilde{q}}(t_{2}) \right] \times \cr
&& \ \ \times \left\{ b_{N0}(0,0)^{*}e^{it_{1N}}\left[ 1 + \epsilon_{\mbox{\tiny pert}}\int_{0}^{t_{2}} dt_{2}' \ \eta(t_{2}') \right] \right. \cr
&& \left. \qquad\qquad + b_{N0}(0,0)e^{-it_{1N}}\left[ 1 - \epsilon_{\mbox{\tiny pert}}\int_{0}^{t_{2}} dt_{2}' \ \eta(t_{2}') \right] \right\} \ , \nonumber
\end{eqnarray}
\begin{eqnarray}
\label{NNuevo}
&& d_{m}(t_{1N},t_{2},t_{3}) \ \simeq \ Y_{m0} \ + \ \epsilon_{\mbox{\tiny pert}} Y_{m1}\ , \cr
&& \cr
&& = \ \epsilon_{\mbox{\tiny pert}} Y_{m1}(t_{1N},t_{2},t_{3}) \ , \cr
&& \cr
&& = \ -i \epsilon_{\mbox{\tiny pert}} \tilde{\tilde{q}}'(t_{2}) \frac{2 \Gamma_{mN}\left[ \tilde{\tilde{q}}(t_{2}) \right] \omega_{N} \left[ \tilde{\tilde{q}}(t_{2}) \right] }{\omega_{m}\left[ \tilde{\tilde{q}}(t_{2}) \right]^{2} - \omega_{N}\left[ \tilde{\tilde{q}}(t_{2}) \right]^{2}} \alpha_{N}\left[ \tilde{\tilde{q}}(t_{2}) \right] \times \cr
&& \cr
&& \ \ \times  \left[ b_{N0}(0,0)^{*}e^{it_{1N}} - b_{N0}(0,0)e^{-it_{1N}} \right] \qquad (m\not= N). \cr
&&
\end{eqnarray}
This follows from (\ref{A3e3p9}), item (a) above, and the results in (\ref{A3e108}), (\ref{A3e114}), and (\ref{A3e125b}).
\\
\\
(c) \ A three-term approximation to $d_{n}(t_{1N},t_{2},t_{3})$ is given by
\begin{eqnarray}
\label{A3e3p53}
d_{n}(t_{1N},t_{2},t_{3}) 
&\simeq& Y_{n0}(t_{1N},t_{2},t_{3}) \ + \ \epsilon_{\mbox{\tiny pert}} Y_{n1}(t_{1N},t_{2},t_{3}) \cr
&&  + \ \epsilon_{\mbox{\tiny pert}}^{2} Y_{n2}(t_{1N},t_{2},t_{3})  \ ,
\end{eqnarray}
where the following terms must be added to the right-hand sides in (\ref{NNuevo})
\begin{eqnarray}
\label{A3p3p53b}
&& \epsilon_{\mbox{\tiny pert}}^{2} Y_{N2}(t_{1N},t_{2},t_{3}) \cr
&& = \ \epsilon_{\mbox{\tiny pert}}^{2} \alpha_{N}\left[ \tilde{\tilde{q}}(t_{2}) \right] \Big\{ \ e^{it_{1N}}\left[ b_{N2}(0,t_{3})^{*} + b_{N0}(0,0)^{*}\psi_{1}(t_{2}) \right]  \  \cr
&& \qquad\qquad\qquad \ \ +  e^{-it_{1N}}\left[ b_{N2}(0,t_{3}) + b_{N0}(0,0) \psi_{3}(t_{2}) \right] \ \Big\} \ , \nonumber
\end{eqnarray}
and for \ $m\not= N$
\begin{eqnarray}
\label{Tres1}
&& \epsilon_{\mbox{\tiny pert}}^{2} Y_{m2}(t_{1N},t_{2},t_{3}) \cr
&& \cr
&& \cr
&=& \epsilon_{\mbox{\tiny pert}}^{2} \alpha_{m}\left[ \tilde{\tilde{q}}(t_{2}) \right]\Big[ a_{m2}(0,t_{3})e^{iW_{m}(t_{2})t_{1N}} \cr
&& \qquad\qquad\qquad\qquad + b_{m2}(0,t_{3})e^{-iW_{m}(t_{2})t_{1N}}  \Big] \cr
&& \cr
&& + \ \epsilon_{\mbox{\tiny pert}}^{2}\left\{ \ e^{it_{1N}}\frac{\omega_{N}\left[ \tilde{\tilde{q}}(t_{2}) \right]^{2} D(t_{2},t_{3})}{\omega_{m}\left[ \tilde{\tilde{q}}(t_{2}) \right]^{2} - \omega_{N}\left[ \tilde{\tilde{q}}(t_{2}) \right]^{2}} \right. \cr
&& \qquad\qquad \left. + e^{-it_{1N}}\frac{\omega_{N}\left[ \tilde{\tilde{q}}(t_{2}) \right]^{2} F(t_{2},t_{3})}{\omega_{m}\left[ \tilde{\tilde{q}}(t_{2}) \right]^{2} - \omega_{N}\left[ \tilde{\tilde{q}}(t_{2}) \right]^{2} } \ \right\} \ . \qquad
\end{eqnarray}
Here $D(t_{2},t_{3})$ and $F(t_{2},t_{3})$ are defined in (\ref{A3e3p20}) and \ $a_{N2}(0,t_{3}) = b_{N2}(0,t_{3})^{*}$ \ and $b_{N2}(0,t_{3})$ are obtained by solving the $\mathcal{O}(\epsilon_{\mbox{\tiny pert}}^{4})$ problem, eliminating secular terms, and applying the initial conditions in (\ref{NNN}). Moreover, (\ref{Tres1}) follows from (\ref{A3e3p9}), item (b) above, and the results in (\ref{A3e125}), (\ref{A3e3p23}), (\ref{A3e3p29}), and (\ref{Nuevo4}). We emphasize that three- and higher-term approximations are deduced with either one of the following two conditions (see the discussion following (\ref{A3e3p45})):
\\
\\
(c.1) $\tilde{\tilde{q}}''(0) = 0$ \ or \ $\Gamma_{mN}[\tilde{\tilde{q}}(0)] = 0$ \ for all \ $m\not= N$:
\\
\\
In this case one has from (\ref{Nuevo2}) 
\begin{eqnarray}
\label{Tres1p1}
a_{m2}(0,t_{3}) \ = \ b_{m2}(0,t_{3}) \ = \ 0 \ , \ \ (t_{3}\geq 0).
\end{eqnarray}
(c.2) $W_{m}'(t_{2}) \simeq 0$ \ for \ $t_{2} \geq 0$:
\\
\\
In this case \ $a_{m2}(0,t_{3}) = b_{m2}(0,t_{3})^{*}$ \ and $b_{m2}(0,t_{3})$ are obtained by solving the $\mathcal{O}(\epsilon_{\mbox{\tiny pert}}^{4})$ problem, eliminating secular terms, and applying the initial conditions in (\ref{A3e3p26}).
\\
\\
The final step is to return to the original variable $\tau$ and obtain an approximation for $c_{n}(\tau)$. In order to do this, one has to take the following sequence of steps: (i) use the definition of the time-scales and the relationship between $c_{n}$ and $d_{n}$ in (\ref{A3e3p1p1})-(\ref{A3e3p2}), (ii) use the approximations for $d_{n}$ in items (a)-(c) above, (iii) use the relationship between $\tilde{q}(\tau)$ and $\tilde{\tilde{q}}[t_{2}(\tau)]$ given in (\ref{4}), and (iv) neglect terms of order $\geq 2$ in $\tilde{q}'(\tau)$ and $\tilde{q}''(\tau)$. The last step has to be taken because the equation governing the evolution of $c_{n}(\tau)$ is correct to first order in $\tilde{q}'(\tau)$ and $\tilde{q}''(\tau)$. This is how the results of Section V are obtained. In particular, to first order in $\tilde{q}'(\tau)$ and $\tilde{q}''(\tau)$ one has the following results:
\begin{eqnarray}
\label{etaF}
\epsilon_{\mbox{\tiny pert}}^{2} \psi_{1}\left[ t_{2}(\tau)\right] 
&=& \epsilon_{\mbox{\tiny pert}}^{2} \psi_{3}\left[ t_{2}(\tau)\right] \ , \cr
&=& \tilde{q}''(\tau)\frac{\omega_{N}'\left[ \tilde{q}(\tau) \right]}{8 \omega_{N}\left[ \tilde{q}(\tau) \right]^{3}} - \tilde{q}''(0)\frac{\omega_{N}'\left[ \tilde{q}(0) \right]}{8 \omega_{N}\left[ \tilde{q}(0) \right]^{3}} \ , \cr
&& \cr
&& \cr
\epsilon_{\mbox{\tiny pert}}^{2} \psi_{2}\left[ t_{2}(\tau)\right]
&=& -i \epsilon_{\mbox{\tiny pert}}\frac{ \tilde{q}'(\tau) }{b_{N0}(0,0)^{*}}\frac{\omega_{N}'\left[ \tilde{q}(\tau) \right]}{4 \omega_{N}\left[ \tilde{q}(\tau) \right]^{2}} \ , \cr
&& \cr
&& \cr
\epsilon_{\mbox{\tiny pert}}^{2} \psi_{4}\left[ t_{2}(\tau)\right]
&=& i \epsilon_{\mbox{\tiny pert}} \frac{ \tilde{q}'(\tau) }{b_{N0}(0,0)}\frac{\omega_{N}'\left[ \tilde{q}(\tau) \right]}{4 \omega_{N}\left[ \tilde{q}(\tau) \right]^{2}} \ , 
\end{eqnarray}
\begin{eqnarray}
\label{etaF2}
&& \epsilon_{\mbox{\tiny pert}}\int_{0}^{t_{2}(\tau)} dt_{2}' \eta (t_{2}') \ = \ -i\tilde{q}'(\tau)\frac{\omega_{N}'\left[ \tilde{q}(\tau) \right]}{4 \omega_{N}\left[ \tilde{q}(\tau) \right]^{2}} \ , \cr
&& \cr
&& \epsilon_{\mbox{\tiny pert}}^{2}\left\{
\begin{array}{c}
D\left[ t_{2}(\tau) , t_{3}(\tau) \right] \cr
F\left[ t_{2}(\tau) , t_{3}(\tau) \right]
\end{array}
\right\}
 \cr
&& \cr
&=& -\tilde{q}''(\tau)\frac{\Gamma_{mN}\left[ \tilde{q}(\tau) \right]}{\omega_{N}\left[ \tilde{q}(\tau) \right]^{2}} 
\left\{
\begin{array}{c}
b_{N0}\left[  t_{2}(\tau) , t_{3}(\tau) \right]^{*} \cr
b_{N0}\left[  t_{2}(\tau) , t_{3}(\tau) \right]
\end{array}
\right\} \times \cr
&& \cr
&& \ \times \frac{\omega_{m}\left[ \tilde{q}(\tau) \right]^{2} + 3 \omega_{N}\left[ \tilde{q}(\tau) \right]^{2}}{\omega_{m}\left[ \tilde{q}(\tau) \right]^{2} - \omega_{N}\left[ \tilde{q}(\tau) \right]^{2}} \ . 
\end{eqnarray}
To end this appendix we establish one final approximation. From item (c) above and (\ref{Tres1p1}) note that the terms $e^{\pm iW_{m}(t_{2})t_{1N}}$ in $\epsilon_{\mbox{\tiny pert}}^{2}Y_{m2}$ do not appear if $\tilde{q}''(0) = 0$ or $\Gamma_{mN}[\tilde{q}(0)]=0$ for all $m\not=N$. Then, one can use \ $W_{m}'(t_{2}) \simeq 0$ \ for both cases in items (c.1) and (c.2) to obtain the following approximation:
\begin{eqnarray}
\label{etaF3}
e^{\pm i W_{m}\left[ t_{2}(\tau) \right] t_{1N}(\tau)} \ \simeq \ \mbox{exp}\left\{ \pm i \int_{0}^{\tau} d\tau ' \ \omega_{m}\left[ \tilde{q}(\tau') \right] \right\} \ . \qquad
\end{eqnarray}
This result is obtained by using the expressions for $t_{1N}(\tau)$ and $W_{m}(t_{2})$ in (\ref{A3e3p1p1}) and (\ref{A2eOp}) and the relation between $\tilde{q}(\tau)$ and $\tilde{\tilde{q}}[t_{2}(\tau)]$ in (\ref{4}) as follows:
\begin{eqnarray}
&& W_{m}'(t_{2}) \ \simeq \ 0 \ , \ \ (t_{2} \geq 0) \ \ \Rightarrow \ \ W_{m}(t_{2}) \ \simeq \mbox{constant}  \cr
&& \cr 
&\Rightarrow& W_{m}\left[ t_{2}(\tau) \right] t_{1N}(\tau) \cr
&& \ \simeq \ \int_{0}^{\tau} d\tau'  W_{m}\left[ t_{2}(\tau') \right] \omega_{N}\left\{ \tilde{\tilde{q}}[t_{2}(\tau')] \right\} \ , \cr
&& \ = \ \int_{0}^{\tau} d\tau ' \ \omega_{m}\left\{ \tilde{\tilde{q}}[t_{2}(\tau')] \right\} \cr
&& \ = \ \int_{0}^{\tau} d\tau ' \ \omega_{m}\left[ \tilde{q}(\tau') \right] \ .
\end{eqnarray}

\end{document}